\documentclass[aps,prl,twocolumn,showpacs,superscriptaddress]{revtex4-2}

\usepackage{amsmath}
\usepackage{graphicx}
\usepackage{lmodern}
\usepackage{amsmath}

\usepackage{color}
\usepackage{amssymb}
\usepackage{bm}
\usepackage{braket}
\usepackage{mathtools}

\DeclareMathOperator{\sgn}{sgn}

\usepackage[dvipsnames]{xcolor}
\usepackage[colorlinks=true]{hyperref}
\hypersetup{
    colorlinks=true,
    linkcolor=blue,
    citecolor=blue,
    urlcolor=blue
} 

\usepackage[normalem]{ulem}

\makeatletter
\def\maketitle{
\@author@finish
\title@column\titleblock@produce
\suppressfloats[t]}
\makeatother

\begin{document}

\title {
Possible Sliding Regimes in Twisted Bilayer WTe$_2$}

\author{Yi-Ming Wu}
\affiliation{Stanford Institute for Theoretical Physics, Stanford University, Stanford, California 94305, USA}
\author{Chaitanya Murthy}
\affiliation{Department of Physics and Astronomy, University of Rochester, Rochester, New York 14627, USA}
\affiliation{Department of Physics, Stanford University, Stanford, California 94305, USA}
\author{Steven A. Kivelson}
\affiliation{Department of Physics, Stanford University, Stanford, California 94305, USA}

\date{\today}

\begin{abstract}
Inspired by the observation of increasingly one-dimensional (1D) behavior with decreasing temperature in small-angle twisted bilayers of WTe$_2$ (tWTe$_2$), we theoretically explore the exotic sliding regimes that could be realized in tWTe$_2$. At zero displacement field, while hole-doped tWTe$_2$ can be thought of as an array of weakly coupled conventional two-flavor 1D electron gases (1DEGs), the electron-doped regime is equivalent to coupled {\it four}-flavor 1DEGs, due to the presence of an additional ``valley'' degree of freedom. In the decoupled limit, the electron-doped system can thus realize phases with a range of interesting ordering tendencies, including $4k_F$ charge-density-wave and charge-$4e$ superconductivity. Dimensional crossovers and cross-wire transport due to inter-wire couplings of various kinds are also discussed. We find that a sliding Luther--Emery liquid with small inter-wire couplings is  probably most consistent with current experiments on hole-doped tWTe$_2$.
\end{abstract}

\maketitle

{\it Introduction.}---%
One dimensional quantum electronic systems are the best understood examples of non-Fermi liquids, i.e.~metals without coherent electronic quasiparticles. 
Indeed, the low energy elementary excitations in a 1D electron gas (1DEG) are all bosonic, and exhibit phenomena such as power-law scaling of correlation functions and spin-charge separation \cite{emery1979,JVoit_1995,Giamarchibook,gogolin2004bosonization}. 
Besides the 1DEG, there also exist systems with quasi-1D properties in higher dimension, such as organic (super)conductors \cite{Lorenz2002,doi:10.1126/science.252.5012.1509,lebed2008physics} and Cs${}_3$As${}_3$-chain based superconductors \cite{PhysRevX.5.011013,PhysRevLett.114.147004,PhysRevB.91.020506,Tang2015}.
Moreover, while the stripe phases of underdoped cuprates are not literally quasi-1D, it has been suggested that their unusual non-Fermi liquid and non-BCS properties can be obtained by considering adiabatic continuity from an artificial strongly stripe-ordered quasi-1D limit \cite{Tranquada1995,PhysRevLett.86.4362,PhysRevB.62.3422,PhysRevB.56.6120,PhysRevLett.88.137005}.

\begin{figure}
  \includegraphics[width=8.5cm]{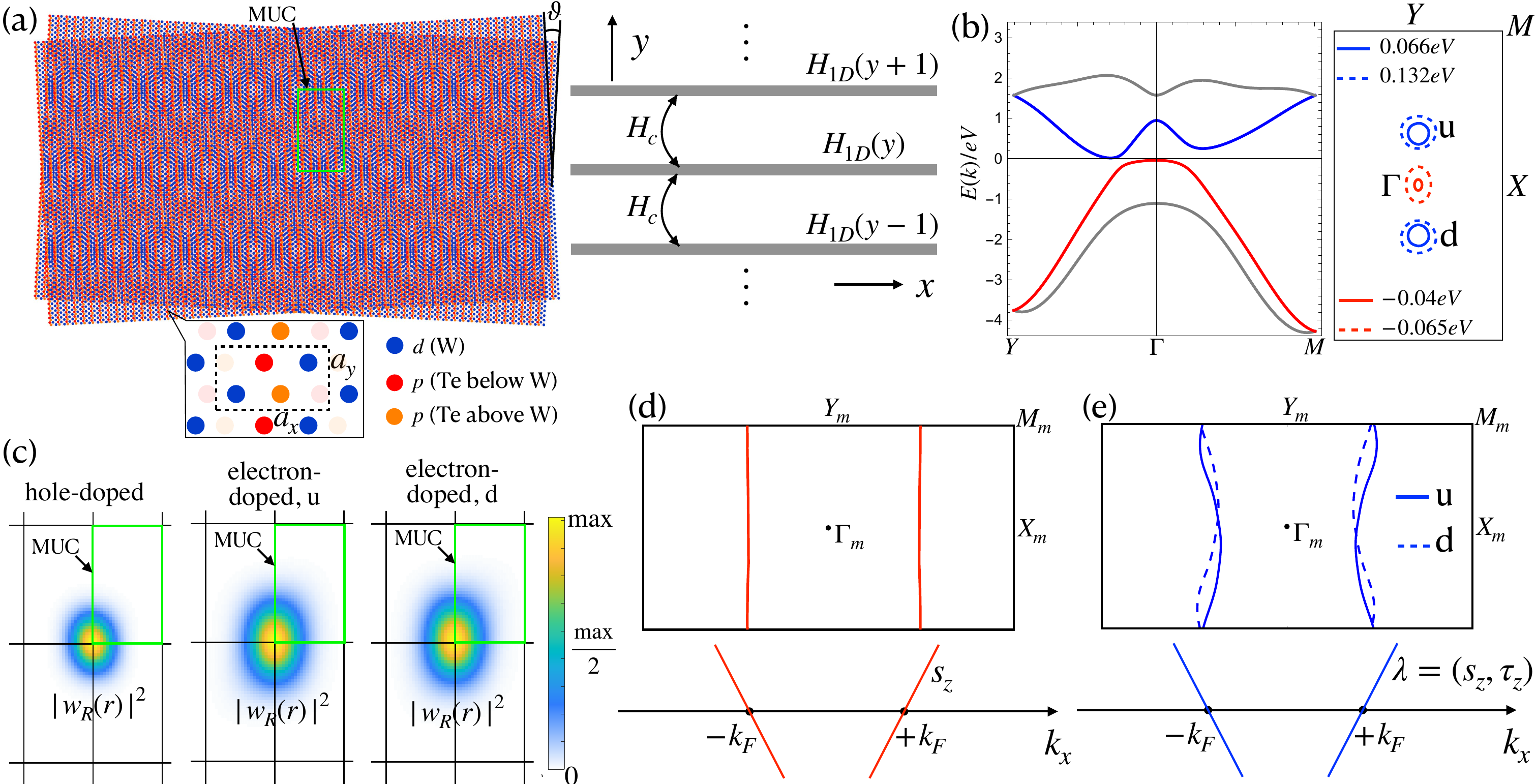}
  \caption{Emergent 1D physics in tWTe$_2$. 
  (a) Moiré pattern shows 1D stripes [the green rectangle is one moiré unit cell (MUC)], which we model as an array of weakly coupled wires. 
  (b) Left panel: 
  low energy band structure of monolayer WTe$_2$; red and blue are 
  valence and conduction bands.  
  Right panel: electron and hole pockets at the indicated Fermi energies for doped WTe$_2$. 
  {(c) 
  Localized Wannier functions for the first moiré bands of tWTe$_2$ at $\vartheta\approx 3^\circ$.}
  Typical Fermi surfaces and reduced 1D models for hole- and electron-doped tWTe$_2$ are shown in (d) and (e) respectively. 
   }\label{fig:model}
\end{figure}

At low enough temperatures, almost all quasi-1D systems ultimately crossover to states with some version of 2D coherence.  
However, there can exist ``sliding phases'' (e.g.~sliding Luttinger or Luther--Emery liquids) \cite{Kivelson1998,PhysRevLett.85.2160,PhysRevB.64.045120,PhysRevLett.86.676,PhysRevB.63.081103,PhysRevB.91.205141,PhysRevB.96.075131,PhysRevB.96.165111,GermanSierra_2003,PhysRevB.66.174424,PhysRevB.91.205141,PhysRevB.90.241101,Fleurov_2018,Begum_2019,PhysRevLett.124.136801,Du2023,hu2023twisted,PhysRevB.108.L100512,doi:10.1126/sciadv.aar8027,Dudy_2013,Chudzinski2017}
with a parametrically broad temperature window, $T_{\text{2D}} < T < T_{\text{1D}}$, within which the intra-chain quantum dynamics is highly coherent but all inter-chain effects are incoherent and perturbative.
One essential question is how far can the sliding metal phase persist down in temperature, and might there exist circumstances in which it truly persists to $T \to 0$? 
Recent experiments on small angle twisted bilayer WTe${}_2$ [tWTe$_2$, see Fig.~\ref{fig:model}(a)] have seemingly evinced an affirmative answer to this question \cite{Wang2022,Yu2023}. 
In common  with other twisted bilayer transition-metal-dichalcogenides (TMDs) \cite{Tang2020,Regan2020,Xu2020,Jin2021,Huang2021,Li2021,Wang2020,Ghiotto2021,Zhang2020,Shabani2021,Weston2020,Cai2023,zeng2023integer,Park_2023,xu2023observation,Li2021b,zhao2022realization,foutty2023mapping,Wufc2018,Wufc2019,Pan2020,Zhang2021,PhysRevLett.130.126001,PhysRevLett.131.136502,PhysRevLett.131.136501,PhysRevB.109.115111,Devakul2021,doi:10.1073/pnas.2112673118}, the properties of tWTe$_2$ are highly tunable by twist angle and gate-controlled doping.
This, combined with its anisotropic electronic structure, make it an ideal system to explore various kinds of sliding phases.

In this paper, we study tWTe${}_2$ from a theoretical perspective. 
As shown in Fig.~\ref{fig:model}(a), the stripe-like moiré pattern of tWTe${}_2$ naturally lends itself to a model of weakly coupled 1D wires.
We first explicitly construct the moiré band structures and show that both hole- and electron-doped tWTe$_2$ indeed host quasi-1D moiré bands.
In the decoupled wire limit, while the hole-regime is a relatively conventional two-flavor 1DEG, the electron-regime realizes a four-flavor 1DEG.
The existence of both spin and pseudospin flavors in the electron-regime results in a rich variety of possible favored correlations, even in the decoupled wire limit, including phases with dominant Wigner-crystal-like $4k_F$ charge-density-wave (CDW) \cite{PhysRev.46.1002,PhysRevB.39.5005,PhysRevLett.102.126402,PhysRevLett.71.1864,PhysRevB.17.494} or charge-$4e$ superconducting \cite{Affleck,Agterberg2008,Berg2009,PhysRevLett.103.010404,Agterberg2011,Fernandes2021,Jian2021,liu2023charge,wu2023dwave,PhysRevB.76.054521,PhysRevLett.95.266404,PhysRevB.82.134511,Babaev2004,BABAEV2004397} correlations. 
Moreover, from an analysis of the hole-doped regime, we conclude that the experimentally established  bound on $T_{\text{2D}}$ reported in Ref.~\cite{Yu2023}, as well as the most salient features of the anisotropic transport at the lowest accessible temperatures, can most readily be understood as reflecting the behavior of a sliding Luther--Emery (LE) liquid.

\vspace{0.4em}
{\it Model.}---%
Monolayer 1T$'$-WTe$_2$ has a rectangular unit cell, and its low energy electrons are from two $d$-orbitals on W atoms and two $p$-orbitals on Te atoms \cite{PhysRevB.99.121105,PhysRevMaterials.3.054206}, as shown in Fig.~\ref{fig:model}(a).
The symmetries of the free-standing monolayer include lattice translations, time-reversal $\mathcal{T}$, three-dimensional inversion $\mathcal{P}$, and a glide mirror $\bar{M}_y$ that sends $(x,y,z) \mapsto (x,-y+a_y/2,z)$, where $a_y$ is the lattice constant in the $y$-direction. 
The combination of $\mathcal{P}$ and $ \mathcal{T}$ ensures a two-fold (Kramers) degeneracy of each band.
Undoped WTe$_2$ is a $\mathbb{Z}_2$ topological insulator due to an $s_z$-conserving spin-orbit-coupling (SOC) \cite{Tang2017,Sanfeng2018,PhysRevX.11.041034}, while with electron doping it becomes superconducting \cite{doi:10.1126/science.aar4426,doi:10.1126/science.aar4642,PhysRevLett.125.097001,doi:10.1073/pnas.2117735119,PhysRevB.108.014509}.
In Fig.~\ref{fig:model}(b) we show the band structure and Fermi pockets at small doping of monolayer WTe$_2$ based on the tight-binding model of Ref.~\cite{PhysRevMaterials.3.054206}. 
Note that for hole doping there is a single Fermi pocket at the $\Gamma$ point, while for electron doping there are two well-separated pockets related by $\bar{M}_y \mathcal{P} \mathcal{T}$, which we label `u' and `d' respectively.

In twisted bilayer WTe$_2$, both $\mathcal{P}$ and $\bar{M}_y$ are broken, but at small twist angle $\vartheta$ they remain approximate symmetries; in this regime the distinction between $\bar{M}_y$ and regular mirror $M_y$ is also unimportant at low energies.
A displacement field breaks $\mathcal{P}$ but preserves $M_y$.
{Following Refs.~\cite{BM,PhysRevResearch.4.043151,Wufc2018,Wufc2019,PhysRevB.105.235421,SM}, we have obtained the moiré band structures of tWTe$_2$ at zero displacement field and constructed layer-averaged exponentially localized Wannier functions (LWFs) for the first moiré bands in both hole- and electron-regimes. 
Fig.~\ref{fig:model}(c) shows typical LWFs 
for $\vartheta\approx 3^\circ$, and Fig.~\ref{fig:model}(d) and (e) show examples of 
corresponding Fermi surfaces and 
reduced 1D dispersions.}

The hole regime of tWTe$_2$ can be described as a weakly-coupled array of conventional spinful 1DEGs, whose properties are characterized by 
charge and spin Luttinger parameters $K_\mathrm{c}$ and $K_\mathrm{s}$. The sliding regimes and dimensional crossovers in hole-doped tWTe$_2$ are therefore readily understood based on existing theoretical results \cite{refId0,Bourbonnais_1988,PhysRevB.42.6623,10.1007/BFb0104636,PhysRevLett.74.968,PismaZhETF.56.523,giamarchi1997mott,PhysRevB.59.12326,tsuchiizu2001commensurate}.
{The electron regime hosts fermions with an additional ``valley'' pseudospin $\tau_z = \mathrm{u}, \mathrm{d}$ (which is conserved at small $\vartheta$ since the u- and d-pockets are decoupled in this limit), and hence realizes an array of four-flavor $\lambda=(s_z, \tau_z)$ 1DEGs. 
That the LWFs are well localized on moiré lattice sites and 
particularly well separated in the $y$-direction justifies a coupled-wire view of the electronic structure; the pseudospin character of the `u' and `d' labels is reflected in the near equivalence of the corresponding LWFs.}
The symmetries  ensure that (at zero displacement field) $k_F$ is identical for all flavors in the decoupled-wire limit.

The electron-doped regime of tWTe$_2$ can in principle realize a rich variety of phases due to its four flavors, and is our main focus. 
To proceed with the analysis, we bosonize each wire via four density operators
\begin{equation}
  \begin{aligned}
    &\hat\rho_{0}(x) = \tfrac{1}{2} \big[
    \hat\rho_{\uparrow,\mathrm{u}}(x)
    +\hat\rho_{\uparrow,\mathrm{d}}(x)
    +\hat\rho_{\downarrow,\mathrm{u}}(x)
    +\hat\rho_{\downarrow,\mathrm{d}}(x) \big], \\
    &\hat\rho_{1}(x) = \tfrac{1}{2} \big[
    \hat\rho_{\uparrow,\mathrm{u}}(x)
    +\hat\rho_{\uparrow,\mathrm{d}}(x)
    -\hat\rho_{\downarrow,\mathrm{u}}(x)
    -\hat\rho_{\downarrow,\mathrm{d}}(x) \big], \\
    &\hat\rho_{2}(x) = \tfrac{1}{2} \big[
    \hat\rho_{\uparrow,\mathrm{u}}(x)
    -\hat\rho_{\uparrow,\mathrm{d}}(x)
    +\hat\rho_{\downarrow,\mathrm{u}}(x)
    -\hat\rho_{\downarrow,\mathrm{d}}(x) \big], \\
    &\hat\rho_{3}(x) = \tfrac{1}{2} \big[
    \hat\rho_{\uparrow,\mathrm{u}}(x)
    -\hat\rho_{\uparrow,\mathrm{d}}(x)
    -\hat\rho_{\downarrow,\mathrm{u}}(x)
    +\hat\rho_{\downarrow,\mathrm{d}}(x) \big].
  \end{aligned}
\end{equation}
Here the subscripts $ 0$--$3$ label, respectively, the total charge, spin, {valley, and spin-valley} 
densities. 
Their fluctuations are represented by bosonic fields $\phi_i(x)$ via $\delta\hat \rho_i(x) =- \partial_x\phi_i(x)/\pi$, with dual fields $\theta_i(x)$ defined such that $\Pi_i(x) = \partial_x\theta_i(x)/\pi$
is the conjugate momentum.

We consider the generic case in which the band filling is incommensurate, so umklapp scattering can be neglected, i.e.~the Hamiltonian is invariant under $\phi_0(x) \to \phi_0(x) + \mathrm{const}$. 
The most general form of the effective Hamiltonian of a decoupled wire, respecting the symmetries of the problem 
{($\mathcal{T}$, $\mathcal{P}$ and $M_y$, as well as conservation of total $s_z$ and $\tau_z$)} is then \cite{SM}
\begin{align}
    H_{1D} &= \sum_{i=0}^3
    \frac{v_i}{2} \int dx \left\{ \pi K_i [\Pi_i(x)]^2 + \frac{1}{\pi K_i}[\partial_x\phi_i(x)]^2 \right\} \nonumber \\
    &- {\sum_{\{ijk\}} \frac{g_i}{(\pi \alpha)^2}}
    \int dx \, \cos[2\phi_j(x)]\cos[2\phi_k(x)], \label{eq:1DH}
\end{align}
where $v_i$ are the velocities of the four bosonic fields, $K_i$ are 
Luttinger parameters incorporating all forward-scattering interactions, $g_i$ are the strengths of the leading back-scattering interactions, $\alpha$ is a short-distance cutoff, and $\{i,j,k\}$ runs over the three cyclic permutations of $\{1,2,3\}$~%
\footnote{Note that this form neglects a variety of higher order (or at least less relevant) terms, including higher-derivative terms and higher-order cosines of various sorts.}.
The values of $v_i$, $K_i$ and $g_i$ are not constrained by the symmetries; 
in particular, the absence of $SU(2)$ symmetry for both spin and 
{valley} implies there is no reason to expect the $K_i$'s to be near $1$.

Of possible inter-wire
{couplings}, a subset are forward-scattering interactions that couple the densities $\hat\rho_i$, or the conjugate currents, on neighboring wires.
In bosonized form, these interactions are marginal, and can be treated exactly; 
{they lead to} a renormalization and proliferation of effective Luttinger parameters,
{but preserve the ``sliding'' symmetry of the system under arbitrary shifts of $\phi_0$ on each wire~\cite{PhysRevLett.85.2160,PhysRevB.64.045120,PhysRevLett.86.676}.}
If large, these 
{terms can} have a qualitative effect on the dimensional crossover. 
However, these interactions are likely small {in current experimental setups} due to the screening by top and bottom gates, and hence can be neglected.

Thus, the crossover to higher-dimensional behavior 
is induced by inter-wire coupling of the form
\begin{equation}
  H_c = -\sum_a J_a \sum_{y} \int dx \ 
  O_a^\dagger(x,y) \, O_a(x,y+1) +\text{h.c.}, \label{eq:couple}
\end{equation}
where $y \in \mathbb{Z}$ labels different wires and $a$ labels different couplings.
If the decoupled wire remains gapless, single-particle hopping $J_{e}= t_\bot$ is usually the dominant process,  
{where} $O_e^\dagger(x,y)\equiv \psi^\dagger(x,y)$ is just the fermion creation operator. 
However, when there are interaction-induced gaps on each wire, $J_e$ is irrelevant at energies smaller than the gap, and higher-order processes are the key actors.
Inter-wire interactions between particle-particle (e.g.~charge-2e SC) or particle-hole (e.g.~$2k_F$ CDW) orders are generated to second order in $t_{\bot}$, giving rise to a ``bare'' value of $J_{2e} \sim J_{2k_F} \sim t_{\bot}^2/w$, where $w$ is the band width~
\footnote{For the case of CDW correlations, but not SDW or SC fluctuations, inter-chain Coulomb interactions can also generate couplings, $J_{2k_F} \sim V$.  However, there are reasons to expect these to be small, not least if the distance to metallic gates is small compared to the distance between wires.}.
We will also be interested in still higher order {terms}, especially charge-$4e$ SC and $4k_F$ CDW, both of which have bare values $J_{4e} \sim J_{4k_F}\sim t_\bot^4/w^3$.

\begin{figure}
  \includegraphics[width=8cm]{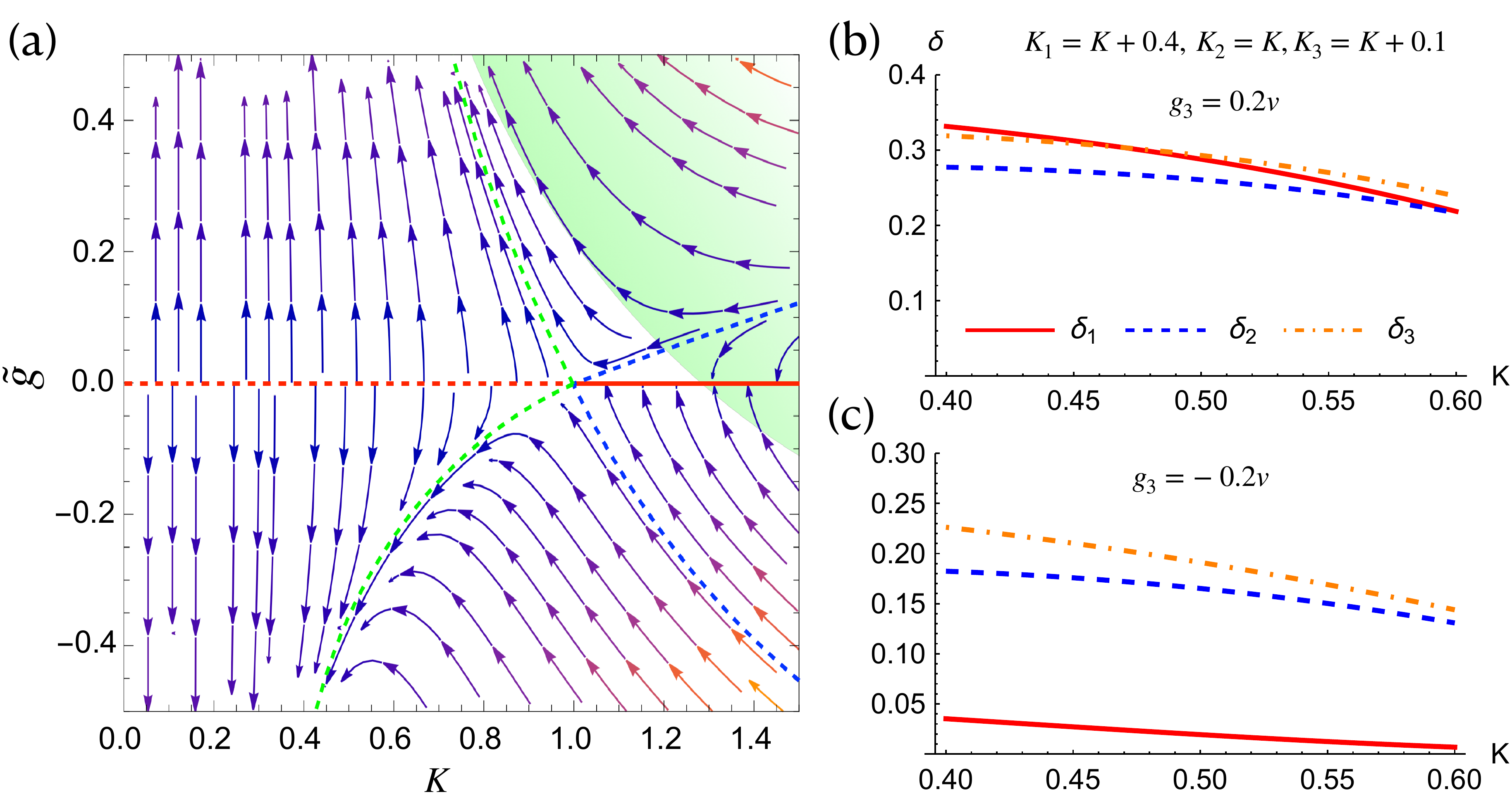}
  \caption{(a) RG flows from Eq.~\eqref{eq:RGflow} in the invariant symmetric subspace $\tilde{g}_j = \tilde{g}$ and $K_j = K$. 
  A fully gapless, stable Luttinger liquid fixed line is shown in solid red. 
  The dashed red is an unstable LL fixed line. 
  Beyond the separatrix (dotted blue) the system flows to strong coupling, indicating a maximally gapped generalized LE phase. 
  Small perturbations that break the flavor symmetries scale away (flows converge on the represented plane) within the green region. 
  (b) 
  Dimensionless gaps $\delta_i$ in the unfrustrated fully gapped case obtained from Eq.~\eqref{eq:selfconsistent}. 
  (c) Hierarchy in $\delta_i$'s in the frustrated fully gapped case.
  For simplicity we take $g_1=0.3v, g_2=0.25v$ and $v_j=v$.
}\label{fig:RG} 
\end{figure}

\vspace{0.4em}
{\it Renormalization group (RG) analysis.}---%
We first analyze the decoupled 1D wires. 
From Eq.~\eqref{eq:1DH} we see that the charge field $\phi_0(x)$ is absent from the cosine terms, and thus remains gapless, as required by the generalized Luttinger theorem \cite{Affleck}. 
If any of the cosine terms is relevant, this typically leads to the opening of a gap in the $\phi_1$, $\phi_2$, and/or $\phi_3$ sector without symmetry breaking, in which case the system can be viewed as a generalization of the LE liquid.
To quadratic order in the dimensionless couplings $\tilde{g}_i \equiv g_i/\bar{v}$, where $\bar{v} \equiv (v_1 + v_2 + v_3)/3$, the RG equations at $T \to 0$ are \cite{SM}
\begin{align}
    dK_1/dl &= -K_1^2 \big[\tilde{g}^2_2(A_3K_3+A_1K_1) + \tilde{g}^2_3(A_2K_2+A_1K_1)\big],\nonumber \\
    d\tilde{g}_1/dl &= [2-(K_2+K_3)]\tilde{g}_1 + A_1'\tilde{g}_2 \tilde{g}_3 K_1,\label{eq:RGflow}
\end{align}
and four other equations obtained by cyclic permutations of the indices $\{1,2,3\}$, where $A_j$ and $A_j'$ are positive constants that depend on the relative velocities $v_j/\bar{v}$ and on the cutoff scheme~%
\footnote{The precise values of both $A_j$ and $A_j'$ depend on the momentum cutoff procedure; see Ref.~\cite{RevModPhys.51.659} for more discussions. Here in our calculation we use $A_j \approx 0.02$ and $A_j' \approx 0.56$; see details in~\cite{SM}.}.
These non-linear equations define extremely complicated flows in a 6-dimensional coupling space, which we have extensively explored in various regimes.
We find three qualitatively distinct sorts of ``long RG-time'' ($l \gg 1$) behavior (generalized basins of attraction) of these flows:

1)
If $K_i+K_j>2$ for all $\{i,j\}$, all the $\tilde g_k$ are irrelevant, meaning that the ground state is a fully gapless Luttinger liquid (LL) with four gapless modes, each of which will generically have a different velocity.
The power law correlations in this phase are governed by four continuously tunable (marginal) parameters, $K_{0,1,2,3}$.  

2) 
If two factors $K_i+K_j>2$ and one is $< 2$, the RG flows go to strong coupling along a trajectory in which one $\tilde{g}_j$ grows while the other two tend to zero, leading to a partially gapped generalized LE phase with two of the $\phi_j$'s gapped.
For example, consider the case in which $K_1+K_2$ and $K_1+K_3 > 2$ but $K_2+K_3 < 2$. 
In this case, both $\tilde{g}_2$ and $\tilde{g}_3$ are irrelevant and only weakly renormalize $K_1$, so we need only consider the flows in the subspace $\tilde{g}_{2,3}=0$ and $K_1= \text{constant}$.
In this subspace, $\epsilon \equiv 1/K_3 - 1/K_2$ is also an invariant. 
For $\epsilon = 0$, the flows are exactly the same as those for the conventional two-flavor 1DEG, while for $\epsilon \neq 0$ they are qualitatively similar.

3) 
We have not found any condition in which only two of the $\tilde{g}_j$'s flow to strong coupling.  
Thus, the third case arises when all three flow to larger values, implying a maximally gapped generalized LE phase, in which only $\phi_0$ remains gapless, and $K_0$ determines the power-law decay of various correlation functions.
The nature of the flow to strong coupling depends qualitatively on $\eta \equiv \sgn(\tilde g_1 \tilde g_2 \tilde g_3)$.
If $\eta = 1$, there exist stable (attractive) trajectories along which all $|\tilde{g}_j|$ are equal, whereas similar trajectories are unstable if $\eta = -1$.
Indeed for $\eta = -1$ the cosine terms in Eq.~\eqref{eq:1DH} resemble antiferromagnetically coupled Ising spins on a triangular lattice, and are thus frustrated, leading to the growth of an initial hierarchy among the $|\tilde{g}_j|$'s under the RG.
In Fig.~\ref{fig:RG}(a) we show the flows within the invariant symmetric subspace $\tilde g_j=\tilde g$ and $K_j=K$, highlighting the stable region in green.

\vspace{0.4em}
{\it Self-consistent {gap equations}.}---%
Complementary to the weak coupling RG analysis, here we estimate the values of the gaps, $\Delta_j$, in the various strong-coupling limits. 
We solve the problem variationally, using a trial Gaussian action with mass terms as variational parameters \cite{SM}. 
In the case without frustration ($\eta=1$), the result is a set of coupled self-consistency conditions for the gaps: 
\begin{equation}
  \delta_1^2 = \frac{4K_1}{\pi v_1} \, \delta_1^{K_1} \! 
  \left(|g_2| \delta_3^{K_3}+ |g_3| \delta_2^{K_2} \right) , 
  \label{eq:selfconsistent}
\end{equation}
and two similar equations obtained by cyclic permutations of the indices $\{1,2,3\}$, where $\delta_i \equiv \Delta_i/(v_i/\alpha)\ll 1$ is the dimensionless gap.
The (always possible) solution to Eq.~\eqref{eq:selfconsistent} with $\delta_1=0$ is physically relevant if $K_1+K_2$ and $K_1+K_3 >2$ (i.e.~if $g_2$ and $g_3$ are both irrelevant) while the non-zero solution is physical otherwise. 
Applying this same logic to all three gap equations we find, consistent with the RG results, that two of the three modes are gapped if only one $g_j$ is relevant; 
for instance, if only $g_1$ is relevant, then $\delta_1 = 0$ and $\delta_2 \sim \delta_3 \sim |g_1|^{1/(2-K_2-K_3)}$ with $\delta_2/\delta_3 =\sqrt{K_2 v_3/K_3 v_2}$.
All three modes are gapped if either two or three of the $g_j$'s is relevant, and the induced gaps $\delta_{1,2,3}$ are generally comparable in magnitude, as shown in Fig.~\ref{fig:RG}(b).
In the case with frustration ($\eta=-1$), the gap equations are the same except that the smallest $|g_j|$ is replaced by $-|g_j|$~
\footnote{The self-consistent Gaussian analysis is more complicated in the (fine-tuned) case in which $\eta = -1$ and the smallest $|g_j|$ is not unique; we exclude this case.}.
Frustration suppresses the gaps unequally, resulting in a hierarchy in which two gaps are parametrically larger than the third [see Fig.~\ref{fig:RG}(c) and the SM for more details].

\begin{figure}
  \includegraphics[width=8.5cm]{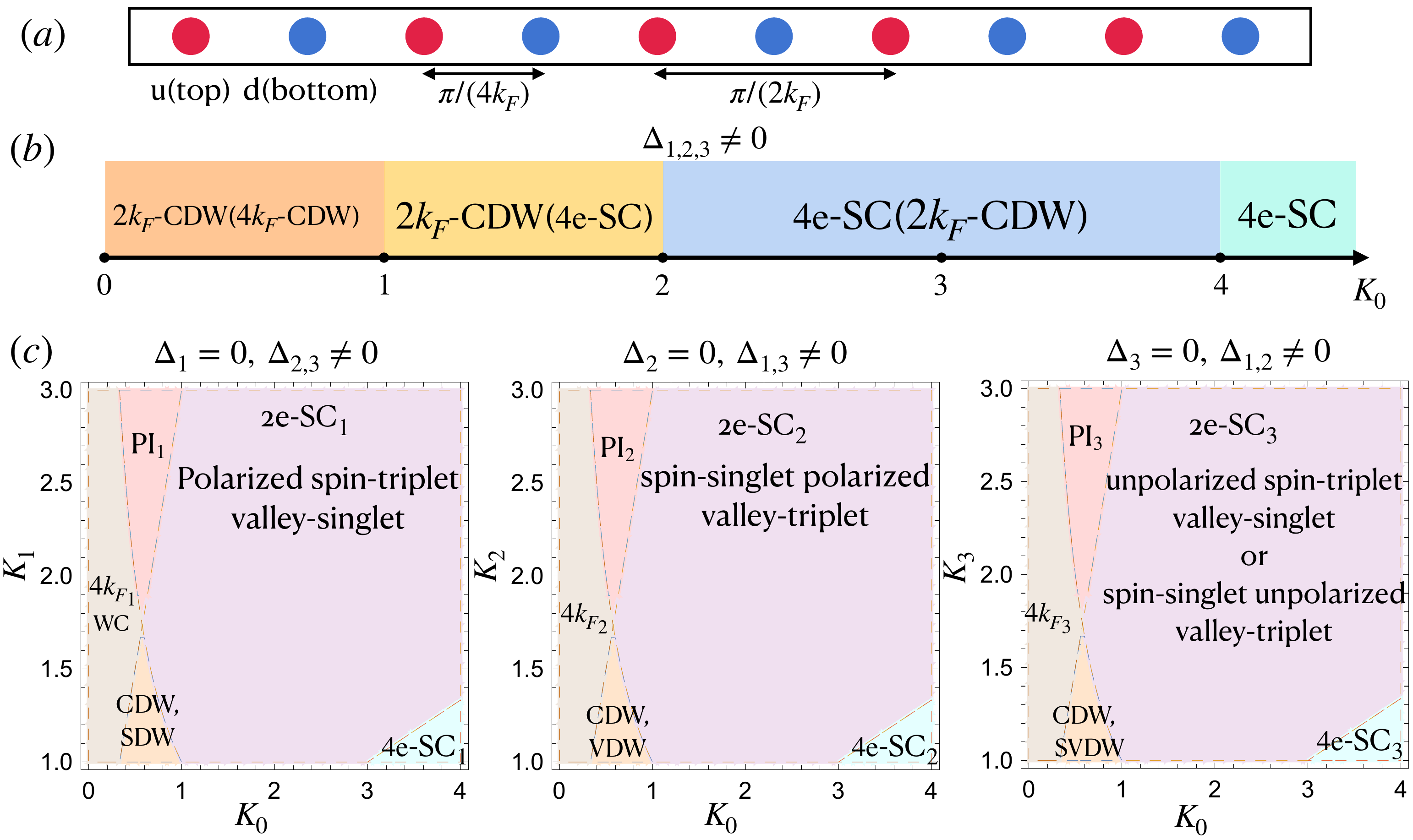}
  \caption{Phases of decoupled wires.  
  (a) Schematic of the spatial order identified as WC, where red and blue represent u and d. 
  (b) Phase diagram 
  in the fully gapped case with phases identified with the most divergent susceptibility as $T\to 0$, with subleading divergences noted in parentheses. 
  (c) Phase diagrams for partially gapped cases when the relevant $g_i>0$. 
  CDW, SDW, VDW and SVDW are abbreviations for charge, spin, valley and spin-valley $2k_F$ density waves respectively.
  }\label{fig:phase}
\end{figure}

\vspace{0.4em}
{\it Order parameters.}---%
We now identify the order parameters with the slowest-decaying correlations, which determine 
the 2D long-range order at $T \to 0$ when inter-wire couplings are included. 
A complete analysis of all such order parameters, up to fourth-order in fermion operators, is presented in the SM \cite{SM}.
In different regimes, we find the leading orders to be various kinds of momentum-$2k_F$ and -$4k_F$ density waves ($O_{2k_F}\sim\psi_R^\dagger\psi_L$, $O_{4k_F}\sim\psi_R^\dagger\psi_R'^\dagger\psi_L\psi_L'$, where $\psi_{R,L}$ are right or left moving fermions with flavor indices omitted), charge-$2e$ and -$4e$ superconductivities ($O_{2e}\sim \psi_R\psi_L$, $O_{4e}\sim O_{2e} O_{2e}'$), and zero momentum (Pomeranchuk) instabilities ($O_{\text{PI}}\sim O_{2e}^\dagger O_{2e}'$).
Some representative examples in bosonized form are
\begin{align}
    &O_{2k_F\text{-CDW}} \propto \left(\textstyle\prod_{j=1}^3\cos\phi_j-i\prod_{j=1}^3\sin\phi_j\right)e^{i\phi_0-i2k_Fx}, \notag \\
    &O_{2e,j}\propto \cos(\phi_k\pm\phi_i)e^{i(\theta_0\pm\theta_j)}, \notag \\
    &O_{{4k_F}_j} \propto  \cos(2\phi_j)e^{i2\phi_0-i4k_Fx}, \notag \\
    &O_{4e,j} \propto \cos(\phi_k+\phi_i)\cos(\phi_k-\phi_i)e^{i2\theta_0} \notag \\
    &O_{\text{PI},j}\propto \cos(\phi_k+\phi_i)\cos(\phi_k-\phi_i)e^{i2\theta_j} .
    \label{eq:order}
\end{align}
Another interesting $4k_F$-order, which we identify as a ``Wigner crystal'' order $O_{\text{WC}} \propto \sin(2\phi_{1})e^{i2\phi_0-i4k_Fx}$, is invariant under the combination of $x\to x+\pi/(4k_F)$ and inversion, as shown in Fig.~\ref{fig:phase}(a).

When $\phi_{1,2,3}$ are all gapped, the connected (fluctuation) correlation functions involving $\phi_j$, and the full correlation functions involving the dual fields $\theta_j$, decay exponentially with distance \cite{Voit1998}. 
Hence, all order parameters containing $e^{i\theta_j}$ become irrelevant, as do those containing $\sin\phi_j$ or $\cos\phi_j$, depending on whether $\langle \phi_j \rangle = 0$ or $\pi/2$, which in turn depends on $\sgn(g_i)$.
Therefore, the only leading order parameters in this case are $2k_F$-DW or $4e$-SC; these have scaling dimensions $K_0/4$ and $1/K_0$ respectively, so that the corresponding susceptibilities are
\begin{equation}
\begin{aligned}
   \chi_{2k_F} &\sim \Delta_1^{K_1/2} \Delta_2^{K_2/2} \Delta_3^{K_3/2} \, T^{K_0/2 - 2} , \\
   \chi_{\text{4e},i} &\sim \Delta_j^{2K_j} \Delta_k^{2K_k} \, T^{2/K_0 - 2} . \label{eq:1Dchi}
 \end{aligned}
\end{equation}
Clearly $\chi_\text{4e}$ is divergent for $K_0>1$, and is more divergent than $\chi_{2k_F}$ for $K_0>2$.  
Similar analysis shows that $\chi_{4k_F} \sim T^{2K_0-2}$ diverges for $K_0<1$, but is always less divergent than $\chi_{2k_F}$. 
The phase diagram for the fully gapped case with all $g_i > 0$ is depicted in Fig.~\ref{fig:phase}(b).

Phase diagrams for the leading orders in partially gapped cases are shown in Fig.~\ref{fig:phase}(c). 
Depending on which $g_i$ flows to strong coupling, there are three different cases, but the phase boundaries
are the same in each case. 
The specific orders, however, indeed depend on which $g_i$ is relevant and also on $\sgn(g_i)$. 
Figure~\ref{fig:phase}(c) presents the leading orders when the revelant $g_{i}>0$, with a complete discussion given in the SM \cite{SM}.

\begin{table*}
  \centering
  \begin{tabular}{lcccccc}
    \hline\hline
      flavor number&~~~~~ $\alpha_{2k_F}$~~~~~&~~~~~ $\alpha_{4k_F}$~~~~~&~~~~~~~$\alpha_{e}$~~~~~&~~~~~$\alpha_{2e}$~~~~&~~~$\alpha_{4e}$\\\hline

    2 (gapless) &$K_{\text{c}}+K_{\text{s}}$ & $4(K_{\text{c}}+K_{\text{s}})$  & $\frac{1}{4}\sum_{i=c,s}(K_i+K_i^{-1})$ & $K_{\text{s}}+K_{\text{c}}^{-1}$ & $4(K_{\text{s}}+K_{\text{c}}^{-1})$ \\
   2 (spin gapped)   &$K_{\text{c}}$ & $4K_{\text{c}}$   & -- & $K_{\text{c}}^{-1}$ & $4K_{\text{c}}^{-1}$ \\ \hline
   4 (gapless) &$\frac{1}{2}\sum_{j=0}^3 K_j$ & $2(K_0+K_j)$  & $\frac{1}{8}\sum_{i=0}^3(K_i+K_i^{-1})$ & $\frac{1}{2}(K_0^{-1}+K_i+K_j^{-1}+K_k)$ & $2(K_0^{-1}+K_i+K_k)$ \\
   4 ($\Delta_{i,j}\sim\Delta, \Delta_k=0$) & $\frac{1}{2}(K_0+K_k)$ & $2(K_0+K_k)$ & -- & $\frac{1}{2}(K_0^{-1}+K_k^{-1})$ &  $2K_0^{-1}$ \\
   4 ($\Delta_{1,2,3}\sim\Delta$)   & $\frac{1}{2}K_0$ & $2K_0$   & -- & -- & $2K_0^{-1}$ \\ \hline
  \end{tabular}
    \caption{Scaling dimensions for various interchain couplings. Those which are irrelevant in the gapped regime
   are not shown. In the 4-flavor cases, the indices $\{i,j,k\}\in \{1,2,3\}$.}
  \label{order}\label{tab:scalingdimension}
  \end{table*}

  \vspace{0.4em}
{\it Interwire effects}---%
Finally, we address the effect of weak inter-wire couplings.  
We are primarily interested in an intermediate range of temperatures, low compared to the {bandwidth of each wire}, $w$, but large compared to a dimensional crossover scale, $T_{2D}$. 
In this ``sliding regime,'' $T_{\text{2D}} < T \ll w$, the inter-wire couplings can be treated perturbatively.
The cross-wire conductivity $\sigma_\bot(T)$ has a power-law dependence on $T$, and can be expressed as a sum of single-particle and $n=$ even particle (Josephson) tunneling processes $J_{ne}$, as $\sigma_\bot = \sigma_{\bot,e} + \sigma_{\bot,2e} + \cdots$, where $\sigma_{\bot,ne} \sim |J_{ne}|^2$.
The crossover scale $T_{\text{2D}}$ marks the point where the 
most relevant coupling $J_a$ times the associated susceptibility $\chi_a(T)$ becomes order 1.
Thus, $T_{\text{2D}}$ is parametrically smaller than $w$ when the inter-wire coupling is weak.
For the cases considered in this paper \cite{Moser1998,PhysRevB.61.16393,PhysRevLett.74.4499,PhysRevB.11.2042},
\begin{equation}
 \begin{aligned}
   \frac {T_{2D}}{w} &\sim 
   \left(\frac{J_a}{w}\right)^{1/(2-\alpha)}
   \left(\frac{\Delta}{w}\right)^{\alpha'/(2-\alpha)}, \\
   \sigma_{\bot,ne}(T)&\sim\left(\frac{J_{ne}}{w}\right)^{2}\left(\frac{\Delta}{w}\right)^{2\alpha'}\left(\frac T {w}\right)^{2\alpha_{ne}-3},
 \end{aligned}\label{eq:TFL}
\end{equation}
where $\alpha=\min\{\alpha_{2k_F},\alpha_{4k_F},\alpha_{e},\alpha_{2e}, \alpha_{4e}, \dots \}$, 
with the appropriate scaling dimensions in various situations listed in Table~\ref{tab:scalingdimension}. 
If the 1D system remains gapless, $\alpha'=0$, and $\sigma_{\bot}(T)$ is usually dominated by single particle tunneling $\sigma_{\bot,e}(T)$. 
The system can crossover to Fermi liquid behavior below $T_{\text{2D}}$ if $J_{e} = t_\perp$ is the most relevant coupling.
When the system is gapped (which in the four-flavor case can mean either partially or fully gapped), $J_e$ is suppressed, and $\sigma_{\bot}(T)$ is dominated by $\sigma_{\bot,2e}$ or in some circumstances $\sigma_{\bot,4e}$. 
In this case, $\alpha' = \alpha_{2e}(\Delta=0)-\alpha_{2e}(\Delta\neq0)$ for
order parameters like $2k_F$-CDW or $2e$-SC, while $\alpha'=\alpha_{4e}(\Delta=0)-\alpha_{4e}(\Delta\neq0)$ for order parameters like $4k_F$-CDW or $4e$-SC. 
This $\alpha'$ ensures that in the gapped case, $\sigma_{\bot,ne}(T)$ at $T\sim \Delta$ smoothly crosses over to its behavior in the gapless case at $T>\Delta$.

In the fully gapped regime of the four-flavor 1DEG, Eq.~\eqref{eq:TFL} applies when all induced gaps are of the same order $\Delta$. 
However, if there is a hierarchy in the induced gaps such that $\Delta_{\text{max}} \gg \Delta_{\text{min}}$ as illustrated in Fig.~\ref{fig:RG}(c), and if in addition $\Delta_{\text{max}} \gg T_{\text{2D}} \gg \Delta_{\text{min}}$, the system cannot be distinguished from a partially gapped one, since at $T > T_{\text{2D}}$ the smaller gap $\Delta_{\text{min}}$ can never be seen. 
But if $\Delta_{\text{min}} \gg T_{\text{2D}}$, the system crosses over from one particular sliding LE liquid at $\Delta_{\text{max}} \gg T \gg \Delta_{\text{min}}$ to another fully gapped sliding LE liquid at $T \ll \Delta_{\text{min}}$.

\vspace{0.4em}
{\it Connection to experiment in hole doped WTe$_2$.}---%
If we assume the gapless (LL) scenario proposed in Ref.~\cite{Yu2023}, the observed $T$ dependence of $\sigma_\bot$ is dominated by $\sigma_{\bot,e}$, which implies $\alpha_{e}\approx2.26>2$ and hence $J_e=t_\bot$ is irrelevant.
Although the estimated $T_{\text{2D}}$ from this scenario might be consistent with experiment, one would need to assume $K_{\text{c}}$ and/or $K_{\text{s}}$ are unusually far from 1 (for instance, for $K_{\text{s}}\approx 1$ one would need $K_{\text{c}} \approx 6.9$ or $0.14$). 
This is difficult to reconcile with short range interactions \cite{schulz1995fermi}, and moreover would imply either $2k_F$-CDW or $2e$-SC would be strongly relevant. 
However, if we assume there is a spin gap, we obtain $K_{\text{c}}\approx0.44$ by comparing $\sigma_{\bot,2e}$ with Ref.~\cite{Yu2023}. 
Taking $J_{2k_F}\approx J_e^2/w\approx 0.04^2w$, $\Delta_{\text{s}}/w=0.05$ and $w\approx 4$ meV for estimation~%
\footnote{Here $w\approx4$ meV is consistent with the moiré band structure calculation in the SM. From both band structure calculation and comparing $\sigma_{\bot,e}$ to experiment (under the assumption that the system is gapless), we find $J_e\sim10^{-2}w$. The spin gap $\Delta_{\text{s}}$ is taken to be the minimal possible value that is above the highest temperature $\sim 2$K in experiment.}, 
we have $\Delta_{\text{s}}\approx2.3$K and, as long as $K_{\text{s}}>1.4$, the crossover temperature $T_{\text{2D}}<50$mK. 
In this scenario consistency with the experiment could be realized with less extreme values of $K_{\text{c}}$ and $K_{\text{s}}$.

\vspace{1em}
{\it Acknowledgements.}---We thank Sid Parameswaran, Trithep Devakul, and Sangfeng Wu for getting us interested in this problem and providing essential guidance. Y.-M.W. and C.M. acknowledge support from the Gordon and Betty Moore Foundation’s EPiQS Initiative through GBMF8686. SAK and C.M. were supported in part by the Department of Energy, Office of Basic Energy Sciences, under contract No.~DEAC02-76SF00515.

\bibliography{tBG}

\end{document}


\clearpage

\begin{widetext}

\begin{center}
  \textbf{SUPPLEMENTARY MATERIAL FOR
  \\[0.2cm]
  \large Possible Sliding Regimes in Twisted Bilayer WTe$_2$}
  \\[0.35cm]
  Yi-Ming Wu,$^{1}$, Chaitanya Murthy,$^{2,3}$ and Steven A. Kivelson$^{3}$
  \\[0.15cm]
  {\small \it $^{1}$Stanford Institute for Theoretical Physics, Stanford University, Stanford, California 94305, USA}
  \\[-1pt]
  {\small \it $^{2}$Department of Physics and Astronomy, University of Rochester, Rochester, New York 14627, USA}
  \\[-1pt]
  {\small \it $^{3}$Department of Physics, Stanford University, Stanford, California 94305, USA}

  \vspace{0.4cm}
  \parbox{0.85\textwidth}{In this Supplemental Material we 
  discuss in detail
  i) the construction of moiré narrow bands of twisted bilayer WTe$_2$ from a continuum model description, 
  ii) explicit bosonization procedures for the four-flavor one-dimensional electron gas, 
  iii) 
  various order parameters and correlation functions, 
  iv) explicit steps to obtain the renormalization group equations and self-consistent gap equations, and 
  v) the calculations for the cross-wire conductivity.} 
\end{center}

\tableofcontents

\section{I. Details of construction of moiré band structure}

In this section we first compare two existing tight-binding models for monolayer WTe$_2$ from Refs.~\cite{PhysRevB.99.121105} and \cite{PhysRevMaterials.3.054206}, and then construct the moiré band structure for small-angle twisted bilayer WTe$_2$. The DFT results in both references are consistent, but the tight binding models are a bit different. Both models have inversion symmetry and time-reversal symmetry, which ensures that each band from both models has two-fold spin degeneracy.

\subsection{A. Comparison between two existing tight-binding models of WTe$_2$} 
\label{sub:comparison_between_two_existing_tight_binding_model_of_}


The first tight binding model (model I) we consider is taken from Ref.~\cite{PhysRevB.99.121105}. The lattice model is depicted in Fig.1(a) in the main text. Note that our choice of unit cell differs from that in Ref.~\cite{PhysRevB.99.121105} by $90^\circ$ rotation.  Each unit cell contains four orbitals for each spin projection. The tight binding model is
\begin{equation}
  H(\bk)=\hat s_0\otimes \hat h(\bk)+V_\text{soc}\hat s_z\otimes\hat\sigma_z\otimes\hat l_y\label{eq:Hk}
\end{equation}
where $\hat s$, $\hat \sigma$ and $\hat l$ acts on spin, sublattice and orbital subspace respectively, and 
\begin{equation}
  \hat h(\bk)=\begin{pmatrix}
    \epsilon_d(\bk) & 0 & \tilde t_d g_{k_y}e^{ik_x a_x} &\tilde t_0 f_{k_y}\\
    0 & \epsilon_p(\bk) & -\tilde t_0 f_{k_y} & \tilde t_p g_{k_y}\\
    \tilde t_d g_{k_y}^*e^{-ik_x a_x} & -\tilde t_0 f_{k_y}^* &\epsilon_d(\bk) & 0\\
    \tilde t_0 f_{k_y}^* & \tilde t_p g_{k_y}^* & 0 & \epsilon_p(\bk)
  \end{pmatrix}
\end{equation}
with $f_{k_y}=1-e^{-ik_y a_y}$, $g_{k_y}=1+e^{-ik_y a_y}$ and 
\begin{equation}
  \epsilon_l=\mu_l+2t_l\cos[k_ya_y]+2t_l'\cos[2k_ya_y] ~~~ \text{for}~l=p,d.
\end{equation}
The parameters used for calculation are (in units of eV)
\begin{equation}
  \begin{aligned}
    &\mu_d=0.4935, ~ t_d=-0.28, ~ t_d'=0.075, \\
    &\mu_p=-1.3265, ~ t_p=0.93, ~ t_p'=0.075, \\
    &\tilde t_d= 0.52 ~, \tilde t_p =0.40, ~ \tilde t_0=1.02, ~ V_\text{soc}=0.115.
  \end{aligned}
\end{equation}
In Fig.\ref{fig:compare}(a)-(c) we show the band structure and the constant energy contours obtained from this model. The presence of the spin-orbit coupling opens a gap along the $\Gamma-$Y line. For slightly hole doped monolayers, there are two hole pockets: one is along $\Gamma$-Y line, the other is located in the opposite position. However, we emphasize that the existence of these two hole pockets does not faithfully represent the DFT results in Ref.~\cite{PhysRevB.99.121105} and the ARPES result ~\cite{Tang2017}, which show the valence band top is still at $\Gamma$ point. For the electron-doped monolayer, there are two electron-pockets located along $\Gamma$-Y line. These two pockets contribute to what we call `u' and `d' fermions, which are related by inversion.

\begin{figure}
  \includegraphics[width=16cm]{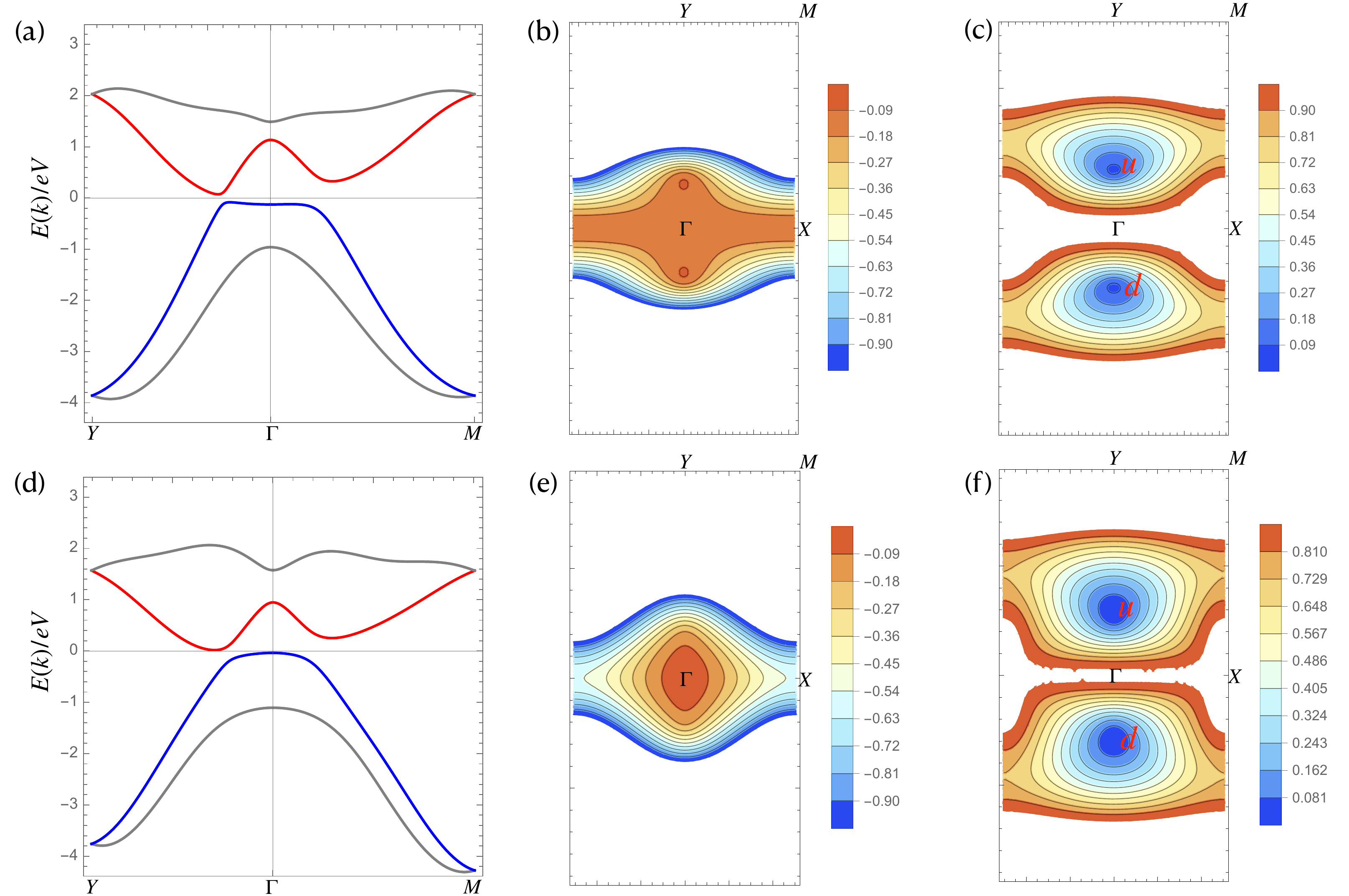}
  \caption{(a)-(c) Band structure of monolayer WTe$_2$ from the tight binding model in Ref.~\cite{PhysRevB.99.121105}. The energy dispersion along a particular trajectory $Y$-$\Gamma$-$M$ is shown in (a), and energy contours for the valence band and conduction band are shown in (b) and (c) respectively.   (d)-(f) Band structure of monolayer WTe$_2$ from the tight binding model in Ref.~\cite{PhysRevMaterials.3.054206}. The energy dispersion along the trajectory $Y$-$\Gamma$-$M$ is shown in (d), and energy contours for the valence band and conduction band are shown in (e) and (f) respectively. The conduction band for these two models is quite similar. For the valence band, the model in Ref.~\cite{PhysRevB.99.121105} indicates that the band top is shifted away from the $\Gamma$ point, while the valence band top obtained from the model in Ref.~\cite{PhysRevMaterials.3.054206} is still at the $\Gamma$ point. This difference will have a significant impact on the hole-doped moiré bands, which are constructed using the valence band dispersion. For the conduction band, both models contain two well-separated electron pockets which we dub as `u' and `d' fermions, and therefore the resulting electron-doped moiré bands from these two models are similar.}\label{fig:compare}
\end{figure}

The second tight-binding model (model II) we consider  is from Ref.~\cite{PhysRevMaterials.3.054206}, which  we also copy for convenience here. The first term is $H_0 s_0$ where 
\begin{equation}
   \begin{aligned}
        H_ {0}  &=[  \frac {\mu _ {p}}{2} +  t_ {px}  \cos  (  ak_ {x}  )+  t_ {py}  \cos ( bk_ {y}  )]  \Gamma_ {1}^ {-} 
+[  \frac {\mu _ {d}}{2}  +  t_ {dx}  \cos  ( ak_ {x}  )]  \Gamma_ {1}^ {+} \\
+&  t_ {dAB} e^ {-ibk_ {y}}  (1+  e^ {iak_ {x}}  )  e^ {i\bk\cdot \Delta _ {1}}  \Gamma_ {2}^ {+}  
+  t_ {pAB}  (1+  e^ {iak_ {x}}  )  e^ {i\bk\cdot \Delta _ {2}}   \Gamma_ {2}^- 
+  t_ {0AB}  (1-  e^ {iak_ {x}}  )  e^ {i\bk\cdot \Delta _ {3}}   \Gamma_ {3}\\  
-&  2it_ {0x}   \sin  (  ak_ {x}  )[  e^ {i\bk\cdot   \Delta_{4}}      \Gamma_ {4}^ {+}  +  e^ {-i\bk\cdot   \Delta_4}   \Gamma_ {4}^ {-}  ]
+  t_ {0ABx}  (  e^ {-iak_ {x}}  -  e^ {2iak_ {x}}  )  e^ {i\bk\cdot \Delta _ {3}}   \Gamma_ {3}+\text{h.c.}
     \end{aligned} \label{eq:modelII1}
\end{equation}
The SOC part is 
\begin{equation}
  \begin{aligned}
    H_{soc}&=[(  \lambda _ {dx}^ {z}   s_ {z}  +  \lambda _ {dx}^ {y}  s_y)  \sin  (  ak_ {x}  )]  \Gamma _ {5}^ {+}  
+[(  \lambda _ {px}^ {z}   s_ {z}  +  \lambda _ {px}^ {y}   s_ {y}  )  \sin  (  ak_ {x}  )]
  \Gamma _ {5}^ {-} \\
-&i  \lambda _ {0AB}^ {y}   s_ {y}  (1+  e^ {iak_ {x}}  )  e^ {i\bk\cdot \Delta _ {3}}   \Gamma_ {6}  
-i(  \lambda _ {0}^ {z}   s_ {z}  +  \lambda _ {0}^ {y}   s_ {y}  )(  e^ {i\bk\cdot   \Delta _ {4}}   {\Gamma_ {4}^ {+}}  -  e^ {-i\bk\cdot \Delta _ {4}}   \Gamma_ {4}^ {-}  )\\
-&i(  \lambda^{'z}_ {0}   s_ {z}  +  \lambda _ {0}^ {'y}   s_ {y}  )
(  e^ {-ibk_ {y}}   e^ {i\bk\cdot   \Delta _ {4}}     \Gamma^+-  e^ {ibk_ {y}}   e^ {-i\bk\cdot \Delta _ {4}}  
\Gamma^-)+\text{h.c.}
  \end{aligned}\label{eq:modelII2}
\end{equation}
Here $a=3.477$\r{A} and $b=6.249$\r{A}. $\Delta_1=\br_{Ad}-\br_{Bd}$, $\Delta_2=\br_{Ap}-\br_{Bp}$, $\Delta_3=\br_{Ad}-\br_{Bp}$ and $\Delta_4=\br_{Ad}-\br_{Ap}$, with $\br_{Ad}=(-0.25a,0.32b)$, $\br_{Bp}=(0.25a,0.07b)$, $\br_{Ap}=(-0.25a,-0.07b)$ and $\br_{Bd}=(0.25a,-0.32b)$.
The $\Gamma$ matrices are defined as 
\begin{equation}
  \begin{aligned}
    \Gamma_0&=\sigma_0l_0, ~ \Gamma_1^\pm=\frac{\sigma_0}{2}(l_0\pm l_3), ~ \Gamma_2^\pm=\frac{1}{4}(\sigma_1+i\sigma_2)(l_0\pm l_3), ~ \Gamma_3=\frac{1}{2}(\sigma_1+i\sigma_2)i l_2\\
    \Gamma_4^\pm&=\frac{1}{4}(\sigma_0\pm\sigma_3)(l_1+il_2), ~ \Gamma_5^\pm=\frac{\sigma_3}{2}(l_0\pm l_3), ~ \Gamma_6=\frac{1}{2}(\sigma_1+i\sigma_2)l_1.
  \end{aligned}
\end{equation}
And the parameters are 
\begin{equation}
  \begin{aligned}
    &\mu_p=-1.75, ~~~&\lambda_{0AB}^y=0.011~~~
    &\mu_d=0.74,~~~ & \lambda_0^y=0.051\\
    &t_{px}=1.13,~~~ & \lambda_0^z=0.012~~~
    &t_{dx}=-0.41,~~~& \lambda_0^{'y}=0.05\\
    &t_{pAB}=0.4,~~~ &\lambda_0^{'z}=0.012~~~
    & t_{dAB}=0.51, ~~~& \lambda_{px}^y=-0.04\\
    &t_{0AB}=0.39, ~~~& \lambda_{px}^z=-0.01~~~
    &t_{0ABx}=0.29,~~~ & \lambda_{dx}^y=-0.031\\
    &t_{0x}=0.14, ~~~& \lambda_{dx}^z=-0.008~~~
    &t_{py}= 0.13
  \end{aligned}
\end{equation}
In Fig.~\ref{fig:compare}(d)-(f) we show the band structure and the constant energy contours obtained from model II. Note in this model, the valence band top is at $\Gamma$ point, which is consistent with the DFT calculations in Ref.~\cite{PhysRevB.99.121105}, and is also consistent with the ARPES data in Ref.~\cite{Tang2017}.   For the electron-doped monolayer, there are two electron-pockets located along $\Gamma$-Y line, which agrees with model I and justifies our division of low energy fermions into `u' and `d' sectors related by inversion. Below, we will use the energy dispersions from the second tight-binding model to construct moiré band structures.

\begin{figure}
  \includegraphics[width=12cm]{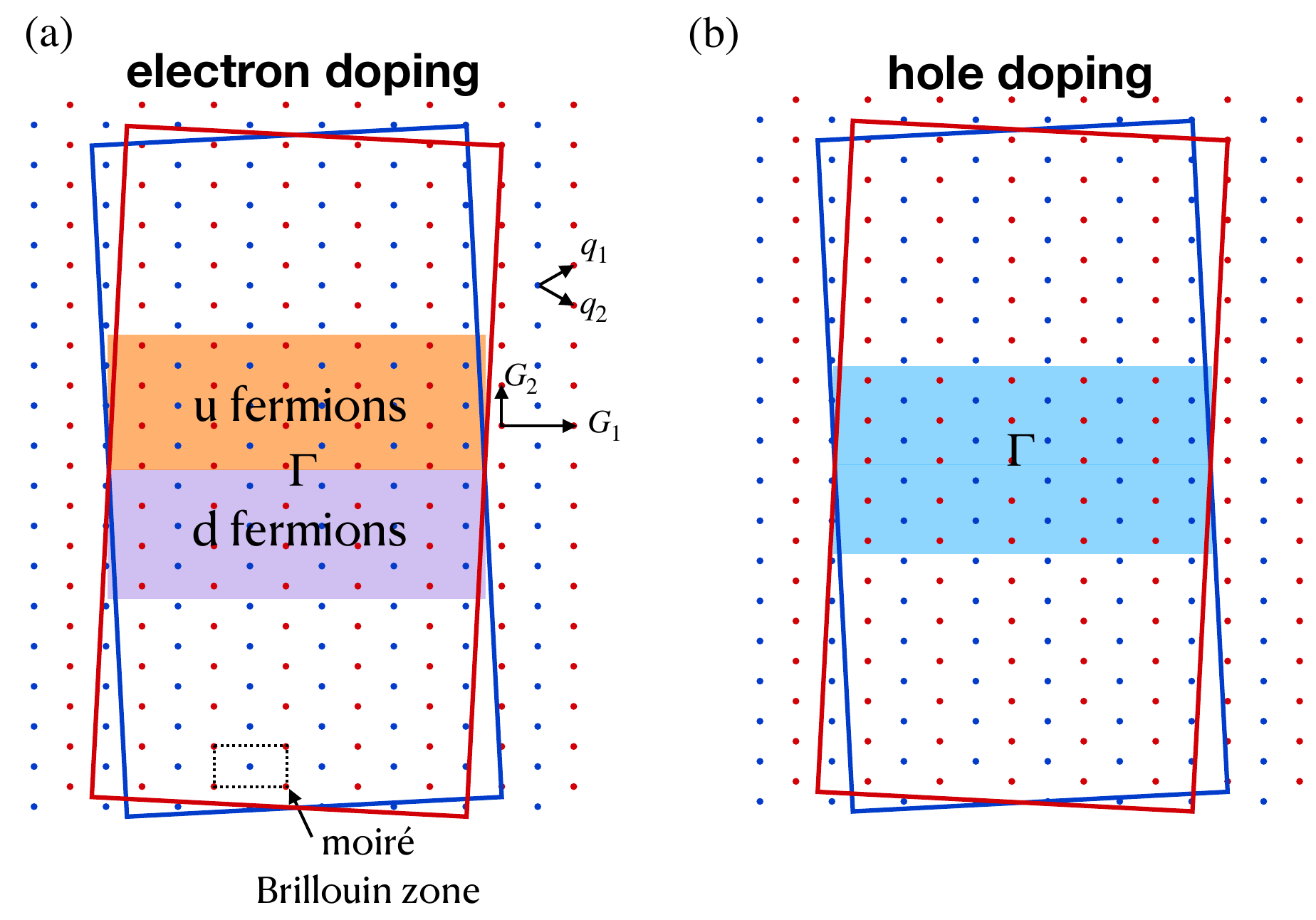}
  \caption{Details of momentum cutoff: only states inside the colored regions are kept for constructing the moiré band structures from the tight binding models. (a) For electron-doped tWTe$_2$, since the low energy fermions are located in two separate positions, we divide the monolayer Brillouin zone into two parts which we call u- and d-fermions. Moiré bands in this case are obtained using u- and d-fermions separately, assuming the coupling between them is negligible. (b) For hole-doped tWTe$_2$, we use the band fermions inside the blue region as a whole to construct moiré bands.
  }\label{fig:SM1}
\end{figure}

\subsection{B. Moiré band structures} 
\label{sub:moiré_band_structures}


We now discuss the band structures for small angle twisted bilayer WTe$_2$, using a continuum model analogous to the {one that has been used for} twisted bilayer graphene~\cite{BM} and other twisted bilayer TMD~\cite{PhysRevLett.121.026402,PhysRevLett.122.086402} systems.
In practice, we need to distinguish between electron-doped and hole-doped tWTe$_2$; 
for the hole-doped system, the low energy fermions are from 
a single hole pocket centered at the $\Gamma$ point, 
while in the  electron-doped system the low energy fermions are from both the u- and d-pockets. If we neglect the coupling between the u and d fermions, we can use each of these electron pockets to construct a moiré band. 
The resulting moiré band fermions acquire an additional pseudospin index $\tau = \text{u}, \text{d}$, which effectively becomes a 
good quantum number in addition to the physical spin, since the particle numbers for u and d fermions are individually conserved. 
In Fig.~\ref{fig:SM1} we show the momentum cutoffs used when constructing the moiré bands. {
Bloch states with momenta within the colored regions are retained for constructing the low-energy moiré bands, while those outside these regions are disregarded. More specifically,}
for hole-doped systems, we consider a large region centered at the $\Gamma$ point, while for electron-doped systems, we separately consider two regions centered at either the u- or d-pockets.

{
To be concrete, we will discuss in a bit more detail the electron-doped system. 
For simplicity we will consider at first only the u-fermions.  
The continuum limit Hamiltonian 
(with spin-degeneracy 
implicit) can be written as
\begin{equation}
  H_{\text{u}}=\begin{pmatrix}
    \varepsilon_{\text{u},t}(\bk)+\Delta_t(\br) & T(\br)\\
    T^\dagger(\br) & \varepsilon_{\text{u},b}(\bk)+\Delta_b(\br)
  \end{pmatrix} ,
\label{eq:moireH}
\end{equation}
where $\epsilon_{\tau,t}(\bk)$ 
and $\epsilon_{\tau,b}(\bk)$ are, respectively, the monolayer band dispersion rotated by $\vartheta/2$ and $-\vartheta/2$ (therefore $\epsilon_{\tau,t}$ and $\epsilon_{\tau,b}$ interchange with each other under the transformation $\vartheta \to -\vartheta$). 
Technically, 
we obtain these by first diagonalizing the tight binding Hamiltonian in Eq.~\eqref{eq:modelII1} and \eqref{eq:modelII2} on a meshgrid, rotating by $\pm\vartheta/2$, and using numerical interpolation to obtain energies at arbitrary $\bk$. The interlayer coupling
\begin{equation}
    T(\br)=2 w[\cos (\bq_1\cdot \br)+\cos(\bq_2\cdot\br)] ,
\end{equation}
where the two vectors $\bq_1$ and $\bq_2$ are shown in Fig.~\ref{fig:SM1}(a), and $w$ is the interlayer tunneling strength. 
The top and bottom layer moiré potentials can be parametrized as 
\begin{equation}
  \begin{aligned}
    \Delta_{t,b}(\br)=2 \delta[\cos (\bm{G}_1\cdot \br\pm\varphi)+\cos(\bm{G}_2\cdot\br\pm\varphi)] ,
  \end{aligned}
\end{equation}
where we have used the convention that $+\varphi$ is for the top layer and $-\varphi$ is for the bottom layer, and 
\begin{equation}
  \bm{G}_1=\bq_1+\bq_2, \qquad \bm{G}_2=\bq_1-\bq_2
\end{equation}
are two reciprocal lattice constants. In this Hamiltonian, $T(\br)$ can be thought of as the lowest harmonic effect due to the moiré pattern, which connects a blue (red) dot to its nearest red (blue) dot and thus represents the interlayer coupling; while $\Delta_{t,b}(\br)$ is the next order harmonic effect of the moiré pattern, which connects a blue (red) dot to its nearest blue (red) dot and thus represents intralayer coupling. The parameter $\varphi$ is the relative phase between the lowest-order and the second-lowest-order harmonic effect.}

\begin{figure}
  \includegraphics[width=18cm]{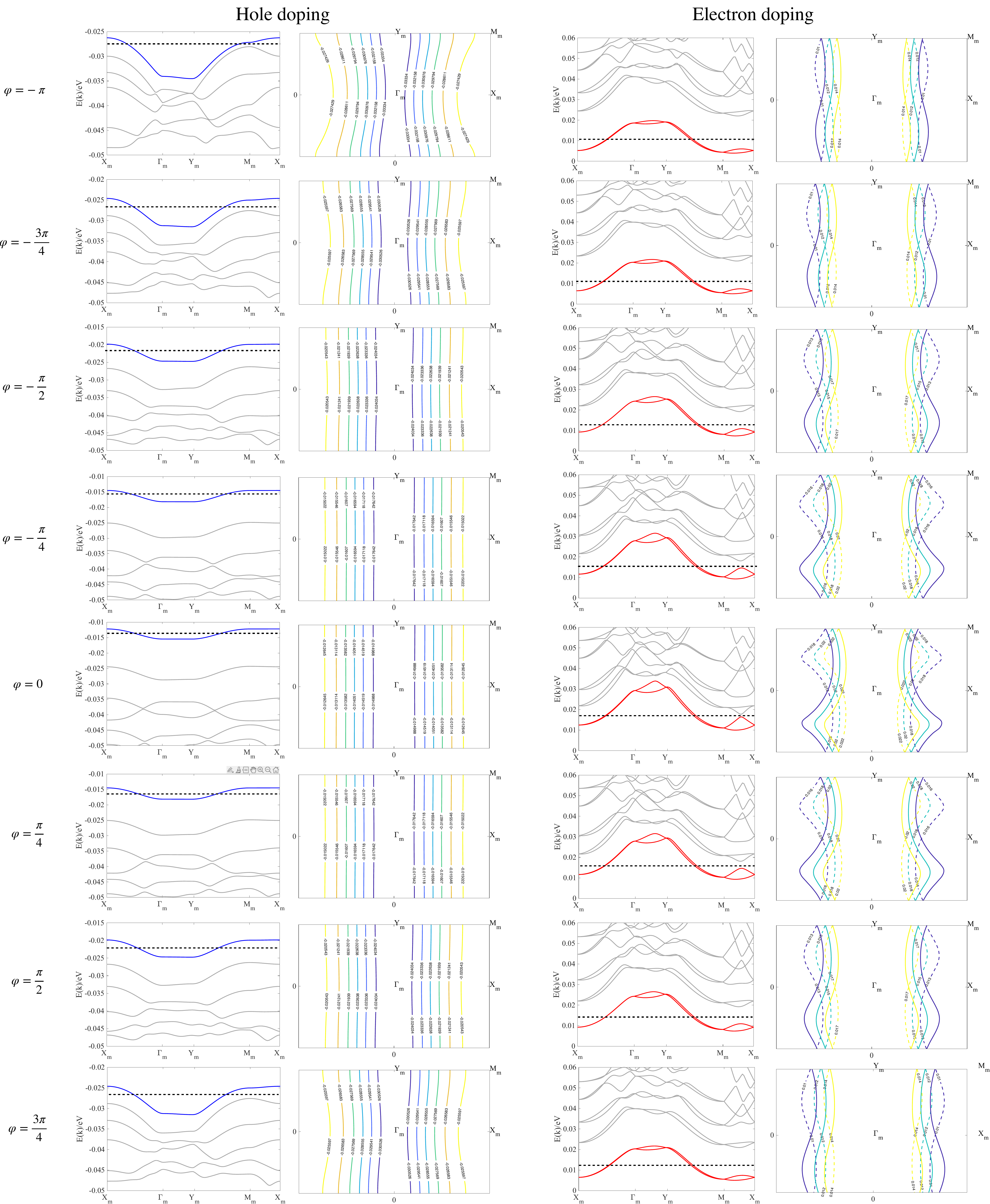}
  \caption{Emergence of quasi-1D moiré band structure from twisted bilayer WTe$_2$, obtained from model II at $\vartheta=2.89^\circ$, $w=-8meV$, $\delta=4meV$ and various values of $\varphi$ for both hole-doping and electron-doping. For each set of parameters, we show both the energy dispersion along $X_m-\Gamma_m-Y_m-M_m-X_m$ and the constant energy contours for the highlighted band in the whole mBZ. The dashed lines in the dispersion plots show examples of Fermi levels at which the system can realize a two flavor quasi-1DEG in the hole-doped regime and a four flavor quasi-1DEG in the electron-doped regime. For the constant energy contours in the electron-doped case, we use solid curves for u-fermions and dashed curves for d-fermions.}\label{fig:moire2}
\end{figure}

\begin{figure}
  \includegraphics[width=16cm]{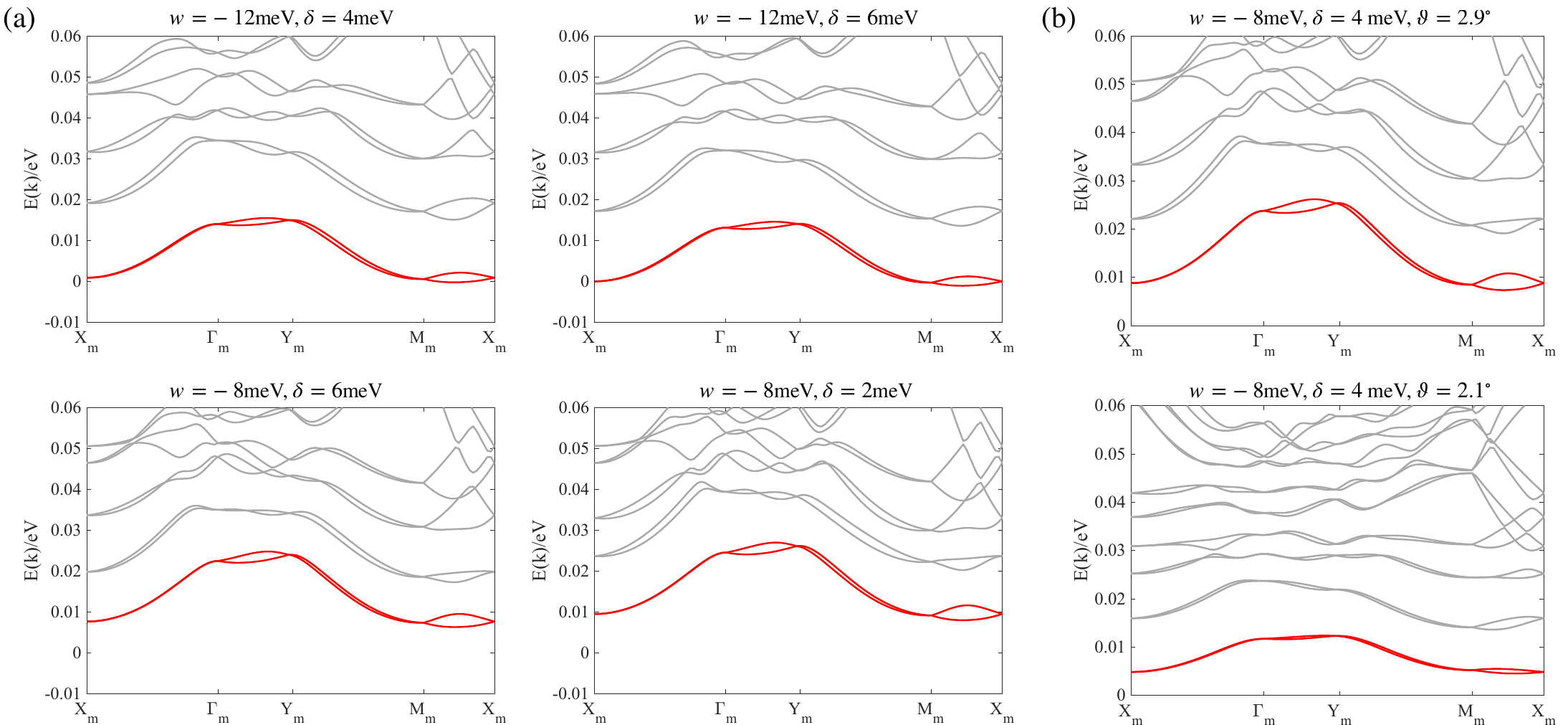}
  \caption{Moiré band structure of tWTe$_2$ for electron-doped cases, with the first two bands (from the u- and d-fermions) highlighted in red. (a) is obtained at a fixed twist angle $\vartheta=2.9^\circ$ and a fixed phase parameter $\varphi=\pi/2$, for several representative parameters $w$ and $\delta$ [see Eq.~\eqref{eq:moireH} for the definition of these parameters]. (b) Comparison between two different twist angles while fixing $w=-8$meV, $\delta=4$meV and $\varphi=\pi/2$. }\label{fig:moire4}
\end{figure}

{The moiré band structures 
of the u-fermions and d-fermions are related by inversion, as can be seen at the Hamiltonian level.
Let us discuss how the Hamiltonian in Eq.~\eqref{eq:moireH} transforms under inversion 
(with the  inversion center located 
between the two layers), which takes $(x,y,z)$ to $(-x,-y,-z)$. 
Clearly $T(\br)$ remains invariant, while
$\Delta_{t}(\br)$ and $\Delta_{b}(\br)$ get interchanged. 
Moreover, since $\bk\to-\bk$, the top layer changes to the bottom layer and $\vartheta$ changes to $-\vartheta$, and hence the dispersion 
$\varepsilon_{\text{u},t}(\bk)\to\varepsilon_{\text{d},t}(-\bk)$ ($\bk$ here is measured from the valley center). Therefore, we have
\begin{equation}
  H_{\text{u}}=\begin{pmatrix}
    \varepsilon_{\text{u},t}(\bk)+\Delta_t(\br) & T(\br)\\
    T^\dagger(\br) & \varepsilon_{\text{u},b}(\bk)+\Delta_b(\br)
  \end{pmatrix} 
  \ \leftrightarrow \
  H_{\text{d}}=\begin{pmatrix}
    \varepsilon_{\text{d},t}(-\bk)+\Delta_b(\br) & T(\br)\\
    T^\dagger(\br) & \varepsilon_{\text{d},b}(-\bk)+\Delta_t(\br)
  \end{pmatrix}
\end{equation}
under the inversion operation. As a result, the quasi-1D energy dispersions obtained from diagonalizing $H_{\text{u}}$ and $H_{\text{d}}$ are related by inversion, as discussed below. 

In Fig.~\ref{fig:moire2} we present band structure results 
for both the hole-doped and the electron-doped moiré bands obtained from model II with a set of representative parameters $w=-8$meV, $\delta=4$meV 
and twist angle $\theta=2.89^\circ$. (It is  natural to assume that $\delta$ is smaller than $w$, since $\delta$ arises from a higher order effect than $w$.) The parameter $\varphi$ is chosen to range from $-\pi$ to $\pi$.  
Similar parameters also show up in the models of twisted bilayer 2H-TMDs. These parameters can be fit 
to first-principles calculations [e.g.~density functional theory (DFT)] at zero twist angle and finite interlayer relative displacement, as was done for twisted homo-bilayer MoTe$_2$ in Ref.~\cite{PhysRevLett.122.086402}
and for the twisted hetero-bilayer WSe$_2$/MoSe$_2$ in Ref.~\cite{PhysRevLett.121.026402}. In both studies, it was found that $\varphi$ is close to $\pi/2$. 
However, another DFT study on WSe$_2$/MoSe$_2$ found different parameters (particularly, $\varphi\approx\pi/4$), see {Ref.~\cite{PhysRevB.102.201115}.}
Here we treat these parameters as phenomenological ones, as was done in the case of twisted homo-bilayer WSe$_2$ in {Ref.~\cite{PhysRevResearch.2.033087}.} In reality they depend on experimental details such as twist angle, coupling to substrate and possible displacement field and can thus presumably be tuned by changing the experimental setting.

For each set of parameters in Fig.~\ref{fig:moire2}, we show both the band dispersion along the trajectory $X_m-\Gamma_m-Y_m-M_m-X_m$ (with the first moiré bands highlighted in blue or red) and several constant energy contours for the highlighted bands. 
We clearly see that the quasi-1D character is ubiquitous in the moiré band structures of WTe$_2$, i.e. for all the chosen parameters the moiré bands are more dispersive in the $x$-direction (e.g.~along $\Gamma_m$-$X_m$) than in the $y$-direction (e.g.~along $\Gamma_m$-$Y_m$). 
In the hole doped regime, the first moiré band (highlighted in blue) can thus be modeled by coupling an array of two-flavor 1DEGs, where the flavor is the physical spin. In the electron doped regime, the first two moiré bands (highlighted in red) can be modeled by coupling an array of four flavor 1DEGs, where the flavor consists both of physical spin and pseudospin (valley), as dictated by the symmetries discussed above. 
By comparing results for different $\varphi$, we see that 
$\varphi$ mainly affects the $y$-direction dispersion, or equivalently the interwire coupling $t_\bot$, but other essential ingredients like the quasi-1D character and the inversion-symmetry-dictated relation between u- and d-fermions remain qualitatively unchanged.

In Fig.~\ref{fig:moire4} we present additional results for the electron-doped moiré band structures at a fixed $\varphi=\pi/2$. In Fig.~\ref{fig:moire4}(a) we fix $\vartheta=2.9^\circ$ and study the effect of varying the inter- and intra-layer couplings $w$ and $\delta$. We see that the bandwidth and bandgap change with varying $w$ and $\delta$, but the overall band structure remains qualitatively the same. In Fig.~\ref{fig:moire4}(b) we fix $w$ and $\delta$ and compare two different twist angles $\vartheta$. As expected, decreasing the twist angle yields a set of flatter bands, but the quasi-1D character remains intact.

}

\begin{figure}
  \includegraphics[width=16cm]{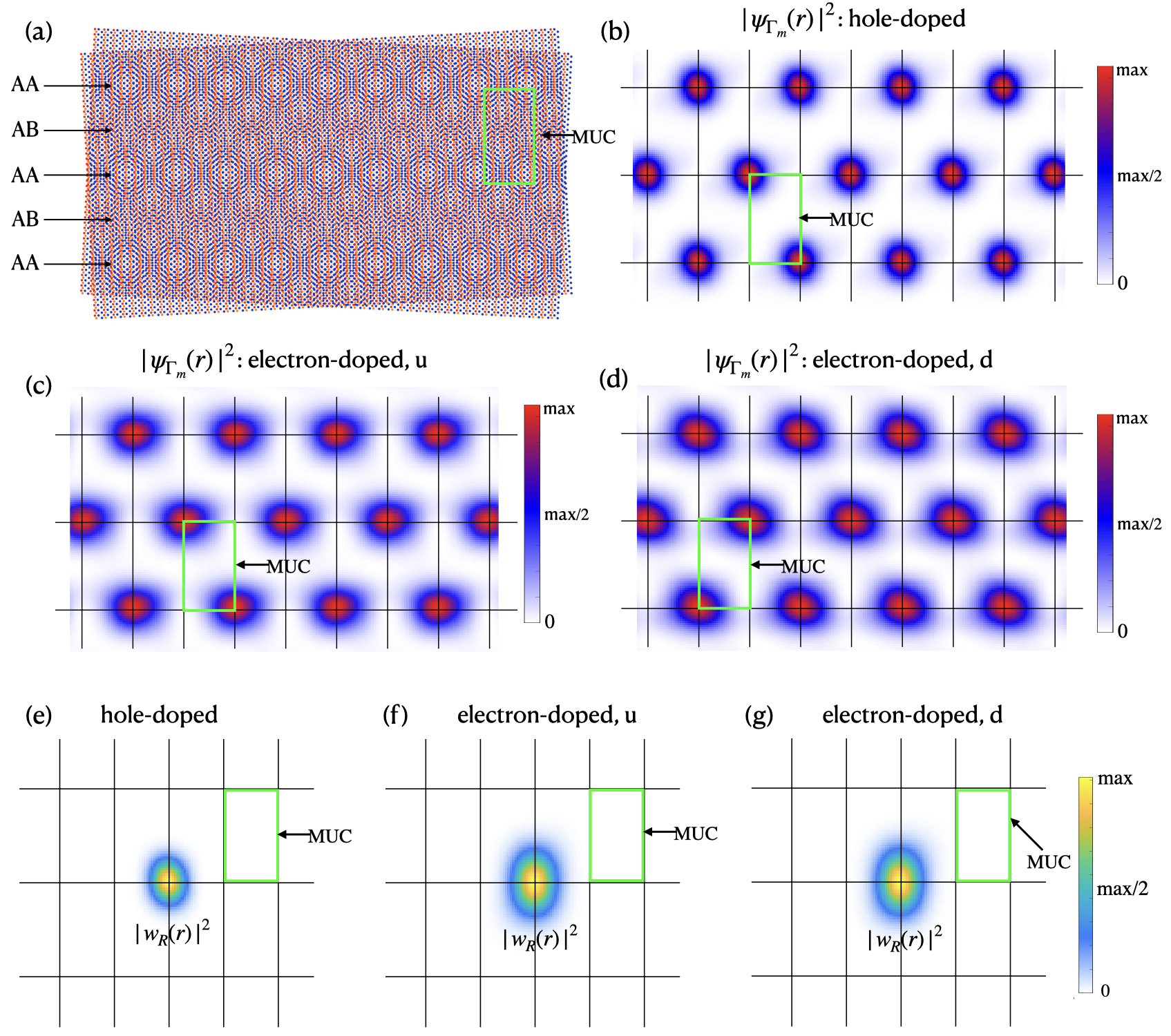}
  \caption{(a) Moiré pattern shows AA and AB stacking stripes. The green rectangle indicates one moiré unit cell (MUC). (b) Spatial configuration of the $\Gamma_m$-point Bloch wave function of the first hole-doped spin-degenerate moiré band. Here each rectangle plaquette is one moiré unit cell and 
  each vertex is a moiré lattice site.  (c) and (d) Spatial configuration of the $\Gamma_m$-point Bloch wave functions of the first two electron-doped spin-degenerate moiré u- and d-bands, respectively. (e-g) Layer-averaged exponentially localized Wannier functions ${w}_{\bm{R}}(\br)$ [see Eq.~\eqref{eq:MLWF} for definition] of the first moiré bands (for one spin projection) for (e) hole-doped regime, (f) electron-doped u-band and (g) electron-doped d-band. Here we plot the spatial configuration of $|w_{\bm{R}}(\br)|^2$ for a particular $\bm{R}$, which shows the localization of $w_{\bm{R}}(\br)$ around $\bm{R}$. The parameters used in this figure are the same as in Fig.~\ref{fig:moire2}, with $\varphi=\pi/2$. 
 }\label{fig:wannier}
\end{figure}


{Complementary to the band structure, the wavefunctions from diagonalizing the Hamiltonian in Eq.~\eqref{eq:moireH} contain information such as band topology {and quantum geometry}. We have explicitly calculated the Chern numbers of the first moiré band(s) for both hole- and electron-doped regimes, and for all the parameters that have been explored, we always 
find that the first moiré bands are {topologically} trivial.  This implies that it is possible to construct a set of 
localized Wannier orbitals from the Bloch wavefunctions.
In Fig.~\ref{fig:wannier}(b-d) we show the spatial configuration of the $\bk=\Gamma_m$ point Bloch wavefunctions $|\psi_{\Gamma_m}(\br)|^2$ corresponding to the first moiré band(s) in both the hole- and electron-doped cases. It is clear that the wavefunctions are extended in space and are mainly distributed on every other lattice site. (Note that different spin projections are degenerate.) 

Unlike the Bloch wavefunctions, the Wannier functions contain information from 
the entire moiré Brillouin zone (mBZ) 
and are thus localized in space. 
Here we introduce the layer-averaged exponentially localized Wannier orbitals $w_{\bm{R}}(\br)$ with $\bm{R}$ the moiré lattice vector, constructed by choosing a smooth gauge for the Bloch wavefunctions of the first moiré bands for both hole- and electron-doped cases, i.e.
\begin{equation}
    \begin{aligned}
       w_{\bm{R}}(\br)=&\frac{1}{2}\sum_{\bk\in \text{mBZ}} e^{i\bk\cdot\bR} \, \psi_{\bk}(\br), 
       \qquad
       \psi_{\bk}(\br)=e^{i\bk\cdot\br}\sum_{\bm{G}_j\in \text{top}}\tilde u_{\bm{G}_j,\bk}(\br)+e^{i\bk\cdot\br}\sum_{\bm{G}_j\in \text{bottom}}\tilde u_{\bm{G}_j,\bk}(\br),\\
        &u_{\bm{G}_j,\bk}(\br)\xrightarrow{\text{smooth gauge}}\tilde{u}_{\bm{G}_j,\bk}(\br)
    \end{aligned}\label{eq:MLWF}
\end{equation}
{where the original $u_{\bm{G}_j,\bk}(\br)$ are the eigenvectors of Eq.~\eqref{eq:moireH}.}
We note that $G_j$ encodes the information of layer indices, as shown in the red and blue dots in Fig.~\ref{fig:SM1}, so that one can also construct Wannier functions for each layer separately. Here we choose to include $G_j$'s from both layers, as shown in Eq.~\eqref{eq:MLWF}, so the resulting Wannier function are the layer-averaged ones which contain information from both layers. The smooth gauge is chosen such that the Berry connection $\bm{A}=-i\braket{\tilde u_{\bm{G}_j,\bk}|\nabla_{\bk}|\tilde u_{\bm{G}_j,\bk}}$ becomes a constant in the $k_x$-direction and varies smoothly in the $k_y$-direction. This 
choice of gauge leads to exponentially localized Wannier functions~\cite{PhysRevB.56.12847,RevModPhys.84.1419,PhysRevLett.107.126803}.
In Fig.~\ref{fig:wannier}(e-g) we plot $|w_{\bm{R}}(\br)|^2$ in both the hole- and electron-doped regimes. 
The plots are for both spin projections, since the system has spin degeneracy at every $\bk$ point. 
It is clear that in the hole-doped regime, the Wannier function spreading is smaller than that in the electron doped regime. For the electron-doped case, we construct the Wannier functions for the u-fermions and d-fermions separately. It is apparent that the u- and d-fermion Wannier functions look almost the same, 
further supporting the notion that the u- and d-flavors serve as a pseudospin degree of freedom, in addition to the physical spin. 
}

\section{II. Bosonization of the four-flavor 1D system} 
As discussed in the main text and in the above section that we model the electron doped regime as in the decoupling limit with multi component 1DEG labeled by $\lambda=(s_z,\tau_z)$, where $s_z=\uparrow,\downarrow$ is the physical spin and $\tau_z=\text{u,d}$ is Fermi line index. 
The spin degeneracy ensures that the spin-up fermions and spin-down fermions have the same Fermi momentum $k_F$. Furthermore, according to the inversion symmetry between u and d fermions discussed above, the u and d fermions should also have the same $k_F$ if we require each wire to respect this inversion symmetry. 
For simplicity, we also assume that $k_F$ is incommensurate so we neglect Umklapp scattering. 
The total Hamiltonian for each wire can be written as 
\begin{equation}
    H_{1D}=H_0+H_I
\end{equation}
where
\begin{equation}
    H_0=\sum_{k,r,\lambda} v_F(rk-k_F)\psi_{r,\lambda}^\dagger (k)\psi_{r,\lambda}(k)=2\pi v_F\sum_{q>0,r,\lambda}\rho_{r,\lambda}(q)\rho_{r,\lambda}(-q)
\end{equation}
Here $L$ is the wire length, $v_F$ is the Fermi velocity and $r=\pm$ denotes right ($+$) and left ($-$) movers. The momentum summation is $\sum_q\equiv\frac{1}{L}\sum_{q}$. The interaction part is written as (assuming short range interaction)
\begin{equation}
    \begin{aligned}
        H_I&=\frac{1}{L}\sum_{k,p,q}\sum_{\lambda\neq\lambda'}V_{1,\lambda\lambda'}\psi^\dagger_{+,\lambda}(k)\psi^\dagger_{-,\lambda'}(p)\psi_{+,\lambda'}(p+q)\psi_{-,\lambda}(k-q)\\
        &+\frac{1}{L}\sum_{k,p,q}\sum_{r,\lambda=\lambda'}V_{1,\lambda\lambda'}\psi^\dagger_{r,\lambda}(k)\psi^\dagger_{-r,\lambda'}(p)\psi_{r,\lambda'}(p+q)\psi_{-r,\lambda}(k-q)\\
        &+\frac{1}{L}\sum_{q,r,\lambda,\lambda'}V_{2,\lambda\lambda'}\rho_{r,\lambda}(q)\rho_{-r,\lambda'}(-q)+\frac{1}{L}\sum_{r,q,\lambda.\lambda'}V_{4,\lambda\lambda'}\rho_{r,\lambda}(q)\rho_{r,\lambda'}(-q).\\
    \end{aligned}
\end{equation}
Here we have adopted the notations from previous literatures such that $V_{1}$ is for backscatterting, and $V_2$ and $V_4$ are two forward scatterings. We assume the system is away from commensurate filling such that $4k_F$ is not a reciprocal lattice vector, so we neglect the umklapp scattering $V_3$.
In analogy to two-component 1D systems, it will be convenient to further distinguish $V_{i,\lambda\lambda}$ according to different spin and pseudospin (valley) configurations. To this end, we define, 
\begin{equation}
    \begin{aligned}
        V_{i,\lambda\lambda'}=V_{i,1},~~~\text{if}~~ s_z=s_z',~ \tau_z=\tau_z';\\
        V_{i,\lambda\lambda'}=V_{i,2},~~~\text{if}~~ s_z\neq s_z',~ \tau_z=\tau_z';\\
        V_{i,\lambda\lambda'}=V_{i,3},~~~\text{if}~~ s_z=s_z',~ \tau_z\neq\tau_z';\\
        V_{i,\lambda\lambda'}=V_{i,4},~~~\text{if}~~ s_z\neq s_z',~ \tau_z\neq\tau_z'.\\
    \end{aligned}\label{eq:gij}
\end{equation}

Like the the spin and charge densities in conventional 1DEG, for the four-component system it's useful to define the following four chiral-densities,
\begin{equation}
   \begin{aligned}
        \rho_{0}&=\frac{1}{2}\left(\rho_{\uparrow \text{u}}+\rho_{\downarrow \text{u}}+\rho_{\uparrow \text{d}}+\rho_{\downarrow \text{d}}\right),\\
        \rho_{1}&=\frac{1}{2}\left(\rho_{\uparrow \text{u}}-\rho_{\downarrow \text{u}}+\rho_{\uparrow \text{d}}-\rho_{\downarrow \text{d}}\right),\\
        \rho_{2}&=\frac{1}{2}\left(\rho_{\uparrow \text{u}}+\rho_{\downarrow \text{u}}-\rho_{\uparrow \text{d}}-\rho_{\downarrow \text{d}}\right),\\
        \rho_{3}&=\frac{1}{2}\left(\rho_{\uparrow \text{u}}-\rho_{\downarrow \text{u}}-\rho_{\uparrow \text{d}}+\rho_{\downarrow \text{d}}\right),
   \end{aligned}\label{eq:rhoi}
\end{equation}
where the $r$-index is omitted for brevity. 
Using Eq.\eqref{eq:gij} and \eqref{eq:rhoi}, we can rewrite $H_0$, the two forward-scattering terms ($V_2$ and $V_{4}$) and the backward-scattering with $\lambda=\lambda'$ ($V_{1,1}$) in $H_I$ as
\begin{equation}
   \begin{aligned}
        \sum_{q} &[(V_{2,1}-V_{1,1})(\rho_{r,0}\rho_{-r,0}+\rho_{r,1}\rho_{-r,1}+\rho_{r,2}\rho_{-r,2}+\rho_{r,3}\rho_{-r,3})+ V_{2,2}(\rho_{r,0}\rho_{-r,0}-\rho_{r,1}\rho_{-r,1}+\rho_{r,2}\rho_{-r,2}-\rho_{r,3}\rho_{-r,3})\\
        +& V_{2,3}(\rho_{r,0}\rho_{-r,0}+\rho_{r,1}\rho_{-r,1}-\rho_{r,2}\rho_{-r,2}-\rho_{r,3}\rho_{-r,3})+ V_{2,4}(\rho_{r,0}\rho_{-r,0}-\rho_{r,1}\rho_{-r,1}-\rho_{r,2}\rho_{-r,2}+\rho_{r,3}\rho_{-r,3})]\\
        +\sum_{r,q} &[(\pi v_F+V_{4,1})(\rho_{r,0}^2+\rho_{r,1}^2+\rho_{r,2}^2+\rho_{r,3}^2)+ V_{4,2}(\rho_{r,0}^2-\rho_{r,1}^2+\rho_{r,2}^2-\rho_{r,3}^2)\\
        +& V_{4,3}(\rho_{r,0}^2+\rho_{r,1}^2-\rho_{r,2}^2-\rho_{r,3}^2)+ V_{4,4}(\rho_{r,0}^2-\rho_{r,1}^2-\rho_{r,2}^2+\rho_{r,3}^2)].
   \end{aligned}\label{eq:gij24}
\end{equation}
The other backward-scattering terms are cast into cosine terms by introducing the following bosonization formula,
\begin{equation}
    \psi_{r,\lambda}(x)=\lim_{\alpha\to 0}\frac{U_{r,\lambda}}{\sqrt{2\pi\alpha}}\exp\left[i\theta_\lambda(x)+ir(k_F x-\phi_\lambda(x))\right]\label{eq:bosonizationF}
\end{equation}
where $\alpha$ is a UV cutoff and $U_{r,\lambda}$ are the Klein factors.
It's also convenient at this point to introduce $\phi_i$ in terms of $\phi_{\lambda}$, completely parallel to Eq.\eqref{eq:rhoi}, 
\begin{equation}
   \begin{aligned}
        \phi_{0}&=\frac{1}{2}\left(\phi_{\uparrow \text{u}}+\phi_{\downarrow \text{u}}+\phi_{\uparrow \text{d}}+\phi_{\downarrow \text{d}}\right),\\
        \phi_{1}&=\frac{1}{2}\left(\phi_{\uparrow \text{u}}-\phi_{\downarrow \text{u}}+\phi_{\uparrow \text{d}}-\phi_{\downarrow \text{d}}\right),\\
        \phi_{2}&=\frac{1}{2}\left(\phi_{\uparrow \text{u}}+\phi_{\downarrow \text{u}}-\phi_{\uparrow \text{d}}-\phi_{\downarrow \text{d}}\right),\\
        \phi_{3}&=\frac{1}{2}\left(\phi_{\uparrow \text{u}}-\phi_{\downarrow \text{u}}-\phi_{\uparrow \text{d}}+\phi_{\downarrow \text{d}}\right).
   \end{aligned}\label{eq:phii}
\end{equation}
The inverse relations are
\begin{equation}
    \begin{aligned}
        \phi_{\uparrow \text{u}}&=\frac{1}{2}\left(\phi_0+\phi_1+\phi_2+\phi_3\right),\\
         \phi_{\downarrow \text{u}}&=\frac{1}{2}\left(\phi_0-\phi_1+\phi_2-\phi_3\right),\\
          \phi_{\uparrow \text{d}}&=\frac{1}{2}\left(\phi_0+\phi_1-\phi_2-\phi_3\right),\\
           \phi_{\downarrow \text{d}}&=\frac{1}{2}\left(\phi_0-\phi_1-\phi_2+\phi_3\right).\\
    \end{aligned}\label{eq:inv_phii}
\end{equation}
Using this relation, it is straightforward to rewrite the back-scattering terms as 
\begin{equation}
  \begin{aligned}
    &\frac{V_{1,2}}{(2\pi \alpha)^2}\int dx\left\{U^\dagger_{R\uparrow u}U^\dagger_{L\downarrow u}U_{R\downarrow u}U_{L\uparrow u}e^{i2(\phi_2+\phi_4)}+sU^\dagger_{R\uparrow d}U^\dagger_{L\downarrow d}U_{R\downarrow d}U_{L\uparrow d}e^{i2(\phi_2-\phi_4)}+\text{h.c.}\right\}\\
    +&\frac{V_{1,3}}{(2\pi \alpha)^2}\int dx\left\{U^\dagger_{R\uparrow u}U^\dagger_{L\uparrow d}U_{R\uparrow d}U_{L\uparrow u}e^{i2(\phi_3+\phi_4)}+sU^\dagger_{R\downarrow u}U^\dagger_{L\downarrow d}U_{R\downarrow d}U_{L\downarrow u}e^{i2(\phi_3-\phi_4)}+\text{h.c.}\right\}\\
    +&\frac{V_{1,4}}{(2\pi \alpha)^2}\int dx\left\{U^\dagger_{R\uparrow u}U^\dagger_{L\downarrow d}U_{R\downarrow d}U_{L\uparrow u}e^{i2(\phi_2+\phi_3)}+sU^\dagger_{R\uparrow d}U^\dagger_{L\downarrow u}U_{R\downarrow u}U_{L\uparrow d}e^{i2(\phi_2-\phi_3)}+\text{h.c.}\right\}
  \end{aligned}\label{eq:option1}
\end{equation}
These Klein factors in thermodynamic limit can be replaced by Majorana fermion operators, given that their change on the particle number can be neglected. So we will introduce $R_{\lambda}$ and $L_{\lambda}$ to denote the Majoranas which satisfy $R_{\lambda}R_{\lambda}=L_{\lambda}L_{\lambda}=1$. They are important in order to properly identify the order parameters, as we will see in the next section. At this point, 
we need to use them to determine $s$. The strategy is to come up with some auxiliary order parameter, for instance (the boson parts are omitted), 
\begin{equation}
  \begin{aligned}
    &R_{\uparrow u}L_{\downarrow u}R_{\downarrow u}L_{\uparrow u}+sR_{\uparrow d}L_{\downarrow d}R_{\downarrow d}L_{\uparrow d}\\
    +&R_{\uparrow u}L_{\uparrow d}R_{\uparrow d}L_{\uparrow u}+sR_{\downarrow d}L_{\downarrow u}R_{\downarrow u}L_{\downarrow d}.
  \end{aligned}
\end{equation}
Suppose we evaluate this operator perturbatively in the presence of $V_{1,4}$, so we need to expand $e^{-\beta H_I}$ to the first order of $V_{1,4}$. In order to maintain the form of the order parameter, we need to set $s=-1$. As a result, the bosonized form of the interactions is given by
\begin{equation}
    -\frac{4V_{1,2}}{(2\pi \alpha)^2}\int dx \sin(2\phi_1)\sin(2\phi_3)-\frac{4V_{1,3}}{(2\pi \alpha)^2}\int dx \sin(2\phi_2)\sin(2\phi_3)-\frac{4V_{1,4}}{(2\pi \alpha)^2}\int dx \sin(2\phi_1)\sin(2\phi_2).\label{eq:sin}
\end{equation}
Note in the main text we have introduced new definitions for these backscattering terms as follows,
\begin{equation}
  g_2:=V_{1,2}, ~~ g_1:= V_{1,3}, ~~ g_3:= V_{1,4}.\label{eq:convention_conversion}
\end{equation}
Alternatively, we can use $-V_{1,2}$, $-V_{1,3}$ and $-V_{1,4}$ in Eq.\eqref{eq:option1}, and similar reasoning leads to $s=1$ instead. Therefore in this convention the interaction is given by
\begin{equation}
    -\frac{4g_2}{(2\pi \alpha)^2}\int dx \cos(2\phi_1)\cos(2\phi_3)-\frac{4g_1}{(2\pi \alpha)^2}\int dx \cos(2\phi_2)\cos(2\phi_3)-\frac{4g_3}{(2\pi \alpha)^2}\int dx \cos(2\phi_1)\cos(2\phi_2),\label{eq:cos}
\end{equation}
where we have used the new convention defined in Eq.\eqref{eq:convention_conversion}.

The field $\phi_i$ is related to $\rho_i$ as follows
\begin{equation}
    -\frac{1}{\pi}\partial_x \phi_i(x)= \rho_{+,i}(x)+\rho_{-,i}(x)-\rho_{0,i}.
\end{equation}
where $\rho_0$ is the average density.
So we can identify 
\begin{equation}
    \begin{aligned}
        \text{Charge density fluctuations}:~ \frac{1}{\pi}\partial_x\phi_0(x),& ~~~~\text{Spin density fluctuations}:~ \frac{1}{\pi}\partial_x\phi_1(x),\\
        \text{Valley density fluctuations}:~ \frac{1}{\pi}\partial_x\phi_2(x),& ~~~~\text{Spin-valley density fluctuations}:~ \frac{1}{\pi}\partial_x\phi_3(x).
    \end{aligned}
\end{equation}
The conjugate field $\Pi_{i}$ is
\begin{equation}
    \Pi_{i}(x)= \rho_{+,i}(x)-\rho_{-,i}(x).
\end{equation}
It is related to the $\theta(x)$ in Eq.\eqref{eq:bosonizationF} via 
\begin{equation}
    \Pi_i(x)=\frac{1}{\pi}\partial_x\theta(x).
\end{equation}
Expressing the fermion density operators in terms of $\Pi_i(x)$ and $\phi_i(x)$ we have, 
\begin{equation}
   \begin{aligned}
        [\rho_{+,i}(x)]^2+[\rho_{-,i}(x)]^2&=\frac{[\frac{1}{\pi}\partial_x \phi_i(x)-\rho_{0,i}]^2+[\Pi_{i}(x)]^2}{2}\\
         \rho_{+,i}(x)\rho_{-,i}(x)+\rho_{-,i}(x)\rho_{+,i}(x)&=\frac{[\frac{1}{\pi}\partial_x \phi_i(x)-\rho_{0,i}]^2-[\Pi_{i}(x)]^2}{2}.
   \end{aligned}
\end{equation}
Substituting these relations to Eq.\eqref{eq:gij24} and combining it with Eq.\eqref{eq:cos}, we arrive at the final bosonized Hamiltonian (up to some constant)
\begin{equation}
    \begin{aligned}
        H=&\sum_{i=0}^3 \int dx \left\{\frac{ \pi v_i K_i}{2}\left[\Pi_i(x)\right]^2+\frac{ v_i}{2\pi K_i}\left[\partial_x\phi_i(x)\right]^2\right\}\\
    &-\frac{4g_2}{(2\pi \alpha)^2}\int dx \cos(2\phi_1)\cos(2\phi_3)-\frac{4g_1}{(2\pi \alpha)^2}\int dx \cos(2\phi_2)\cos(2\phi_3)-\frac{4g_3}{(2\pi \alpha)^2}\int dx \cos(2\phi_1)\cos(2\phi_2),
    \end{aligned}\label{eq:H}
\end{equation}
where $\phi_i\equiv \phi_i(x)$ and the renormalized velocities $u_i$ and Luttinger parameters $K_i$ are given by
\begin{equation}
    v_i=\sqrt{(v_F+G_i/\pi)(v_F+U_i/\pi)}, ~~K_i=\sqrt{\frac{\pi v_F +U_i}{\pi v_F+G_i}}.\label{eq:vK}
\end{equation}
Here $U_i$ and $G_i$ are determined by $V_{i,j}$:
\begin{equation}
    \begin{aligned}
        G_0=V_{4,1}+V_{4,2}+V_{4,3}+V_{4,4}+V_{2,1}-V_{1,1}+V_{2,2}+V_{2,3}+V_{2,4},\\
        U_0=V_{4,1}+V_{4,2}+V_{4,3}+V_{4,4}-V_{2,1}+V_{1,1}-V_{2,2}-V_{2,3}-V_{2,4},\\
        G_1=V_{4,1}-V_{4,2}+V_{4,3}-V_{4,4}+V_{2,1}-V_{1,1}-V_{2,2}+V_{2,3}-V_{2,4},\\
        U_1=V_{4,1}-V_{4,2}+V_{4,3}-V_{4,4}-V_{2,1}+V_{1,1}+V_{2,2}-V_{2,3}+V_{2,4},\\
        G_2=V_{4,1}+V_{4,2}-V_{4,3}-V_{4,4}+V_{2,1}-V_{1,1}+V_{2,2}-V_{2,3}-V_{2,4},\\
        U_2=V_{4,1}+V_{4,2}-V_{4,3}-V_{4,4}-V_{2,1}+V_{1,1}-V_{2,2}+V_{2,3}+V_{2,4},\\
        G_3=V_{4,1}-V_{4,2}-V_{4,3}+V_{4,4}+V_{2,1}-V_{1,1}-V_{2,2}-V_{2,3}+V_{2,4},\\
        U_3=V_{4,1}-V_{4,2}-V_{4,3}+V_{4,4}-V_{2,1}+V_{1,1}+V_{2,2}+V_{2,3}-V_{2,4}.
    \end{aligned}
\end{equation}
In a special case when all $V_{i,j}=V_{i,j'}\equiv V_{i}$ for every $i$, $j$ and $j'$, and neglecting the backscattering term $V_{1,i}$, we obtain
\begin{equation}
     \begin{aligned}
         &G_0=4(V_4+V_2), ~~ U_0= 4(V_4-V_2),\\
         &G_1=U_1=G_2=U_2=G_3=U_3=0.\\
     \end{aligned}
 \end{equation} 
Then it is easy to see from Eq.\eqref{eq:vK} that $v_1=v_2=v_3=v_F$ and $K_1=K_2=K_3=1$. When the backscattering is included, as we will see below that for repulsive interaction, the system  still flows to this fixed point, while for attraction, the system tends to gap opening. 

Without the backscattering (cosine) terms, Eq.\eqref{eq:H} remains unchanged via 
\begin{equation}
    \theta_i(x) \leftrightarrow \phi_i(x) ~~\text{and} ~~ K_i \leftrightarrow \frac{1}{K_i}\label{eq:dual}
\end{equation}
which is called T-duality. This observation will be used later to evaluate the correlation functions. 

The Hamiltonian in Eq.\eqref{eq:H} leads to the following action:
\begin{equation}
    S_{tot}=\sum_{i=0}^3S_i + S_I
\end{equation}
where 
\begin{equation}
    \begin{aligned}
        S_i&=\frac{1}{2\pi K_i}\int dxd\tau \left\{\frac{1}{v_i}\left(\partial_\tau\phi_i\right)^2+v_i\left(\partial_x\phi_i\right)^2\right\}\\
        &=\frac{1}{2\pi K_i}\int dxdy \left\{\tilde v_i\left(\partial_x\phi_i\right)^2+\frac{1}{\tilde v_i}\left(\partial_y\phi_i\right)^2\right\}\\
    \end{aligned}\label{eq:Si24}
\end{equation}
with $y:=\bar v \tau$,  $\tilde v_i=v_i/\bar v$ and $\bar v=(v_1+v_2+v_3)/3$. The integration limits will be the same in zero temperature limit. Similarly, 
\begin{equation}
    S_I=-\frac{4}{(2\pi\alpha)^2}\int d^2\br\left[ \frac{g_2}{\bar v}\cos(2\phi_1)\cos(2\phi_3) +\frac{g_1}{\bar v}\cos(2\phi_2)\cos(2\phi_3) + \frac{g_3}{\bar v}\cos(2\phi_1)\cos(2\phi_2) \right].\label{eq:SI25}
\end{equation}
Without the cosine terms, the regularized Green's function is given by
\begin{equation}
    G(x,t):=\braket{\phi_j(x,t)\phi_j(0,0)}=-\frac{K_j}{4}\ln\left[\frac{x^2+y^2+\alpha^2}{R^2}\right]_{y\to iv_j t+\alpha}=-\frac{K_j}{4}\ln\left[\frac{x^2+(iv_j t+\alpha)^2}{R^2}\right]\label{eq:Green1}
\end{equation}
where $\alpha\to0$ and $R\to\infty$ are the UV and IR cutoff respectively. According to the T-duality we have 
\begin{equation}
    G_{\theta\theta}(x,t):=\braket{\theta_j(x,t)\theta_j(0,0)}=-\frac{1}{4K_j}\ln\left[\frac{x^2+(iv_j t+\alpha)^2}{R^2}\right].\label{eq:Green2}
\end{equation}
The Green's function between $\phi_j$ field and $\theta_j$ field is calculated from the original action when both of these two fields are present, 
\begin{equation}
    G_{\phi\theta}(x,t):=\braket{\phi_j(x,t)\theta_j(0,0)}=-\frac{1}{4}\ln\frac{i(v_j t+x)+\alpha}{i(v_j t-x)+\alpha}.\label{eq:Green3}
\end{equation}

On the other hand, if the boson becomes massive, namely
\begin{equation}
    S_i=\frac{1}{2\pi K_i}\int dxdy \left\{\left(\partial_x\phi_i\right)^2+\left(\partial_y\phi_i\right)^2+m^2 \phi_i^2\right\},\label{eq:mass27}
\end{equation}
then the correlation function is given by
\begin{equation}
    \braket{\phi_i(\br)\phi_i(0)}=\frac{K_i}{2}K_0(m|\br|)
\end{equation}
where $K_0(...)$ is the modified Bessel function of the second kind, which decays exponentially at large distance,
\begin{equation}
     K_0(m|\br|)\sim e^{-m|\br|}.
 \end{equation} 
 and behaves as logarithmic at short distances
 \begin{equation}
     K_0(m|\br|) \sim -\frac{1}{2}\ln (m^2 \br^2).
 \end{equation}
 Therefore, we can identify the correlation length as $\xi=1/m$ which is finite. The correlation function in space-time coordinates in the massive case is thus 
 \begin{equation}
     G(x,t)=\frac{K_j}{2}K_0\left(\frac{\sqrt{x^2+(iv_j t+\alpha)^2}}{\xi}\right). \label{eq:massGreen}
 \end{equation}

\section{III. Order parameters and correlation functions at Gaussian fixed point} 
\label{sub:order_parameters}
To obtain the correlation functions, one can make use of the general formula for the vertex operator correlation functions for a massless boson model (let's at this point neglect the logarithmic corrections due to finite backscattering in the limit when all $K_i\to1$),
\begin{equation}
     \begin{aligned}
         \braket{e^{i\beta_1\phi(\br_1)}...e^{i\beta_N\phi(\br_N)}}&=\exp\left(-\sum_{i>j}\beta_i\beta_jG(\br_i,\br_j)\right)\exp\left(-\frac{1}{2}\sum_i\beta_i^2G(\br_i,\br_i)\right)\\
     \end{aligned}
 \end{equation} 
 This equation can be proved by using the cumulant expansion formula for Gaussian action. 

\subsection{A. Massless boson}
If the boson model is massless, we expect that the expectation value of the vertex operator vanishes. This is because
\begin{equation}
    \braket{e^{i\beta\phi_j(x)}}=\exp\left[-\frac{1}{2}\beta^2G(0)\right]=\left(\frac{\alpha}{R}\right)^{\frac{\beta^2K_j}{4}},
\end{equation}
where $G(0)$ is the Green's function in Eq.\eqref{eq:Green1} in the IR limit. Apparently, this results vanishes in the limit when $R\to\infty$. This results is consistent with the fact that in the massless case the expectation value of order paramters remains zero. 

For multi-point correlation functions, we use the explicit formula for the Green's function in Eq.\eqref{eq:Green1} in the massless case and obtain
\begin{equation}
    \begin{aligned}
        \braket{e^{i\beta_1\phi(\br_1)}...e^{i\beta_N\phi(\br_N)}}&=\prod_{i>j}\left[\frac{(x_i-x_j)^2+(y_i-y_j)^2+\alpha^2}{R^2}\right]^{\frac{\beta_i\beta_jK}{4}}\prod_i\left[\frac{\alpha^2}{R^2}\right]^{\frac{\beta_i^2K}{8}}\\
         &=\prod_{i>j}\left[\frac{(x_i-x_j)^2+(y_i-y_j)^2+\alpha^2}{\alpha^2}\right]^{\frac{\beta_i\beta_jK}{4}}\left[\frac{\alpha}{R}\right]^{\frac{K}{8}(\sum_i\beta_i)^2}.
    \end{aligned}
\end{equation}
The correlation function does not vanish in the limit $R\to \infty$ only if $\sum_i\beta_i=0$. From this general expression for correlations functions, it is easy to obtain 
 \begin{equation}
     \begin{aligned}
         \braket{e^{i\phi_j(x,t)-i\phi_j(0)}}&=\left(\frac{\alpha^2}{x^2+(iv_j t+\alpha)^2}\right)^{\frac{K_j}{4}},\\
         \braket{\cos\phi_j(x,t)\cos\phi_j(0,0)}&=\braket{\sin\phi_j(x,t)\sin\phi_j(0,0)}=\frac{1}{2}\left(\frac{\alpha^2}{x^2+(iv_j t+\alpha)^2}\right)^{\frac{K_j}{4}},\\
           \braket{\cos\phi_j(x,t)\sin\phi_j(0,0)}&=\braket{\sin\phi_j(x,t)\cos\phi_j(0,0)}=0.
     \end{aligned}
 \end{equation}
 The correlation functions for the $\theta$ field can be found by the duality relation in Eq.\eqref{eq:dual}.
 \begin{equation}
     \begin{aligned}
         \braket{e^{i\theta_j(x,t)-i\theta_j(0)}}&=\left(\frac{\alpha^2}{x^2+(iv_j t+\alpha)^2}\right)^{\frac{1}{4K_j}},\\
         \braket{\cos\theta_j(x,t)\cos\theta_j(0,0)}&=\braket{\sin\theta_j(x,t)\sin\theta_j(0,0)}=\frac{1}{2}\left(\frac{\alpha^2}{x^2+(iv_j t+\alpha)^2}\right)^{\frac{1}{4K_j}},\\
           \braket{\cos\theta_j(x,t)\sin\theta_j(0,0)}&=\braket{\sin\theta_j(x,t)\cos\theta_j(0,0)}=0.
     \end{aligned}
 \end{equation}

\subsection{B. Massive boson}
 If we use the massive action with the Green's function given by Eq.\eqref{eq:massGreen}, and if the expectation value of $\phi_j$ is $\braket{\phi_j}=0$, we obtain for the expectation value
 \begin{equation}
     \braket{e^{i\beta\phi_j(x)}}=\exp\left[-\frac{1}{2}\beta^2G(0)\right]=\left(\frac{\alpha}{\xi}\right)^{\frac{\beta^2K_j}{4}},
 \end{equation}
 which remains finite as long as $\xi=1/m$ is finite. 

 The two point correlation function $\braket{e^{i\phi_j(\br)-i\phi_j(0)}}$ is of long range, and the fluctuation 
 \begin{equation}
    \begin{aligned}
         &\braket{e^{i\phi_j(\br)-i\phi_j(0)}}-\braket{e^{i\phi_j(\br)}}\braket{e^{-i\phi_j(0)}}\\
         &=e^{G(\br)-G(0)}-e^{-G(0)}\approx e^{-G(0)}G(\br)=\left(\frac{\alpha}{\xi}\right)^{\frac{K_j}{2}}\frac{K_j}{2}e^{-\frac{r}{\xi}}
    \end{aligned}
 \end{equation}
  has exponential decaying behavior at large distances, due to the presence of the mass term. 

 When a mass term of $\phi_j(x)$ field develops, the correlation functions of $e^{i(...)\theta_j(x)}$ has exponentially decaying behavior. A direct calculation based on the effective $\theta$-only action is difficult, but some arguments based on symmetry and continuity all supports this result\cite{Voit1998}.

\subsection{C. Order parameters}
With the additional pseudo-spin degrees of freedom, there exist many order parameters which we list below (now setting $s_z=\pm$ and $\tau_z=\pm$):
\begin{enumerate}
    \item Charge-density-wave: 
    \begin{equation}
       \begin{aligned}
            O_{\text{CDW}}&=\sum_{\lambda}\psi^\dagger_{+,\lambda}\psi_{-,\lambda}\propto e^{-i 2k_F x}\left(R_{\uparrow u}L_{\uparrow u}e^{i(\phi_0+\phi_1+\phi_2+\phi_3)}+R_{\downarrow u}L_{\downarrow u}e^{i(\phi_0-\phi_1+\phi_2-\phi_3)}\right.\\
            &~~~~~~~~~~~+\left.R_{\uparrow d}L_{\uparrow d}e^{i(\phi_0+\phi_1-\phi_2-\phi_3)}+R_{\downarrow d}L_{\downarrow d}e^{i(\phi_0-\phi_1-\phi_2+\phi_3)}\right)\\
       \end{aligned}
    \end{equation}

    Using the convention of Eq.\eqref{eq:option1} with $s=1$, which leads to Eq.\eqref{eq:cos}, we find that the Klein factors are consistent with the following convention for the CDW order parameter:
    \begin{equation}
       \begin{aligned}
            O_{\text{CDW}}&=\sum_{\lambda}\psi^\dagger_{+,\lambda}\psi_{-,\lambda}\propto e^{-i 2k_F x}\left(e^{i(\phi_0+\phi_1+\phi_2+\phi_3)}+e^{i(\phi_0-\phi_1+\phi_2-\phi_3)}+e^{i(\phi_0+\phi_1-\phi_2-\phi_3)}+e^{i(\phi_0-\phi_1-\phi_2+\phi_3)}\right)\\
            &=4e^{-i 2k_F x}\left(\cos\phi_1\cos\phi_2\cos\phi_3-i\sin\phi_1\sin\phi_2\sin\phi_3\right)e^{i\phi_0}
       \end{aligned}
    \end{equation}
    Below we just list the final results after properly removing the Klein factors.

    \item Spin-density-wave ($z$-component): 

    \begin{equation}
        \begin{aligned}
             O_{\text{SDW}}=\sum_{\lambda}s_z\psi^\dagger_{+,\lambda}\psi_{-,\lambda}\propto4ie^{-i2k_F x}\left(\sin\phi_1\cos\phi_2\cos\phi_3+i\cos\phi_1\sin\phi_2\sin\phi_3\right)e^{i\phi_0}
        \end{aligned}
    \end{equation}

    \item Pseudospin(valley)-density-wave ($z$-component):  
    \begin{equation}
        O_{\text{VDW}}=\sum_{\lambda}\tau_z\psi^\dagger_{+,\lambda}\psi_{-,\lambda}\propto4ie^{-i2k_Fx}\left(\cos\phi_1\sin\phi_2\cos\phi_3+i\sin\phi_1\cos\phi_2\sin\phi_3\right)e^{i\phi_0}
    \end{equation}

    \item Spin-Pseudospin(valley)-density-wave ($zz$-component): 
    \begin{equation}
        O_{\text{SVDW}}=\sum_{s_z,\tau_z}s_z\tau_z\psi^\dagger_{+,s_z,\tau_z}\psi_{-,s_z,\tau_z}\propto 4i e^{- i2  k_Fx} \left(
           \cos \phi _1 \cos \phi _2 \sin \phi _3+i\sin \phi _1 \sin \phi _2 \cos \phi _3
        \right)e^{i\phi_0}
    \end{equation}

    \item Spin-singlet Pseudospin-singlet superconductivity: 
    \begin{equation}
       \begin{aligned}
          O_{SS}&=\sum_{s_z,\tau_z}s_z\tau_z\psi_{+,s_z,\tau_z}\psi_{-,-s_z,-\tau_z}\propto  \left(R_{\uparrow u}L_{\downarrow d}e^{i(\theta_0+\theta_3-\phi_1-\phi_2)}-R_{\uparrow d}L_{\downarrow u}e^{i(\theta_0-\theta_3-\phi_1+\phi_2)}\right.\\
            &~~~~~~~~~~~-\left.R_{\downarrow u}L_{\uparrow d}e^{i(\theta_0-\theta_3+\phi_1-\phi_2)}+R_{\downarrow d}L_{\uparrow u}e^{i(\theta_0+\theta_3+\phi_1+\phi_2)}\right)\\
            &\propto\left(e^{i(\theta_0+\theta_3-\phi_1-\phi_2)}-e^{i(\theta_0-\theta_3-\phi_1+\phi_2)}+e^{i(\theta_0-\theta_3+\phi_1-\phi_2)}-e^{i(\theta_0+\theta_3+\phi_1+\phi_2)}\right)\\
        &=4\left(
        \sin \phi _1 \cos \phi _2 \sin \theta _3
    -i  \cos \phi _1 \sin \phi _2 \cos \theta _3
        \right) e^{i\theta_0}
       \end{aligned}
    \end{equation}

    The following superconducting order parameters are obtained in similar way. 

    \item Spin-singlet Pseudospin-triplet superconductivity:
    \begin{equation}
        \begin{aligned}
            O_{ST_1}&=\sum_{s_z}s_z\psi_{+,s_z,+}\psi_{-,-s_z,+}\propto 2\cos(\phi_1+\phi_3)e^{i(\theta_0+\theta_2)},\\
            O_{ST_{-1}}&=\sum_{s_z}s_z\psi_{+,s_z,-}\psi_{-,-s_z,-}\propto 2\cos(\phi_1-\phi_3)e^{i(\theta_0-\theta_2)},\\
            O_{ST_0}&=\sum_{s_z,\tau_z}s_z\psi_{+,s_z,\tau_z}\psi_{-,-s_z,-\tau_z}\propto 4\left(
            \cos \phi _1 \cos \phi _2 \cos \theta _3 
        -i  \sin \phi _1 \sin \phi _2 \sin \theta _3 
            \right) e^{i\theta_0}.
        \end{aligned}
    \end{equation}

    \item Spin-triplet Pseudospin-singlet superconductivity:
    \begin{equation}
        \begin{aligned}
            O_{T_1S}&=\sum_{\tau_z}\tau_z\psi_{+,+,\tau_z}\psi_{-,+,-\tau_z}\propto 2\cos(\phi_2+\phi_3)e^{i(\theta_0+\theta_1)},\\
            O_{T_{-1}S}&=\sum_{\tau_z}\tau_z\psi_{+,-,\tau_z}\psi_{-,-,-\tau_z}\propto 2\cos(\phi_2-\phi_3)e^{i(\theta_0-\theta_1)},\\
            O_{T_0S}&=\sum_{s_z,\tau_z}\tau_z\psi_{+,s_z,\tau_z}\psi_{-,-s_z,-\tau_z}\propto 4i\left(
            \cos \phi _1 \cos \phi _2 \sin \theta _3 
        +i  \sin \phi _1 \sin \phi _2 \cos \theta _3 
            \right) e^{i\theta_0}.\label{eq:orderparameter_TS}
        \end{aligned}
    \end{equation}

    \item Spin-triplet Pseudospin-triplet superconductivity:
    \begin{equation}
        \begin{aligned}
            O_{T_0 T_0} & = \sum_{s_z,\tau_z}\psi_{+,s_z,\tau_z}\psi_{-,-s_z,-\tau_z} \propto -4i\left(
            \sin \phi _1 \cos \phi _2 \cos \theta _3 
        +i  \cos \phi _1 \sin \phi _2 \sin \theta _3 
            \right) e^{i\theta_0},\\
             O_{T_0 T_{1}} & =\sum_{s_z}\psi_{+,s_z,+}\psi_{-,-s_z,+} \propto  2  \sin \left(\phi _1+\phi _3\right)e^{i \left(\theta _0+\theta _2\right)}\\
             O_{T_0 T_{-1}} & =\sum_{s_z}\psi_{+,s_z,-}\psi_{-,-s_z,-} \propto 2  \sin \left(\phi _1-\phi _3\right)e^{i \left(\theta _0-\theta _2\right)} \\
             O_{T_1 T_0} & = \sum_{\tau_z}\psi_{+,+,\tau_z}\psi_{-,+,-\tau_z} \propto  2  \sin \left(\phi _2+\phi _3\right)e^{i \left(\theta _0+\theta _1\right)}\\
              O_{T_1 T_{1}} & = \psi_{+,+,+}\psi_{-,+,+} \propto e^{i \left(\theta _0+\theta _1+\theta _2+\theta _3\right)}\\
             O_{T_1 T_{-1}} & = \psi_{+,+,-}\psi_{-,+,-} \propto e^{i \left(\theta _0+\theta _1-\theta _2-\theta _3\right)}\\
             O_{T_{-1} T_0} & =  \sum_{\tau_z}\psi_{+,-,\tau_z}\psi_{-,-,-\tau_z} \propto  2 \sin \left(\phi _2-\phi _3\right)e^{i \left(\theta _0-\theta _1\right)} \\
              O_{T_{-1} T_{1}} & = \psi_{+,-,+}\psi_{-,-,+} \propto e^{i \left(\theta _0-\theta _1+\theta _2-\theta _3\right)} \\
             O_{T_{-1} T_{-1}} & = \psi_{+,-,-}\psi_{-,-,-} \propto e^{i \left(\theta _0-\theta _1-\theta _2+\theta _3\right)} \\
        \end{aligned}
    \end{equation}

\end{enumerate}

In addition to these two particle order parameters, we can also consider four-fermion order parameters. One apparent example is the charge-4e superconductivity, whose order parameters can be
\begin{equation}
  \begin{aligned}
    O_{4e,1}&=O_{T_1 S} O_{T_{-1} S}=\sum_{\tau_z,\tau_z'}\tau_z\tau_z'\psi_{+,+,\tau_z}\psi_{-,+,-\tau_z}\psi_{+,-,\tau_z'}\psi_{-,-,-\tau_z'}\propto4\cos(\phi_2+\phi_3)\cos(\phi_2-\phi_3)e^{i2\theta_0}\\
    O_{4e,2}&=O_{S T_{1}} O_{S T_{-1}}=\sum_{s_z,s_z'}s_z s_z'\psi_{+,s_z,+}\psi_{-,-s_z,+}\psi_{+,s_z',-}\psi_{-,-s_z',-}\propto4\cos(\phi_1+\phi_3)\cos(\phi_1-\phi_3)e^{i2\theta_0},\\
    O_{4e,3}&=\frac{1}{4}(O_{S T_{0}}+O_{T_{0} S })(O_{S T_{0}}-O_{T_{0} S }) \propto4\cos(\phi_1+\phi_2)\cos(\phi_1-\phi_2)e^{i2\theta_0}\\
    O_{4e,1'}&=O_{T_1 T_0} O_{T_{-1} T_0}\propto4\sin(\phi_2+\phi_3)\sin(\phi_2-\phi_3)e^{i2\theta_0}\\
    O_{4e,2'}&=O_{T_0 T_1 } O_{T_0T_{-1} }\propto4\sin(\phi_1+\phi_3)\sin(\phi_1-\phi_3)e^{i2\theta_0},\\
    O_{4e,3'}&=\frac{1}{4}(O_{S S}+O_{T_{0} T_{0} })(O_{S S}-O_{T_{0}T_{0} }) \propto4\sin(\phi_1+\phi_2)\sin(\phi_1-\phi_2)e^{i2\theta_0}.
  \end{aligned}
\end{equation}
In the density-wave channel, there are also various kinds of $4k_F$ density waves, such as 
\begin{equation}
  \begin{aligned}
  &O_{4k_F,1}=\sum_{s_z,\tau_z}\psi^\dagger_{+,s_z,\tau_z}\psi^\dagger_{+,s_z,-\tau_z}\psi_{-,s_z,-\tau_z}\psi_{-,s_z,\tau_z}\propto 4 e^{-i4k_Fx} \cos(2\phi_1)e^{i2\phi_0},\\
    &O_{4k_F,2}=\sum_{s_z,\tau_z}\psi^\dagger_{+,s_z,\tau_z}\psi^\dagger_{+,-s_z,\tau_z}\psi_{-,-s_z,\tau_z}\psi_{-,s_z,\tau_z}\propto 4 e^{-i4k_Fx} \cos(2\phi_2)e^{i2\phi_0},\\
    &O_{4k_F,3}=\sum_{s_z,\tau_z}\psi^\dagger_{+,s_z,\tau_z}\psi^\dagger_{+,-s_z,-\tau_z}\psi_{-,-s_z,-\tau_z}\psi_{-,s_z,\tau_z}\propto 4 e^{-i4k_Fx} \cos(2\phi_3)e^{i2\phi_0}.\\
  \end{aligned}
\end{equation}
There are of course many other charge-4e and $4k_F$-density wave orders, but they all contain $e^{i2\theta_0}$ and $e^{i2\phi_0}$.

We will also be interested in some zero-charge, zero momentum orders, which we call Pomeranchuk instabilities (PI). Some representative examples are given by fusions of charge-$2e$ orders as well:
\begin{equation}
  \begin{aligned}
    O_\text{PI,1}&=O_{T_1 S}^\dagger O_{T_{-1} S}\propto4\cos(\phi_2+\phi_3)\cos(\phi_2-\phi_3)e^{-i2\theta_1}\\
    O_\text{PI,2}&=O_{S T_{1}}^\dagger O_{S T_{-1}}\propto4\cos(\phi_1+\phi_3)\cos(\phi_1-\phi_3)e^{-i2\theta_2},\\
    O_\text{PI,3}&=\frac{1}{4}(O_{S T_{0}}^\dagger+O_{T_{0} S }^\dagger)(O_{S T_{0}}-O_{T_{0} S }) \propto4\cos(\phi_1+\phi_2)\cos(\phi_1-\phi_2)e^{-i2\theta_3}\\
    O_{\text{PI,1}'}&=O_{T_1 T_0}^\dagger O_{T_{-1} T_0}\propto4\sin(\phi_2+\phi_3)\sin(\phi_2-\phi_3)e^{-i2\theta_1}\\
    O_{\text{PI,2}'}&=O_{T_0 T_{1}}^\dagger O_{T_0 T_{-1}}\propto4\sin(\phi_1+\phi_3)\sin(\phi_1-\phi_3)e^{-i2\theta_2},\\
    O_{\text{PI,3}'}&=\frac{1}{4}(O_{S S}^\dagger+O_{T_{0} T_{0} }^\dagger)(O_{S S}-O_{T_{0}} O_{T_{0} }) \propto4\sin(\phi_1+\phi_2)\sin(\phi_1-\phi_2)e^{-i2\theta_3}.
  \end{aligned}
\end{equation}

In table \ref{tab:orders} we discuss the leading order parameters in the partially gapped casese. Here the leading orders are the ones which have the most divergent susceptibility, and their correlation functions have the slowest decay rate. In the partically gapped case, only one of the couplings is relevant. The leading orders depend on the which $g_i$ flows to strong coupling, and also on the sign of the relevant $g_i$. Table \ref{tab:orders} summarizes different cases. 

\begin{table*}
  \centering
  \begin{tabular}{lccccccccccc}
    \hline\hline
     & & $\Delta_1=0$ & & && $\Delta_2=0$&&&&$\Delta_3=0$ &\\
    &$g_1>0$& & $g_1<0$ &&$g_2>0$&&$g_2<0$&&$g_3>0$ && $g_3<0$\\\hline
    $2k_F$ &$O_\text{CDW}, O_\text{SDW}$& & $O_\text{VDW}, O_\text{SVDW}$ &&$O_\text{CDW}, O_\text{VDW}$&&$O_\text{SDW}, O_\text{SVDW}$&&$O_\text{CDW}, O_\text{SVDW}$ && $O_\text{SDW}, O_\text{VDW}$\\
    $4k_F$ &$O_{4k_F,1}$& & $O_{4k_F,1}$ &&$O_{4k_F,2}$&&$O_{4k_F,2}$&&$O_{4k_F,3}$ && $O_{4k_F,3}$\\
    $2e$ &$O_{T_{\pm1}S}$& & $O_{T_{\pm1}T_0}$ &&$O_{ST_{\pm1}}$&&$O_{T_0T_{\pm1}}$&&$\frac{1}{2}(O_{ST_0}\pm O_{T_0S})$ && $\frac{1}{2}(O_{SS}\pm O_{T_0T_0})$\\
    $4e$ &$O_{4e,1}$& & $O_{4e,1'}$ &&$O_{4e,2}$&&$O_{4e,2'}$&&$O_{4e,3}$ && $O_{4e,3'}$\\
    PI &$O_{\text{PI},1}$& & $O_{\text{PI},1'}$ &&$O_{\text{PI},2}$&&$O_{\text{PI},2'}$&&$O_{\text{PI},3}$ && $O_{\text{PI},3'}$\\
   \hline
  \end{tabular}
    \caption{Leading order parameters for the partially gapped case when one of the three $\phi_{1,2,3}$ remains gapless but the other two develop finite gaps.}\label{tab:orders}
  \end{table*}


\section{IV. Renormalization group analysis} 

To proceed with the RG analysis, we reformulate the action as follows
\begin{equation}
    \begin{aligned}
        S&=\sum_i \frac{1}{2}\int d^2\br \left(\tilde v_i(\partial_x\phi_i)^2+\frac{1}{\tilde v_i}(\partial_y\phi_i)^2\right)\\
        & + \int d^2\br \left[u_{2}\cos(\beta_1\phi_1)\cos(\beta_3\phi_3)+u_{1}\cos(\beta_2\phi_2)\cos(\beta_3\phi_3)+u_{3}\cos(\beta_1\phi_1)\cos(\beta_2\phi_2)\right]\\
        &=S_0[\phi_i]+S_I[\phi_i]
    \end{aligned}\label{eq:cosineterms}
\end{equation}
where we have introduced
\begin{equation}
     u_{i}=\frac{-g_{i}}{\pi^2\alpha^2 \bar v},  ~~ \beta_i=\sqrt{4\pi K_i}\label{eq:betai}
\end{equation}
Each of these fields $\phi_i(x)$ can be separated into slow mode $\phi_i^s(x)$ and fast mode $h_i(x)$
\begin{equation}
    \phi_i(x)=\phi_i^s(x)+h_i(x)
\end{equation}
and the Gaussian part of the action is also separable 
\begin{equation}
    S_0[\phi_i]= S_0[\phi_i^s] +  S_0[h_i].
\end{equation}
The partition function is then 
\begin{equation}
    \begin{aligned}
        Z &= \int\prod_i\mathcal{D}[\phi_i^s]e^{-S_0[\phi_i^s]}\prod_i\mathcal{D}[h_i]e^{-S_0[h_i]} e^{-S_I[\phi_i^s+h_i]}\\
        &=Z_h \int\prod_i\mathcal{D}[\phi_i^s]e^{-S_0[\phi_i^s]} \braket{e^{-S_I[\phi_i^s+h_i]}}_h\\
        &=Z_h \int\prod_i\mathcal{D}[\phi_i^s]e^{-S_{\text{eff}}[\phi_i^s]}
    \end{aligned}
\end{equation}
where   
\begin{equation}
    S_{\text{eff}}[\phi_i^s]=S_0[\phi_i^s]+ \braket{S_I[\phi_i^s+h_i]}_h-\frac{1}{2}\left(\braket{S_I^2[\phi_i^s+h_i]}_h-\braket{S_I[\phi_i^s+h_i]}_h^2\right)
\end{equation}
and $Z_h$ is the partition function coming from fast mode which is unimportant, and $\braket{...}_h$ means average over all the fast mode under $e^{-S_0[h_i]}$.

We begin with the first order corrections, i.e. the term $\braket{S_I[\phi_i^s+h_i]}_h$. As an example, we discuss only one of the three interacting terms,
\begin{equation}
     \braket{u_{2}\int d^2\br\cos[\beta_1(\phi_1^s+h_1)]\cos[\beta_3(\phi_3^s+h_3)]}_h=u_{2}\int d^2\br\cos(\beta_1\phi_1^s)\braket{e^{i\beta_1h_1}}_{h_1}\cos(\beta_3\phi_3^s)\braket{e^{i\beta_3h_3}}_{h_3}.\label{eq:12}
 \end{equation} 
 The average of the vertex $e^{i\beta_1h_1}$ and $e^{i\beta_3h_3}$ is evaluated as follows,
 \begin{equation}
     \braket{e^{i\beta_1h_1}}_{h_1}\approx1-\frac{1}{2}\beta_1^2\braket{h_1^2}_{h_1}
 \end{equation}
 and in the $T\to0$ limit,
 \begin{equation}
  \begin{aligned}
      \braket{h_1^2}_{h_1}&=\int_{\Lambda'<\sqrt{k^2+(\omega/\bar v)^2}<\Lambda}\frac{dk d\omega}{(2\pi)^2}\frac{1}{v_1k^2+\omega^2/v_1}\\
      &=\int_{\Lambda'<\sqrt{k^2+(\omega/\bar v)^2}<\Lambda}\frac{dk d(\omega/\bar v)}{(2\pi)^2}\frac{v_1}{\bar v}\frac{1}{(v_1/\bar v)^2k^2+(\omega/\bar v)^2}\\
      &=\int_{\Lambda'<p<\Lambda}\frac{p dp d\theta}{(2\pi)^2}\frac{v_1}{\bar v}\frac{1}{(v_1/\bar v)^2p^2\cos^2\theta+p^2\sin^2\theta}=\frac{dl}{2\pi}\frac{v_1}{\bar v}\int\frac{d\theta}{2\pi}\frac{1}{(v_1/\bar v)^2\cos^2\theta+\sin^2\theta}=\frac{dl}{2\pi}.
  \end{aligned}\label{eq:zero_correlation}
 \end{equation}
 where we use $dl=d\Lambda/\Lambda>0$. Note this result is independent of $v_1$ in the zero temperature limit, so for other sectors it remains the same.  Using this result, Eq.\eqref{eq:12} becomes
 \begin{equation}
     u_{2}\left(1-\frac{\beta_1^2+\beta_3^2}{4\pi}dl\right)\int d^2\br\cos(\beta_1\phi_1^s)\cos(\beta_3\phi_3^s)\label{eq:u122}
 \end{equation}
 and other terms in the first order correction are obtained in exactly the same way. 
 To compare with the original action, we need to rescale the real-space $\phi_i^s$ back to $\phi$. From the Gaussian level, we see $\phi_i^s$ does not change, but since the small distance cutoff is now $1+dl$ larger than the last step, we need to multiply $(1+dl)^2\approx (1+2dl)$ to Eq.\eqref{eq:u122}, so the resulting RG equation for the interactions are [making use of Eq.\eqref{eq:betai} we see that $\beta_j^2/(4\pi)=K_j$]
 \begin{equation}
     \begin{aligned}
         \frac{du_{2}}{dl}&=\left[2-(K_1+K_3)\right]u_{2},\\
          \frac{du_{1}}{dl}&=\left[2-(K_2+K_3)\right]u_{1},\\
           \frac{du_{3}}{dl}&=\left[2-(K_1+K_2)\right]u_{3},
     \end{aligned}\label{eq:dudl}
 \end{equation}

In the next order corrections, we first evaluate 
\begin{equation}
    \begin{aligned}
        \braket{S_I[\phi_i^s+h_i]}_h^2=\int d^2\br_1d^2\br_2\left\{u_{2}^2(1-2B_{13}dl)\cos[\beta_1\phi_1^s(\br_1)]\cos[\beta_1\phi_1^s(\br_2)]\cos[\beta_3\phi_3^s(\br_1)]\cos[\beta_3\phi_3^s(\br_2)]\right.\\
    +u_{1}^2(1-2B_{23}dl)\cos[\beta_2\phi_2^s(\br_1)]\cos[\beta_2\phi_2^s(\br_2)]\cos[\beta_3\phi_3^s(\br_1)]\cos[\beta_3\phi_3^s(\br_2)]\\
    +u_{3}^2(1-2B_{12}dl)\cos[\beta_1\phi_1^s(\br_1)]\cos[\beta_1\phi_1^s(\br_2)]\cos[\beta_2\phi_2^s(\br_1)]\cos[\beta_2\phi_2^s(\br_2)]\\
    +2u_{2}u_{1}[1-(B_{13}+B_{23})dl]\cos[\beta_1\phi_1^s(\br_1)]\cos[\beta_3\phi_3^s(\br_1)]\cos[\beta_2\phi_2^s(\br_2)]\cos[\beta_3\phi_3^s(\br_2)]\\
    +2u_{2}u_{3}[1-(B_{13}+B_{12})dl]\cos[\beta_1\phi_1^s(\br_1)]\cos[\beta_3\phi_3^s(\br_1)]\cos[\beta_1\phi_1^s(\br_2)]\cos[\beta_2\phi_2^s(\br_2)]\\
    +\left.2u_{1}u_{3}[1-(B_{12}+B_{23})dl]\cos[\beta_2\phi_2^s(\br_1)]\cos[\beta_3\phi_3^s(\br_1)]\cos[\beta_1\phi_1^s(\br_2)]\cos[\beta_2\phi_2^s(\br_2)]\right\}\\
    \end{aligned}\label{eq:33}
\end{equation}
where we have used 
\begin{equation}
    B_{ij}=\frac{\beta_i^2+\beta_j^2}{4\pi}=K_i+K_j.
\end{equation}
Similarly we can obtain
\begin{equation}
    \begin{aligned}
        \braket{S_I^2[\phi_i^s+h_i]}_h~~~~~~~~~~~~~~~~~~~~~&\\
        =\int d^2\br_1d^2\br_2 \{\frac{u_{2}^2}{4}(1-2B_{13}dl)&[\cos\beta_1(\phi_1^s(\br_1)+\phi_1^s(\br_2))e^{-\beta_1^2\braket{h_1(\br_1)h_1(\br_2)}}+\cos\beta_1(\phi_1^s(\br_1)-\phi_1^s(\br_2))e^{\beta_1^2\braket{h_1(\br_1)h_1(\br_2)}}]\\
        &[\cos\beta_3(\phi_3^s(\br_1)+\phi_3^s(\br_2))e^{-\beta_3^2\braket{h_3(\br_1)h_3(\br_2)}}+\cos\beta_3(\phi_3^s(\br_1)-\phi_3^s(\br_2))e^{\beta_3^2\braket{h_3(\br_1)h_3(\br_2)}}]\\
        +\frac{u_{1}^2}{4}(1-2B_{23}dl)&[\cos\beta_2(\phi_2^s(\br_1)+\phi_2^s(\br_2))e^{-\beta_2^2\braket{h_2(\br_1)h_2(\br_2)}}+\cos\beta_2(\phi_2^s(\br_1)-\phi_2^s(\br_2))e^{\beta_2^2\braket{h_2(\br_1)h_2(\br_2)}}]\\
        &[\cos\beta_3(\phi_3^s(\br_1)+\phi_3^s(\br_2))e^{-\beta_3^2\braket{h_3(\br_1)h_3(\br_2)}}+\cos\beta_3(\phi_3^s(\br_1)-\phi_3^s(\br_2))e^{\beta_3^2\braket{h_3(\br_1)h_3(\br_2)}}]\\
        +\frac{u_{3}^2}{4}(1-2B_{12}dl)&[\cos\beta_1(\phi_1^s(\br_1)+\phi_1^s(\br_2))e^{-\beta_1^2\braket{h_1(\br_1)h_1(\br_2)}}+\cos\beta_1(\phi_1^s(\br_1)-\phi_1^s(\br_2))e^{\beta_1^2\braket{h_1(\br_1)h_1(\br_2)}}]\\
        &[\cos\beta_2(\phi_2^s(\br_1)+\phi_2^s(\br_2))e^{-\beta_2^2\braket{h_2(\br_1)h_2(\br_2)}}+\cos\beta_2(\phi_2^s(\br_1)-\phi_2^s(\br_2))e^{\beta_2^2\braket{h_2(\br_1)h_2(\br_2)}}]\\
        +\frac{u_{2}u_{1}}{2}(1-(B_{13}+B_{23})dl)&[\cos(\beta_1\phi_1^s(\br_1)+\beta_2\phi_2^s(\br_2))+\cos(\beta_1\phi_1^s(\br_1)-\beta_2\phi_2^s(\br_2))]\\
        &[\cos\beta_3(\phi_3^s(\br_1)+\phi_3^s(\br_2))e^{-\beta_3^2\braket{h_3(\br_1)h_3(\br_2)}}+\cos\beta_3(\phi_3^s(\br_1)-\phi_3^s(\br_2))e^{\beta_3^2\braket{h_3(\br_1)h_3(\br_2)}}]\\
        +\frac{u_{2}u_{3}}{2}(1-(B_{13}+B_{12})dl)&[\cos(\beta_3\phi_3^s(\br_1)+\beta_2\phi_2^s(\br_2))+\cos(\beta_3\phi_3^s(\br_1)-\beta_2\phi_2^s(\br_2))]\\
        &[\cos\beta_1(\phi_1^s(\br_1)+\phi_1^s(\br_2))e^{-\beta_1^2\braket{h_1(\br_1)h_1(\br_2)}}+\cos\beta_1(\phi_1^s(\br_1)-\phi_1^s(\br_2))e^{\beta_1^2\braket{h_1(\br_1)h_1(\br_2)}}]\\
        +\frac{u_{1}u_{3}}{2}(1-(B_{12}+B_{23})dl)&[\cos(\beta_3\phi_3^s(\br_1)+\beta_1\phi_1^s(\br_2))+\cos(\beta_3\phi_3^s(\br_1)-\beta_1\phi_1^s(\br_2))]\\
        &[\cos\beta_2(\phi_2^s(\br_1)+\phi_2^s(\br_2))e^{-\beta_2^2\braket{h_2(\br_1)h_2(\br_2)}}+\cos\beta_2(\phi_2^s(\br_1)-\phi_2^s(\br_2))e^{\beta_2^2\braket{h_2(\br_1)h_2(\br_2)}}]\}.\\
    \end{aligned}\label{eq:35}
\end{equation}
It is now clear that we can easily combine Eq.\eqref{eq:33} and \eqref{eq:35}, and this requires the evaluation of $e^{-\beta_j^2\braket{h_j(\br_1)h_j(\br_2)}}$. In  the case of an abrupt cut-off, the correlation $\braket{h_j(\br)h_j(0)}=F_j(r)dl$. However, when a smooth cutoff is used, $F_j(r)$ can be short range, i.e. $F_j(r)$ is negligible when $r>1/\Lambda$, which validate the approximation that $\br_1\approx \br_2$ in Eq.\eqref{eq:33} and \eqref{eq:35} so we can expand the results in terms of the relative coordinate $\br=\br_1-\br_2$.
Like in Eq.\eqref{eq:zero_correlation}, the correlation at finite distance is calculated as follows
\begin{equation}
    \begin{aligned}
        \braket{h_j(\br)h_j(0)}_{h_j}&=\int_{\Lambda'<p<\Lambda}\frac{dp/p}{2\pi}\int \frac{d\theta}{2\pi}\frac{e^{ip r \cos\theta}}{(v_j/\bar v)\cos^2\theta+\sin^2\theta (\bar v/v_j)}\\
        &=\int_{\Lambda'<p<\Lambda}\frac{dp/p}{2\pi}\int \frac{d\theta}{2\pi}\frac{e^{ip r \cos\theta}}{1+\delta_j \cos2\theta}
    \end{aligned}
\end{equation}
where we have introduced $\delta_j=(v_j-\bar v)/\bar v$. In weak coupling limit, $\delta$ is a small number, so we can approximately have
\begin{equation}
    \begin{aligned}
        \braket{h_j(\br)h_j(0)}_{h_j}\approx\int_{\Lambda'<p<\Lambda}\frac{dp/p}{2\pi}\int \frac{d\theta}{2\pi}(1-\delta_j \cos2\theta)e^{ip r \cos\theta}=\frac{dl}{2\pi}(J_0(\Lambda r)+\delta_j J_2(\Lambda r))\equiv F_j(r)dl
    \end{aligned}
\end{equation}
for all $j$ and $r<1/\Lambda$. Note $F_j$ can be different for different sectors since they have different velocities.
It will be shown later that what are useful in deriving the RG equations are not these correlations, but some skewed form instead. The skewed correlations are defined as 
\begin{equation}
 \begin{aligned}
    \tilde F_j(r)dl= \braket{h_j(\sqrt{\tilde v_j}x, {y}/{\sqrt{\tilde v_j}})h_j(0)}_{h_j}&=\int_{\Lambda'<k<\Lambda}\frac{dk_x dk_y}{(2\pi)^2}\frac{e^{i(\sqrt{\tilde v_j}k_x x +k_y y /\sqrt{\tilde v_j})}}{\tilde v_j k_x^2 + k_y^2/{\tilde v_j}}\\
    &=\int_{\text{ellipse shell}}\frac{ds}{(2\pi)^2}\frac{e^{ikr\cos\theta}}{k^2}
 \end{aligned}
\end{equation}
where the ellipse shell integration is shown in Fig.\ref{fig:ellipseRG}. We can use the polar coordinate formula for the ellipse shell. Next we assume $\delta_j=\tilde v_j-1$ is small, so we have
\begin{equation}
  \tilde F_j(r)dl=\int_0^{2\pi}\frac{d\theta}{2\pi}\int_{\Lambda'(\theta)}^{\Lambda(\theta)}\frac{dk k}{2\pi}\frac{\cos[kr\cos\theta]}{k^2}=\frac{dl}{2\pi}\int_0^{2\pi}\frac{d\theta}{2\pi}\cos[\Lambda(\theta)r\cos\theta]
\end{equation}
where 
\begin{equation}
  \Lambda(\theta)=\frac{\Lambda}{\sqrt{(1+\delta_j)\cos^2\theta+\sin^2\theta/(1+\delta_j)}}.
\end{equation}
Expanding to the leading order in $\delta_j$ we obtain
\begin{equation}
  \tilde F_j(r)dl=\frac{dl}{2\pi}\left[J_0(\Lambda r)+\delta_j\left(\frac{\Lambda r}{2}J_1(\Lambda r)-J_2(\Lambda r)\right)\right]\label{eq:tildeFJ}
\end{equation}

\begin{figure}
  \includegraphics[width=8cm]{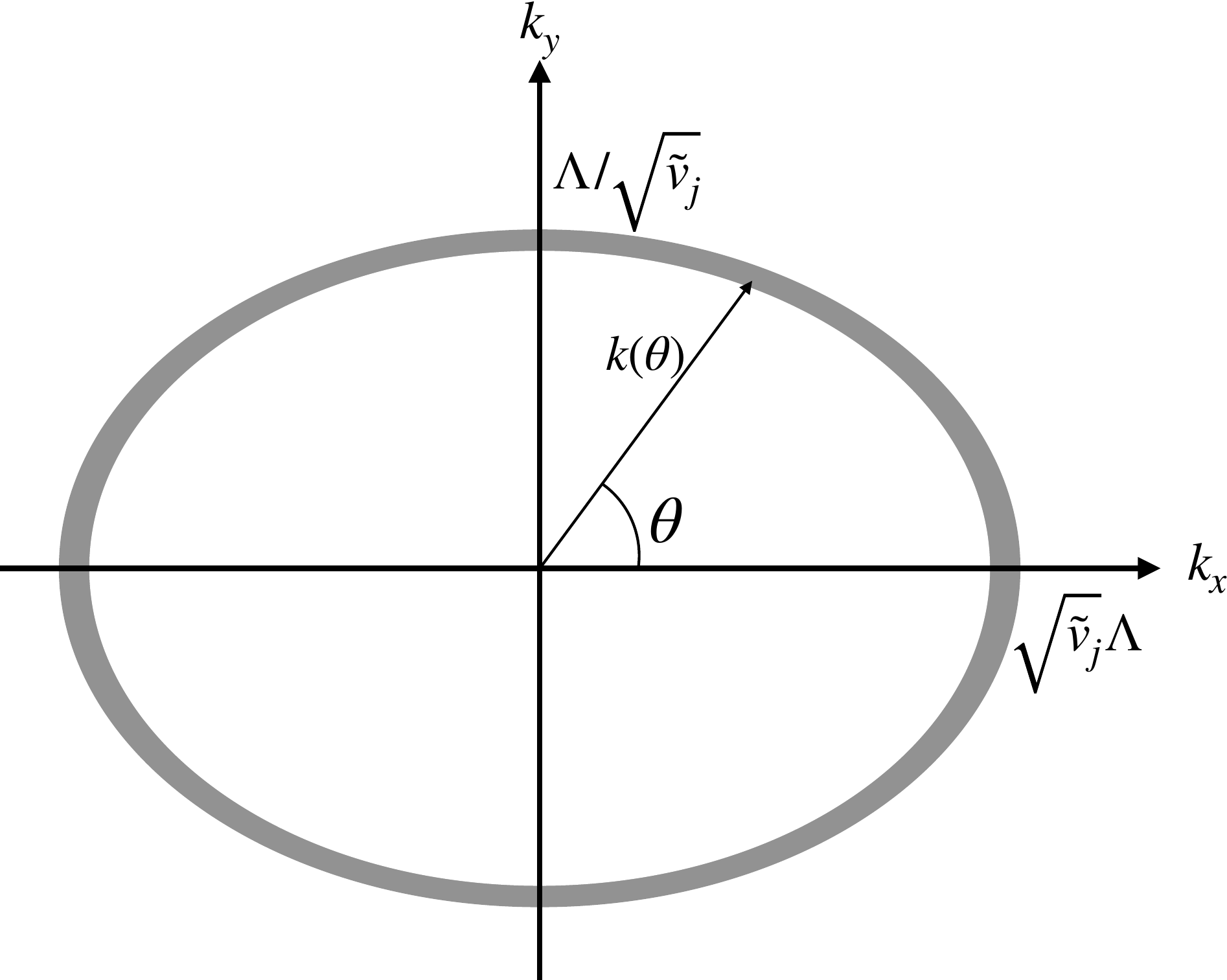}
  \caption{Ellipse momentum shell integration arsing from $v_j\neq \bar v$.}\label{fig:ellipseRG}
\end{figure}

Based on this, it is straightforward to see (defining $\bm{R}=(\bm{r}_1+\bm{r}_2)/2$)
\begin{equation}
    \begin{aligned}
        \braket{S_I^2[\phi_i^s+h_i]}_h-\braket{S_I[\phi_i^s+h_i]}_h^2=\frac{u_{2}^2}{4}\int d^2\bm{R}d^2\br&(1-2B_{13}dl)\{\cos(2\beta_1\phi_1^s(\bR))\cos(2\beta_3\phi_3^s(\bR))(-\beta_1^2F_1(\br)-\beta_3^2F_3(\br))dl\\
        +&\cos(2\beta_1\phi_2^s(\bR))\cos(\beta_3\br\cdot\nabla_R\phi_3^s(\bR))(\beta_3^2F_3(\br)-\beta_1^2F_1(\br))dl\\
        +&\cos(2\beta_3\phi_3^s(\bR))\cos(\beta_1\br\cdot\nabla_R\phi_1^s(\bR))(\beta_1^2F_1(\br)-\beta_3^2F_3(\br))dl\\
        +&\cos(\beta_1\br\cdot\nabla_R\phi_1^s(\bR))\cos(\beta_3\br\cdot\nabla_R\phi_3^s(\bR))(\beta_1^2F_1(\br)+\beta_3^2F_3(\br))dl\}\\
    +& u_{1}^2[...]+u_{3}^2[...]\\
    +\frac{u_{2}u_{1}}{2}(1-(B_{13}+B_{23})dl)\int d^2\bR d^2\br\{\cos[\beta_1\phi_1^s(\bR)+&\beta_2\phi_3^s(\bR)+\beta_1\frac{\br}{2}\cdot\nabla_R\phi_1^s(\bR)-\beta_2\frac{\br}{2}\cdot\nabla_R\phi_2^s(\bR)]\cos[2\beta_3\phi_3^s(\bR)](-\beta_3^2)F_3(\br)dl\\
    +\cos[\beta_1\phi_1^s(\bR)+\beta_2\phi_2^s(\bR)+&\beta_1\frac{\br}{2}\cdot\nabla_R\phi_1^s(\bR)-\beta_2\frac{\br}{2}\cdot\nabla_R\phi_2^s(\bR)]\cos[\beta_3\br\cdot\nabla_R\phi_3^s(\bR)]\beta_3^2F_3(\br)dl\\
    +\cos[\beta_1\phi_1^s(\bR)-\beta_2\phi_2^s(\bR)+&\beta_1\frac{\br}{2}\cdot\nabla_R\phi_1^s(\bR)+\beta_2\frac{\br}{2}\cdot\nabla_R\phi_2^s(\bR)]\cos[2\beta_3\phi_3^s(\bR)](-\beta_3^2)F_3(\br)dl\\
    +\cos[\beta_1\phi_1^s(\bR)-\beta_2\phi_2^s(\bR)+&\beta_1\frac{\br}{2}\cdot\nabla_R\phi_1^s(\bR)+\beta_2\frac{\br}{2}\cdot\nabla_R\phi_2^s(\bR)]\cos[\beta_3\br\cdot\nabla_R\phi_3^s(\bR)]\beta_3^2F_3(\br)dl\}\\
    &+u_{2}u_{3}[...]+u_{1}u_{3}[...].
    \end{aligned}\label{eq:long1}
\end{equation}

From the above expressions, we need to identify the ones which renormalize the gradient terms, i.e. $S_0[\phi_i]$. To this end, we assume we are close the the critical point, and all terms containing $\cos[2\beta_j\phi_j^s]$ are irrelevant. It's easy to see only the squared terms ($u_{1}^2$, $u_{2}^2$ and $u_{3}^2$) contribute to the renormalization of $S_0$. 
In comparing to $S_0$ we note that $\nabla_R\phi_j=(\partial_{R_x}\phi_j,\partial_{R_y}\phi_j)$ is symmetric with respect to interchanging $R_x$ and $R_y$. However, we see from above that with $v_j\neq \bar v$ we have anisotropy. Thus, we will skew the coordinates of $\br$ to $\tilde \br$, such that 
\begin{equation}
  \br\cdot \nabla_R\phi_j=x\partial_{R_x}\phi_j+y\partial_{R_y}\phi_j=\frac{x}{\sqrt{\tilde v_j}}\sqrt{\tilde v_j}\partial_{R_x}\phi_j+y\sqrt{\tilde v_j}\frac{\partial_{R_y}\phi_j}{\sqrt{\tilde v_j}}:=\tilde x \sqrt{\tilde v_j}\partial_{R_x}\phi_j+\tilde{y}\frac{\partial_{R_y}\phi_j}{\sqrt{\tilde v_j}}
\end{equation}
The measure is invariant under the change from $\br $ to $\tilde \br$ so we have, for instance, 
\begin{equation}
  \int d^2\br (\br\cdot \nabla_R\phi_j)^2 F_j(\br) = \int d^2\tilde \br (\tilde\br\cdot \tilde{\nabla_R\phi_j})^2 \tilde F_j(\tilde \br)
\end{equation}
where $\tilde F_j$ is given in Eq.\eqref{eq:tildeFJ}.
Introducing $\tilde g_i=g_i/\bar v$ such that $u_i=\tilde g_i/(\pi^2\alpha^2)$, we have for the gradient terms after a straightforward manipulation
\begin{equation}
    \begin{aligned}
        &\frac{1}{2}\left[1+\left({\tilde g}_{2}^2(A_1\beta_1^2+A_3\beta_3^2)+{\tilde g}_{3}^2(A_1\beta_1^2+A_2\beta_2^2)\right)\frac{\beta_1^2}{8}dl\right]\int d^2\br \left(\tilde v_1(\partial_x\phi_i)^2+\frac{1}{\tilde v_1}(\partial_y\phi_i)^2\right)\\
        +& \frac{1}{2}\left[1+\left({\tilde g}_{1}^2(A_2\beta_2^2+A_3\beta_3^2)+{\tilde g}_{3}^2(A_1\beta_1^2+A_2\beta_2^2)\right)\frac{\beta_2^2}{8}dl\right]\int d^2\br \left(\tilde v_2(\partial_x\phi_i)^2+\frac{1}{\tilde v_2}(\partial_y\phi_i)^2\right)\\
        +& \frac{1}{2}\left[1+\left({\tilde g}_{2}^2(A_1\beta_1^2+A_3\beta_3^2)+{\tilde g}_{1}^2(A_2\beta_2^2+A_3\beta_3^2)\right)\frac{\beta_3^2}{8}dl\right]\int d^2\br \left(\tilde v_3(\partial_x\phi_i)^2+\frac{1}{\tilde v_3}(\partial_y\phi_i)^2\right)
    \end{aligned}
\end{equation}
Transforming this correction to the cosine terms (by rescaling the fields $\phi_j$), we obtain the following RG equations 
\begin{equation}
    \begin{aligned}
        \frac{d\beta_1^2}{dl}&=-\left({\tilde g}_{2}^2(A_1\beta_1^2+A_3\beta_3^2)+{\tilde g}_{3}^2(A_1\beta_1^2+A_2\beta_2^2)\right)\frac{\beta_1^4}{8}\\
         \frac{d\beta_3^2}{dl}&=-\left({\tilde g}_{1}^2(A_2\beta_2^2+A_3\beta_3^2)+{\tilde g}_{3}^2(A_1\beta_1^2+A_2\beta_2^2)\right)\frac{\beta_2^4}{8}\\
         \frac{d\beta_3^2}{dl}&=-\left({\tilde g}_{2}^2(A_1\beta_1^2+A_3\beta_3^2)+{\tilde g}_{1}^2(A_2\beta_2^2+A_3\beta_3^2)\right)\frac{\beta_3^4}{8}\\
    \end{aligned}
\end{equation}
In terms of the Luttinger parameters, 
\begin{equation}
{
    \begin{aligned}
        \frac{d K_1}{dl}&=-2\pi^2K_1^2\left[{\tilde g}_{2}^2(A_1K_1+A_3K_3)+{\tilde g}_{3}^2(A_1K_1+A_2K_2)\right],\\
        \frac{d K_2}{dl}&=-2\pi^2K_2^2\left[{\tilde g}_{1}^2(A_2K_2+A_3K_3)+{\tilde g}_{3}^2(A_1K_1+A_2K_2)\right],\\
        \frac{d K_3}{dl}&=-2\pi^2K_3^2\left[{\tilde g}_{2}^2(A_1K_1+A_3K_3)+{\tilde g}_{1}^2(A_2K_2+A_3K_3)\right].\\
    \end{aligned} }
\end{equation}

Also note that the cross terms in Eq.\eqref{eq:long1}, give rise to additional renormalization to the interactions in addition to Eq.\eqref{eq:dudl}.
 \begin{equation}
 {
     \begin{aligned}
     \frac{d{\tilde g}_{1}}{dl}&=\left[2-(K_2+K_3)\right]{\tilde g}_{1}+4\pi A_1' {\tilde g}_{2}{\tilde g}_{3}K_1,\\
         \frac{d{\tilde g}_{2}}{dl}&=\left[2-(K_1+K_3)\right]{\tilde g}_{2}+4\pi A_2' {\tilde g}_{1}{\tilde g}_{3}K_2,\\
           \frac{d{\tilde g}_{3}}{dl}&=\left[2-(K_1+K_2)\right]{\tilde g}_{3}+4\pi A_3' {\tilde g}_{2}{\tilde g}_{1}K_3.
     \end{aligned} }
 \end{equation}
 where
\begin{equation}
    A_j\approx \frac{\pi }{\pi^4\alpha^4}\int_0^{1/\Lambda} dr r^3 \tilde F_j(r), ~~~ A_j'\approx\frac{2\pi}{\pi^2\alpha^2} \int_0^{1/\Lambda} dr r \tilde F_j(r).
\end{equation}
We can absorb the constants relating $\pi$ into the new definition of $A_j$ and $A_j'$, so we end up with
\begin{equation}
  \boxed{
    \begin{aligned}
        \frac{d K_1}{dl}&=-K_1^2\left[{\tilde g}_{2}^2(A_1K_1+A_3K_3)+{\tilde g}_{3}^2(A_1K_1+A_2K_2)\right],\\
        \frac{d K_2}{dl}&=-K_2^2\left[{\tilde g}_{1}^2(A_2K_2+A_3K_3)+{\tilde g}_{3}^2(A_1K_1+A_2K_2)\right],\\
        \frac{d K_3}{dl}&=-K_3^2\left[{\tilde g}_{2}^2(A_1K_1+A_3K_3)+{\tilde g}_{1}^2(A_2K_2+A_3K_3)\right],\\
        \frac{d{\tilde g}_{1}}{dl}&=\left[2-(K_2+K_3)\right]{\tilde g}_{1}+ A_1' {\tilde g}_{2}{\tilde g}_{3}K_1,\\
         \frac{d{\tilde g}_{2}}{dl}&=\left[2-(K_1+K_3)\right]{\tilde g}_{2}+ A_2' {\tilde g}_{1}{\tilde g}_{3}K_2,\\
           \frac{d{\tilde g}_{3}}{dl}&=\left[2-(K_1+K_2)\right]{\tilde g}_{3}+ A_3' {\tilde g}_{2}{\tilde g}_{1}K_3.
    \end{aligned} }
\end{equation}
where the constants $A_j$ and $A_j'$ are now given by
\begin{equation}
    A_j\approx \frac{2\pi^3 }{\pi^4\alpha^4}\int_0^{1/\Lambda} dr r^3 \tilde F_j(r), ~~~ A_j'\approx\frac{8\pi^2}{\pi^2\alpha^2} \int_0^{1/\Lambda} dr r \tilde F_j(r).
\end{equation}
In zero temperature limit, we can use Eq.\eqref{eq:tildeFJ} and obtain (setting $\alpha\Lambda=1$)
\begin{equation}
  A_j=\frac{1}{\pi^2(\alpha\Lambda)^4}\int_0^1 dx x^3 \tilde F_j(x)\approx 0.02+0.002\delta_j, ~~ A_j'=\frac{4}{\pi (\alpha \Lambda)^2}\int_0^1 dx x \tilde F_j(x)\approx0.56+0.035\delta_j.
\end{equation}
where we remind that $\delta_j=\tilde v_j-1$.

In Fig.\ref{fig:RG1} we show an example of the RG flows obtained by setting $\tilde g_{2}(0)=\tilde g_{3}(0)=0$ and $K_2(0)=K_3(0)$. In this case, $\tilde g_2$ and $\tilde g_3$ remain zero, $K_1$ does not flow, $K_2$ and $K_3$ remain identical which we denote by $K$. In the vicinity of $K\approx 1$ and $\tilde g_{1}\to0$, this is the Kosterlitz-Thouless RG flow.
\begin{figure}
    \includegraphics[width=8cm]{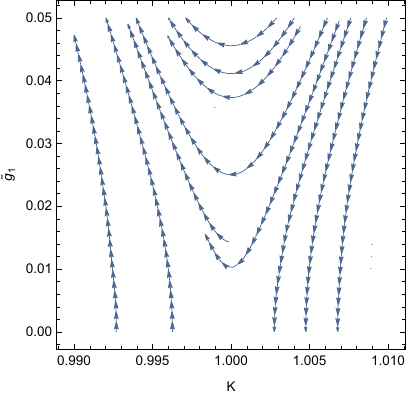}
    \caption{RG flows in the $\tilde g_1$-$K$ plane with the condition $K_2=K_3=K$ and $\tilde g_{2}=\tilde g_{3}=0$. }\label{fig:RG1}
\end{figure}

\section{V. Self-consistent Gaussian approximation for the massive phase} 
\label{sub:self_consistent_gaussian_approximation_for_the_massive_phase}
We now discuss when the three double-cosine terms in Eq.\eqref{eq:cosineterms} are relevant, the gap can be determined self-consistently in a variational approach, also known as self-consistent Gaussian approximation. Let's use the notations formulated in Eq.\eqref{eq:Si24} and\eqref{eq:SI25}. 

Deep inside the massive phase, we can approximate the action as some Gaussian form, expecting the soliton modes are excluded in the low energy limit. To this end, we write the variational Gaussian action as (for $i=1,2,3$)
\begin{equation}
    S_\text{var}[\phi_i]=\frac{T}{2L}\sum_{n,q}\sum_{i,j} G^{-1}_{ij}(q,\omega_n)\phi_i^*(q,\omega_n)\phi_j(q,\omega_n)
\end{equation}
where   
\begin{equation}
 G^{-1}(\bq)=
   \begin{pmatrix}
     \frac{v_1}{\pi K_1}\left(\frac{\omega_n^2}{v_1^2}+q^2+\frac{\Delta_1^2}{v_1^2}\right) & \Delta_{12} & \Delta_{13}\\
     \Delta_{21} & \frac{v_2}{\pi K_2}\left(\frac{\omega_n^2}{v_2^2}+q^2+\frac{\Delta_2^2}{v_2^2}\right)& \Delta_{23}\\
     \Delta_{31} & \Delta_{32} & \frac{v_3}{\pi K_3}\left(\frac{\omega_n^2}{v_3^2}+q^2+\frac{\Delta_3^2}{v_3^2}\right)
   \end{pmatrix}\label{eq:gap82}
\end{equation}
and $\Delta_i$ is the gap for the $i$-th sector.
We will assume that $\Delta_{ij}=\Delta_{ji}$ so $G^{-1}$ is a symmetric matrix which can be diagonalize by an orthogonal matrix $U$:
\begin{equation}
  \phi^\dagger G^{-1}\phi = \tilde \phi^\dagger G_d^{-1} \tilde \phi
\end{equation}
where 
\begin{equation}
  \phi = U \tilde \phi, ~~ G^{-1}=U G_d^{-1} U^{-1}
\end{equation}
and $G_d$ is now a diagonal matrix. 

The variational free energy to be minimized is,
\begin{equation}
    F_{\text{var}}=-T\ln Z_{\text{var}}+T\braket{S-S_\text{var}}_{\text{var}}\label{eq:83}
\end{equation}
where $Z_{\text{var}}$ is the partition function evaluated using $S_\text{var}$ and  the original $S$ is given in Eqs.\eqref{eq:Si24} and \eqref{eq:SI25}. The first term in Eq.\eqref{eq:83} can be evaluated directly, 
\begin{equation}
    -T\ln Z_\text{var}=-T\text{Tr}\ln G(q,\omega_n).
\end{equation}
Note in the second term, $\braket{S_\text{var}}_\text{var}$ does not contain dependence on the variational parameter, so it can be dropped.
We are then left with evaluating $\braket{S}_\text{var}$, which involves in evaluating the average of the  product of two cosine terms $g_i\cos(2\phi_j)\cos(2\phi_k)$ using the Gaussian distribution function. This requires one first identify the classical ground state of these cosine terms. In the case when there is no frustration, i.e. when $\sgn(g_1g_2g_3)>0$, the classical ground state is easy to find. When all $g_i>0$, all $\phi_i$ should be around $\phi_i=0$. when $g_i>0$ and $g_{j}, g_k<0$, we should have $\phi_{j,k}=0$ and $\phi_i=\pm\pi/2$. Since $\phi_i$ is around $\pm\pi/2$ we can shift $\phi_i$ by $\pm\pi/2$ such that $\cos2\phi_i$ picks up a minus sign, which can be absorbed in $g_j$ and $g_k$. Therefore, in the case without frustration, we can use the absolute value of $g_{i,j,k}$ and then average all the cosine terms at $\phi_{i,j,k}$ around $0$. 
\begin{equation}
    \begin{aligned}
        &\braket{S}_\text{var}=\sum_{i=1}^3\frac{v_i}{2\pi K_i}\sum_{n,q}\left(\frac{\omega_n^2}{v_i^2}+q^2\right)G_{i,i}(q,\omega_n)\\
        &-\frac{1}{2\pi^2\alpha^2}\frac{L}{T}\left(|g_{1}|\braket{e^{i2(\phi_2+\phi_3)}+e^{i2(\phi_2-\phi_3)}}_\text{var}+|g_{2}|\braket{e^{i2(\phi_3+\phi_1)}+e^{i2(\phi_3-\phi_1)}}_\text{var}+|g_{3}|\braket{e^{i2(\phi_1+\phi_2)}+e^{i2(\phi_1-\phi_2)}}_\text{var}\right).
    \end{aligned}\label{eq:109}
\end{equation}
Here we average the $\cos(2\phi_i)$ around $\phi_i=0$. Going to the $\tilde\phi$ basis, we have
\begin{equation}
  \braket{e^{i2(\phi_2\pm\phi_3)}}_\text{var}= \braket{e^{i2\sum_j(U_{2j}\pm U_{3j})\tilde\phi_j}}_\text{var}=e^{-\frac{2T}{L}\sum_{n,q}\sum_j(U_{2j}\pm U_{3j})^2 G_{d,j}(q,\omega_n)}.
\end{equation}
Since $G_d$ is diagonal, and $U^T=U^{-1}$, we have
\begin{equation}
  \braket{e^{i2(\phi_2\pm \phi_3)}}_\text{var}=e^{-\frac{2T}{L}\sum_{n,q}(G_{22}+G_{33}\pm2G_{23})}
\end{equation}
and similarly for other two terms in Eq.\eqref{eq:109}. 
Therefore
\begin{equation}
    \begin{aligned}
        &\braket{S}_\text{var}=\sum_{i=1}^3\frac{v_i}{2\pi K_i}\sum_{n,q}\left(\frac{\omega_n^2}{v_i^2}+q^2\right)G_{i,i}(q,\omega_n)-\frac{1}{\pi^2\alpha^2}\frac{L}{T}\times\\
        &\times\left(|g_{1}|e^{-\frac{2T}{L}\sum_{n,q}(G_{22}+G_{33})}\cosh[4\text{Tr}G_{23}]+|g_{2}|e^{-\frac{2T}{L}\sum_{n,q}(G_{11}+G_{33})}\cosh[4\text{Tr}G_{13}]+|g_{3}|e^{-\frac{2T}{L}\sum_{n,q}(G_{11}+G_{22})}\cosh[4\text{Tr}G_{12}]\right).
    \end{aligned}\label{eq:1}
\end{equation}
Here $\text{Tr}G_{ij}=\frac{T}{L}\sum_{q,n}G_{ij}(q,\omega_n)$.
To obtain $G$, we need to inverse the matrix $G^{-1}$. In the case when all $\Delta_{ij}$ are close to zero, we can approximately have 
\begin{equation}
 G(\bq)\approx
   \begin{pmatrix}
     \frac{\pi K_1}{v_1}\left(\frac{\omega_n^2}{v_1^2}+q^2+\frac{\Delta_1^2}{v_1^2}\right)^{-1} &  -\frac{\pi^2 K_1K_2}{v_1v_2}\frac{\Delta_{12}}{(\frac{\omega_n^2}{v_1^2}+q^2+\frac{\Delta_1^2}{v_1^2})(\frac{\omega_n^2}{v_2^2}+q^2+\frac{\Delta_2^2}{v_2^2})} & -\frac{\pi^2 K_1K_3}{v_1v_3}\frac{\Delta_{13}}{(\frac{\omega_n^2}{v_1^2}+q^2+\frac{\Delta_1^2}{v_1^2})(\frac{\omega_n^2}{v_3^2}+q^2+\frac{\Delta_3^2}{v_3^2})}\\
     -\frac{\pi^2 K_1K_2}{v_1v_2}\frac{\Delta_{12}}{(\frac{\omega_n^2}{v_1^2}+q^2+\frac{\Delta_1^2}{v_1^2})(\frac{\omega_n^2}{v_2^2}+q^2+\frac{\Delta_2^2}{v_2^2})} & \frac{\pi K_2}{v_2}\left(\frac{\omega_n^2}{v_2^2}+q^2+\frac{\Delta_2^2}{v_2^2}\right)^{-1}& -\frac{\pi^2 K_2K_3}{v_2v_3}\frac{\Delta_{23}}{(\frac{\omega_n^2}{v_2^2}+q^2+\frac{\Delta_2^2}{v_2^2})(\frac{\omega_n^2}{v_3^2}+q^2+\frac{\Delta_3^2}{v_3^2})}\\
     -\frac{\pi^2 K_1K_3}{v_1v_3}\frac{\Delta_{13}}{(\frac{\omega_n^2}{v_1^2}+q^2+\frac{\Delta_1^2}{v_1^2})(\frac{\omega_n^2}{v_3^2}+q^2+\frac{\Delta_3^2}{v_3^2})} & -\frac{\pi^2 K_2K_3}{v_2v_3}\frac{\Delta_{23}}{(\frac{\omega_n^2}{v_2^2}+q^2+\frac{\Delta_2^2}{v_2^2})(\frac{\omega_n^2}{v_3^2}+q^2+\frac{\Delta_3^2}{v_3^2})} & \frac{\pi K_3}{v_3}\left(\frac{\omega_n^2}{v_3^2}+q^2+\frac{\Delta_3^2}{v_3^2}\right)^{-1}
   \end{pmatrix}\label{eq:gap82}
\end{equation}
Setting the functional derivative $\delta F_{\text{var}}/\delta G_{ij}(q,\omega_n)$ to zero, we arrive at the following coupled equations
\begin{equation}
    \begin{aligned}
        G_{11}^{-1}(q,\omega_n)&=\frac{v_1}{\pi K_1}\left(\frac{\omega_n^2}{v_1^2}+q^2\right)+\frac{4}{\pi^2\alpha^2}\left(|g_{2}|e^{-\frac{2T}{L}\sum_{n,q}[G_{11}+G_{33}]}\cosh[4\text{Tr}G_{13}]+|g_{3}|e^{-\frac{2T}{L}\sum_{n,q}[G_{11}+G_{22}]}\cosh[4\text{Tr}G_{12}]\right),\\
        G_{22}^{-1}(q,\omega_n)&=\frac{v_2}{\pi K_2}\left(\frac{\omega_n^2}{v_2^2}+q^2\right)+\frac{4}{\pi^2\alpha^2}\left(|g_{1}|e^{-\frac{2T}{L}\sum_{n,q}[G_{22}+G_{33}]}\cosh[4\text{Tr}G_{23}]+|g_{3}|e^{-\frac{2T}{L}\sum_{n,q}[G_{11}+G_{22}]}\cosh[4\text{Tr}G_{12}]\right),\\
        G_{33}^{-1}(q,\omega_n)&=\frac{v_3}{\pi K_3}\left(\frac{\omega_n^2}{v_3^2}+q^2\right)+\frac{4}{\pi^2\alpha^2}\left(|g_{2}|e^{-\frac{2T}{L}\sum_{n,q}[G_{11}+G_{33}]}\cosh[4\text{Tr}G_{13}]+|g_{1}|e^{-\frac{2T}{L}\sum_{n,q}[G_{22}+G_{33}]}\cosh[4\text{Tr}G_{23}]\right),\\
        G_{12}^{-1}=\Delta_{12}&=\frac{4}{\pi^2\alpha^2}|g_{3}|e^{-\frac{2T}{L}\sum_{n,q}[G_{11}+G_{22}]}\sinh[4\text{Tr}G_{12}],\\
        G_{13}^{-1}=\Delta_{13}&=\frac{4}{\pi^2\alpha^2}|g_{2}|e^{-\frac{2T}{L}\sum_{n,q}[G_{11}+G_{33}]}\sinh[4\text{Tr}G_{13}],\\
        G_{23}^{-1}=\Delta_{23}&=\frac{4}{\pi^2\alpha^2}|g_{1}|e^{-\frac{2T}{L}\sum_{n,q}[G_{22}+G_{33}]}\sinh[4\text{Tr}G_{23}].\\
    \end{aligned}
\end{equation}
From the above equations, it is apparent that $\Delta_{ij}=0$ is always satisfied. Below we adopt this solution, and the equations then become
\begin{equation}
    \begin{aligned}
        G_{1}^{-1}(q,\omega_n)&=\frac{v_1}{\pi K_1}\left(\frac{\omega_n^2}{v_1^2}+q^2\right)+\frac{4}{\pi^2\alpha^2}\left(|g_{2}|e^{-\frac{2T}{L}\sum_{n,q}[G_1(q,\omega_n)+G_3(q,\omega_n)]}+|g_{3}|e^{-\frac{2T}{L}\sum_{n,q}[G_1(q,\omega_n)+G_2(q,\omega_n)]}\right)\\
        G_{2}^{-1}(q,\omega_n)&=\frac{v_2}{\pi K_2}\left(\frac{\omega_n^2}{v_2^2}+q^2\right)+\frac{4}{\pi^2\alpha^2}\left(|g_{1}|e^{-\frac{2T}{L}\sum_{n,q}[G_2(q,\omega_n)+G_3(q,\omega_n)]}+|g_{3}|e^{-\frac{2T}{L}\sum_{n,q}[G_1(q,\omega_n)+G_2(q,\omega_n)]}\right)\\
        G_{3}^{-1}(q,\omega_n)&=\frac{v_3}{\pi K_3}\left(\frac{\omega_n^2}{v_3^2}+q^2\right)+\frac{4}{\pi^2\alpha^2}\left(|g_{2}|e^{-\frac{2T}{L}\sum_{n,q}[G_1(q,\omega_n)+G_3(q,\omega_n)]}+|g_{1}|e^{-\frac{2T}{L}\sum_{n,q}[G_2(q,\omega_n)+G_3(q,\omega_n)]}\right)\\
    \end{aligned}
\end{equation}
We can insert Eq.\eqref{eq:gap82} to obtain equations for $\Delta_i$. We do this in the zero temperature limit, where
\begin{equation}
    -\frac{2T}{L}\sum_{n,q}G_i(q,\omega_n)=-2\pi K_i\int \frac{d^2\bq}{(2\pi)^2}\frac{1}{\bq^2+m_i^2}\approx -\frac{K_i}{2}\ln\left(1+\frac{\Lambda^2}{m_i^2}\right)
\end{equation}
where $\Lambda$ is some UV cutoff for the momentum. Making use of this result, the equations for $\Delta_i$'s are
\begin{equation}
    \begin{aligned}
        \delta_1^2&=\frac{4K_1}{\pi v_1(\alpha\Lambda)^2}\left[|g_{2}|\left(\frac{\delta_1^2}{1+\delta_1^2}\right)^{\frac{K_1}{2}}\left(\frac{\delta_3^2}{1+\delta_3^2}\right)^{\frac{K_3}{2}}+|g_{3}|\left(\frac{\delta_1^2}{1+\delta_1^2}\right)^{\frac{K_1}{2}}\left(\frac{\delta_2^2}{1+\delta_2^2}\right)^{\frac{K_2}{2}}\right]\\
        \delta_2^2&=\frac{4K_2}{\pi v_2(\alpha\Lambda)^2}\left[|g_{1}|\left(\frac{\delta_2^2}{1+\delta_2^2}\right)^{\frac{K_2}{2}}\left(\frac{\delta_3^2}{1+\delta_3^2}\right)^{\frac{K_3}{2}}+|g_{3}|\left(\frac{\delta_1^2}{1+\delta_1^2}\right)^{\frac{K_1}{2}}\left(\frac{\delta_2^2}{1+\delta_2^2}\right)^{\frac{K_2}{2}}\right]\\
        \delta_3^2&=\frac{4K_3}{\pi v_3(\alpha\Lambda)^2}\left[|g_{2}|\left(\frac{\delta_1^2}{1+\delta_1^2}\right)^{\frac{K_1}{2}}\left(\frac{\delta_3^2}{1+\delta_3^2}\right)^{\frac{K_3}{2}}+|g_{1}|\left(\frac{\delta_2^2}{1+\delta_2^2}\right)^{\frac{K_2}{2}}\left(\frac{\delta_3^2}{1+\delta_3^2}\right)^{\frac{K_3}{2}}\right]\\
    \end{aligned}
\end{equation}
where we have used $\delta_i=\Delta_i/(v_i\Lambda)$. 
We usually deal with the case that $\delta_i\ll1$, so the above equations become (setting $\alpha\Lambda=1$)
\begin{equation}
    \begin{aligned}
        \delta_1^2&=\frac{4K_1}{\pi v_1}|\delta_1|^{K_1}\left(|g_{2}||\delta_3|^{K_3}+|g_{3}||\delta_2|^{K_2}\right),\\
        \delta_2^2&=\frac{4K_2}{\pi v_2}|\delta_2|^{K_2}\left(|g_{1}||\delta_3|^{K_3}+|g_{3}||\delta_1|^{K_1}\right),\\
        \delta_3^2&=\frac{4K_3}{\pi v_3}|\delta_3|^{K_3}\left(|g_{2}||\delta_1|^{K_1}+|g_{1}||\delta_2|^{K_2}\right).\\
    \end{aligned}\label{eq:withoutFrustration}
\end{equation}
It is apparent from these equations that $\Delta_i=0$ is always a trivial solution. 

The cases with frustration, i.e. when $\sgn(g_1g_2g_3)<0$, are more involved, due to the possible presence of classically degenerate ground states. For instance, in the case when $g_1=g_2=g_3<0$, there exist several lines in the parameter space which correspond to the degenerate classical ground state. However, as we see from the RG analysis, this accidental symmetric case $g_1=g_2=g_3$ is not protected by any symmetry and is unstable, so we generally do not expect that they are all equal. Therefore, without loss of generality, we assume $|g_1|>|g_2|>|g_3|$. In this case, the classical ground state is unique. For instance, when all $g_{1,2,3}<0$, the classical ground state is $\phi_{1,2}=0$ and $\phi_3=\pm\pi/2$. After shifting $\phi_3$ by $\pm\pi/2$ we can still average every field around $0$. The derivation of the gap equations is parallel to outlined above, with the only difference being the replacement of $|g_3|$ with $-|g_3|$. Therefore, we obtain the gap equations for the case with frustration
\begin{equation}
    \begin{aligned}
        \delta_1^2&=\frac{4K_1}{\pi v_1}|\delta_1|^{K_1}\left(|g_{2}||\delta_3|^{K_3}-|g_{3}||\delta_2|^{K_2}\right),\\
        \delta_2^2&=\frac{4K_2}{\pi v_2}|\delta_2|^{K_2}\left(|g_{1}||\delta_3|^{K_3}-|g_{3}||\delta_1|^{K_1}\right),\\
        \delta_3^2&=\frac{4K_3}{\pi v_3}|\delta_3|^{K_3}\left(|g_{2}||\delta_1|^{K_1}+|g_{1}||\delta_2|^{K_2}\right).\\
    \end{aligned}\label{eq:withFrustration}
\end{equation}

\begin{figure}
  \includegraphics[width=18cm]{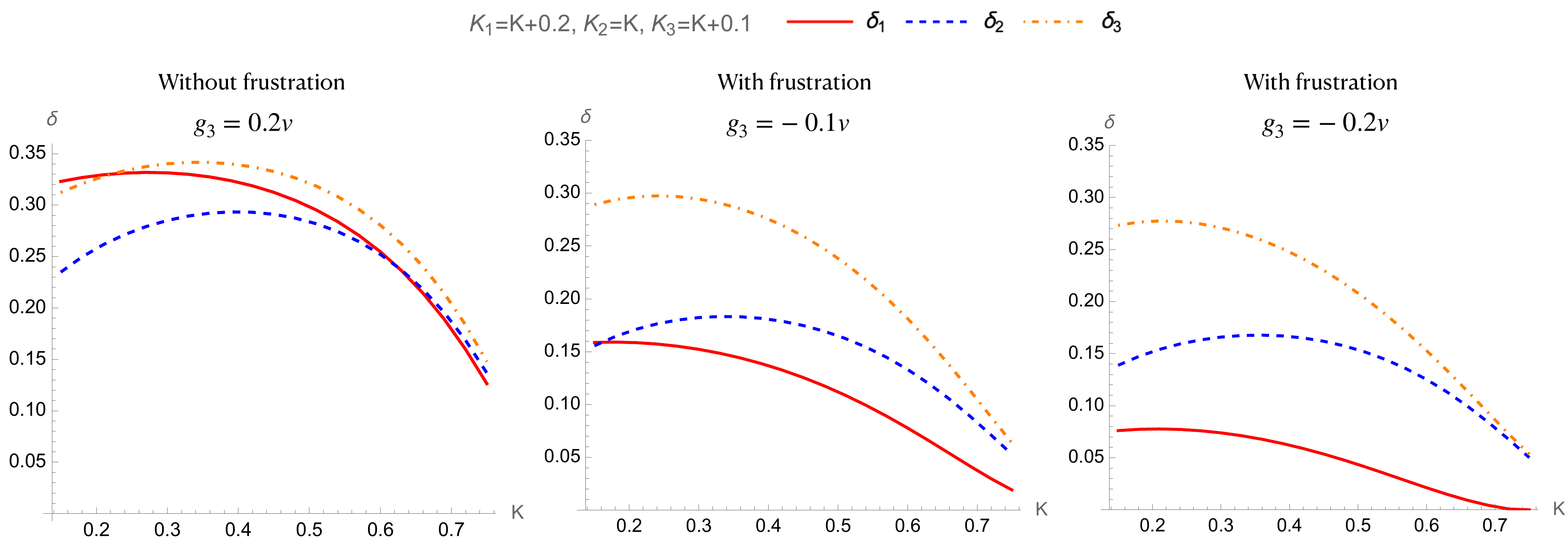}
  \caption{Solution of the gap equations with and without frustration. Here we choose $v_{1,2,3}=v$ and $g_1=0.3v$, $g_2=0.25v$. The Luttinger parameters are shown in the figure.   In the case without frustration (left panel and solved from Eq.\eqref{eq:withoutFrustration}), $\Delta_{1,2,3}$ are comparable to each other. In the case with frustration (middle and right panels, solved from Eq.\eqref{eq:withFrustration}), all gaps get suppressed compared to the  case without frustration. In particular,  $\Delta_1$ gets suppressed the most and when $|g_3|$ gets larger, it becomes the smallest compared to $\Delta_2$ and $\Delta_3$.  }\label{fig:frustration}
\end{figure}

In Fig.\eqref{fig:frustration}, we shown the comparison between the solutions of Eq.\eqref{eq:withoutFrustration} and Eq.\eqref{eq:withFrustration}. We choose the parameters $v_{1,2,3}=v$, $g_1=0.3v$, $g_2=0.25v$, $K_1-0.2=K_3-0.1=K_2=K$ such that in the case without frustration all induced gaps are comparable. We clearly see from the middle and right panels of Fig.\eqref{fig:frustration} that in the presence of frustration, all gaps get suppressed compared to the case without frustration. $\Delta_1$ gets suppressed the most and when $|g_3|$ gets as large as $0.2v$, it becomes much smaller than the other two and therefore a gap hierarchy develops. 

\section{VI. Transport properties} 
\label{sub:transport_property}
We first show if only the CDW order is considered in the inter-wire coupling, there is no transverse current. To see this, we note the inter-wire coupling Hamiltonian for CDW order is  
\begin{equation}
   \begin{aligned}
        H_c&=J\sum_{\braket{ij}} \int dx O_{\text{CDW}}^\dagger(i,x)O_{\text{CDW}}(j,x)+\text{h.c.}\\
        &= J\sum_{\braket{ij},\lambda\lambda'}\int dx \psi^\dagger_{i,-,\lambda}(x)\psi_{i,+,\lambda}(x)\psi^\dagger_{j,+,\lambda'}(x)\psi_{j,-,\lambda'}(x) + (i\leftrightarrow j).
   \end{aligned}\label{eq:interchainH}
\end{equation}
which is the most relevant one when the system is in the massive phase with gaps in all $\phi_2,\phi_3,\phi_4$ sectors and in the presence of strong repulsion.  
From this Hamiltonian, one can calculate the current density operator from the continuity equation $\partial_y j_y =i[\rho(i,x),H_c]=0$ and find the current vanishes.  Note the vanishing is local, i.e. it's not because two current operators on $i$ and $j$ cancel each other, but rather the current on each wire vanishes. 
Therefore, we see that by only turning on the CDW coupling is not enough for obtaining a finite cross-wire conductivity, we must have other couplings. 

The most trivial one, will be the single particle tunneling (keeping only the nearest neighbor),
\begin{equation}
    H_t=t_\bot\sum_{i,rr',\lambda} \int dx \psi_{r,\lambda}^\dagger(i,x)\psi_{r',\lambda}(i+1,x)+\text{h.c.}
\end{equation}
The tunneling charge current due to the presence of this Hamiltonian is given by the continuity equation, and we obtain
\begin{equation}
    \begin{aligned}
        j_y(i,x)&=i c t_\bot\sum_{r,\lambda}\left(\psi^\dagger_{r,\lambda}(i,x)\psi_{r,\lambda}(i+1,x)-\psi^\dagger_{r,\lambda}(i+1,x)\psi_{r,\lambda}(i,x)\right),\\
        j_y(q_y,x)&=-2 c t_\bot\sum_{r,\lambda}\int_{-\pi}^\pi dk_y \psi^\dagger_{r,\lambda} (k_y-\frac{q_y}{2},x)\psi_{r,\lambda}(k_y+\frac{q_y}{2},x)\sin k_y e^{i\frac{q_y}{2}},
    \end{aligned}
\end{equation}
where $c$ is the spacing between adjacent wires, and since $j_y(i,x)$ is Hermitian, $j_y(q_y,x)^\dagger=j_y(-q_y,x).$
Below, we will evaluate the current-current correlation in the Hamiltonian without $H_t$, similar to the approach in Ref.\cite{Moser1998}. However, we will keep $H_c$ to the first order as in a perturbation theory. According to Kubo formula, the real part of conductivity is given by
\begin{equation}
    \text{Re}\sigma_{\bot}(\omega)=-\frac{e^2}{\omega}\text{Im}\left[\lim_{q_y\to0}\int dx \Pi_{\bot}^R(q_y,x,\omega)\right]
\end{equation}
where $\Pi_{\bot}^R(q_y,x,\omega)$ is the Fourier transform of the retarded current-current correlation function, which can be obtained from 
\begin{equation}
    \Pi_{\bot}(q_y,x,i\omega_n)=\braket{j_y(q_y,x,i\omega_n)j_y(-q_y,0,-i\omega_n)}
\end{equation}
by analytic continuation. Here we have assumed translation symmetry. To the first order in $H_c$, we have
\begin{equation}
    \begin{aligned}
        &\lim_{q_y\to0}\int dx \Pi_{\bot}(q_y,x,i\omega_n) = \frac{1}{\beta}\sum_{ik_n}\sum_{r,\lambda}\int\frac{ dk_x dk_y}{(2\pi)^2} 4c^2 t_\bot^2 \sin^2k_y G_{r,\lambda}(k_x,ik_n)G_{r,\lambda}(k_x,ik_n+i\omega_n)\\
        +&\frac{J}{\beta^2}\sum_{ik_n,ip_n}\sum_{r,\lambda}\int \frac{d^2\bk d^2\bp}{(2\pi)^4} 4c^2t_\bot^2 \sin^2k_y\sin^2p_yG_{r,\lambda}(k_x,ik_n)G_{r,\lambda}(k_x,ik_n+i\omega_n)G_{-r,\lambda}(p_x,ip_n)G_{-r,\lambda}(p_x,ip_n+i\omega_n).\label{eq:jjPi1}
    \end{aligned}
\end{equation}
Here we used the fact that all the Green's functions are evaluated in the decoupled Hamiltonian, so these are essentially 1D Hamiltonians and do not depend on $k_y$ or the wire index. Note in the second line of Eq.\eqref{eq:jjPi1} the difference in the $r$-index between these Green's functions are due to the nature of the CDW interaction $H_c$. Upon expressing the Green's functions using their spectral representations
\begin{equation}
  G(k_x,ik_n)=\frac{1}{2\pi}\int d\varepsilon \frac{A(k_x,\varepsilon)}{ik_n-\varepsilon}, ~~A(k_x,\varepsilon)=-2\text{Im}G^R(k_x,\varepsilon)
\end{equation}
and analytically continuing on the real frequency axis by changing $i\omega_n$ to $\omega+i0^+$, we obtain (after summing over $ik_n$)
\begin{equation}
    \begin{aligned}
        &\text{Re}\sigma_{\bot}(\omega)=\frac{e^2ct_\bot^2}{2\pi}\sum_{r,\lambda}\int dx \int d\varepsilon \frac{n_F(\varepsilon)-n_F(\varepsilon+\omega)}{\omega}A_{r,\lambda}(x,\varepsilon)A_{r,\lambda}(-x,\varepsilon+\omega)\\
        &+\frac{e^2Jt_\bot^2}{8\pi^3}\sum_{r,\lambda}\int dx dx'P\int d\epsilon d\epsilon'\frac{n_F(\epsilon)-n_F(\epsilon')}{\omega+\epsilon-\epsilon'}A_{r,\lambda}(x,\epsilon)A_{r,\lambda}(-x,\epsilon')\int d\varepsilon \frac{n_F(\varepsilon)-n_F(\varepsilon+\omega)}{\omega}A_{-r,\lambda}(x',\varepsilon)A_{-r,\lambda}(-x',\varepsilon+\omega),
    \end{aligned}
\end{equation}
where in the second line we use $P\int$ to denote the principle value integral.
To proceed with obtaining the temperature , we need the spectral function for the decoupled 1D system. But here we can estimate the magnitude of $\sigma_\bot$ from the its prefactors (after restoring $\hbar$ everywhere),
\begin{equation}
   \sigma_\bot(T)=\frac{e^2}{h} 4(ack_F^2)\left(\frac{t_\bot}{E_F}\right)^2 F\left(\frac{T}{E_F}\right)\approx 18\frac{e^2}{h} \left(\frac{t_\bot}{E_F}\right)^2 F\left(\frac{T}{E_F}\right) ~~\text{for}~~ k_F\approx \frac{\pi}{2a}\label{eq:singleP_current}
 \end{equation} 
 where $F$ is some dimensionless function that comes from the integral involving the spectral functions. 

\subsection{A. Single-particle tunneling in the gapless case}
We begin by discussing the conductivity due to the single particle tunneling process in the gapless case, i.e. all the four $\phi_j$ fields are gapless. As shown above, this requires the calculation of the single particle spectral function. 
By definition, the retarded single particle Green's function is given by
\begin{equation}
    G^R_{r,\lambda}(x,t)=-i\Theta(t)\braket{\{\psi_{r,\lambda}(x,t),\psi^\dagger_{r,\lambda}(0,0)\}}=-i\Theta(t)(G_{r,\lambda}(x,t)+G_{r,\lambda}(-x,-t))
\end{equation}
where
\begin{equation}
    G_{r,\lambda}(x,t)=\lim_{\alpha\to0}\frac{e^{irk_Fx}}{2\pi\alpha}\braket{e^{i\theta_\lambda(x,t)-ir\phi_\lambda(x,t)}e^{-i\theta_\lambda(0,0)+ir\phi_\lambda(0,0)}}
\end{equation}
according to Eq.\eqref{eq:bosonizationF}. Without loss of generality, we take $r=+$ and $\lambda=\uparrow u$. Using Eq.\eqref{eq:inv_phii} we have
\begin{equation}
    G_{r,\lambda}(x,t)=\lim_{\alpha\to0}\frac{e^{irk_Fx}}{2\pi\alpha}
    \prod_{j=0}^3\braket{e^{\frac{i}{2}[\theta_j(x,t)-\phi_j(x,t)]}e^{-\frac{i}{2}[\theta_j(0,0)-\phi_j(0,0)]}}_j
\end{equation}
For Gaussian models we have $\braket{e^{iA}e^{-iB}}=\exp\left(-\frac{1}{2}\braket{A^2}-\frac{1}{2}\braket{B^2}+\braket{AB}\right)$, thus
\begin{equation}
    \braket{e^{\frac{i}{2}[\theta_j(x,t)-\phi_j(x,t)]}e^{-\frac{i}{2}[\theta_j(0,0)-\phi_j(0,0)]}}_j=\exp\left[\frac{1}{4}\braket{\left(\theta_j(x,t)-\phi_j(x,t)\right)\left(\theta_j(0,0)-\phi_j(0,0)\right)}_j-\frac{1}{4}\braket{(\theta_j-\phi_j)^2}_j\right].
\end{equation}
Using Eq.\eqref{eq:Green1}, \eqref{eq:Green2} and \eqref{eq:Green3}, it is easy to see
\begin{equation}
     \braket{e^{\frac{i}{2}[\theta_j(x,t)-\phi_j(x,t)]}e^{-\frac{i}{2}[\theta_j(0,0)-\phi_j(0,0)]}}_j=\left(\frac{a}{a+i(v_jt-x)}\right)^{1/4}\left(\frac{a^2}{x^2+(iv_jt+a)^2}\right)^{\frac{K_j+\frac{1}{K_j}-2}{16}}.
\end{equation}
As a result, we have,
\begin{equation}
     G_{+,\uparrow A}(x,t)=\frac{e^{ik_Fx}}{2\pi}\prod_{j=0}^3\left(\frac{1}{a+i(v_jt-x)}\right)^{1/4}\left(\frac{a^2}{x^2+(iv_jt+a)^2}\right)^{\gamma_j/2}, ~~~ \gamma_j=\frac{K_j+\frac{1}{K_j}-2}{8}\label{eq:totalGreen}
\end{equation}
In fact, this form does not depend on $\lambda$. For $r=-$, the only changes is to replace $x$ with $-x$ in the above expression. 

For comparison, we note in the conventional two-component case, we have
\begin{equation}
  G(x,t)\sim\prod_{j=c,s}\left(\frac{1}{a+i(v_jt-x)}\right)^{1/2}\left(\frac{a^2}{x^2+(iv_jt+a)^2}\right)^{\gamma_j/2}, ~~~ \gamma_j=\frac{K_j+\frac{1}{K_j}-2}{4}.
\end{equation}

Using this result, and if we neglect the correction due to the CDW order coupling, we have 
\begin{equation}
  \begin{aligned}
    \text{Re}\sigma_{\bot}(\omega\to0, T)&\sim\int dx \int d\varepsilon \frac{1}{4T}\text{sech}^2\left(\frac{\varepsilon}{2T}\right)A(x,\varepsilon)A(-x,\varepsilon+\omega)\\
    &\sim \int dx \int d\varepsilon \frac{1}{4T}\text{sech}^2\left(\frac{\varepsilon}{2T}\right) F(x,\varepsilon)
  \end{aligned}
\end{equation}
where $F(x,\varepsilon)$ obeys
\begin{equation}
  F\left(a x, \frac{\varepsilon}{a}\right)=\frac{1}{a^{2\gamma}}F(x,\varepsilon), ~~~ \gamma=\sum_j \gamma_j.
\end{equation}
The spectral function $A(x,\varepsilon)$ has the following properties:
\begin{equation}
  A(x=0,\varepsilon)=N(\varepsilon)\sim |\varepsilon|^\gamma, ~~~ A(x\gg 1/\varepsilon,\varepsilon) \sim \frac{\cos{\varepsilon x}}{x^{\gamma}},
\end{equation}
so the function $F(x,\varepsilon)$ behaves like
\begin{equation}
  \begin{aligned}
    F(x,\varepsilon)&\sim \varepsilon^{2\gamma}~\text{for}~ x\ll 1/\varepsilon,\\
    &\sim \frac{1}{x^{2\gamma}}~\text{for}~ x\gg 1/\varepsilon.
  \end{aligned}
\end{equation}
Therefore
\begin{equation}
  \text{Re}\sigma_{\bot}(\omega\to0, T)\sim  \int d\varepsilon \frac{1}{4T}\text{sech}^2\left(\frac{\varepsilon}{2T}\right) \left(\int_{0}^{1/\varepsilon} dx \varepsilon^{2\gamma}+\int_{1/\varepsilon}^{+\infty} dx \frac{\cos{2\varepsilon x}}{x^{2\gamma}}\right)\sim T^{2\gamma-1}.\label{eq:sigma1}
\end{equation}
In the main text, we introduce the scaling dimension of the single particle hopping $\alpha_e$, which is related to $\gamma$ simply by $\gamma=\alpha_e-1$.

\subsection{B. Single-particle tunneling in the gapped case}
When there is a gap opening in either one or more of the $\phi_j$ fields, the total single particle density of states has a gap structure. This is because the total real pace Green's function is a product of the four component, as shown in Eq.\eqref{eq:totalGreen}. Consequently, the spectral function (and hence the density of states),  which is the Fourier transform of Eq.\eqref{eq:totalGreen}, is a convolution of different components. It is this convolution that leads to a gap in the total spectral function. For two component 1DEG this was demonstra{}ted in Ref.\cite{Voit1998}
If there is a gap in the total density of states, the conductivity becomes zero when $T$ is below the gap energy scale. However, when $T$ is well above the energy scale set by the gap, Eq.\eqref{eq:sigma1} still applies.

\subsection{C. Josephson tunneling in the gapped case}
Unlike the single particle tunneling, the pair or quadruple hopping process does not depend on the single particle density of states, and can contribute to the cross wire conductivity even in the gapped phase and when $T$ not well above the gap energy scale. 
Below we discuss this Josephson tunneling due to charge-2e and charge-4e orders.
\subsubsection{1. charge-2e superconductivity}
We take spin-triplet pseudospin-triplet pairing order parameter in Eq.\eqref{eq:orderparameter_TS} as an example. Assuming we are in the partially gapped phase when $\phi_{2,3}$ are gapped.  The tunneling is governed by the Hamiltonian 
\begin{equation}
     \begin{aligned}
         H_J&=J_s\sum_{\braket{y,y'}} \int dx O_{T_1,S}^\dagger(x,y)O_{T_1,S}(x,y')+\text{h.c.}\\
         &= \tilde J_s\sum_{\braket{y,y'}} \int dx \cos\left[\theta_0(x,y)+\theta_1(x,y)-\theta_0(x,y')-\theta_1(x,y')\right],
     \end{aligned}
 \end{equation} 
 where $\tilde J_s=J_s \delta_2^{K_2/2}\delta_3^{K_3/2}$ and $J_s\sim t_\bot^2/E_F$.
 On the other hand, the local charge density operator is given by $\rho_0(x,y)=\frac{1}{\pi}\partial_x\phi_0(x,y)$. Then according to the continuity equation, the transverse current density is found to be 
 \begin{equation}
     j_y(x)=\tilde J_s c \sin[\theta_0(x,y+1)+\theta_1(x,y+1)-\theta_0(x,y)-\theta_1(x,y)]
 \end{equation}
 which is the Josephson current. To evaluate the current-current correlation in the decoupled 1D system, we note that only the correlation involving two current operators with the same $y$ survives. Thus we have
 \begin{equation}
    \begin{aligned}
         \Pi_{\bot}(x,y,\tau)&=\tilde J_s^2c^2\braket{\sin[\theta_0(x,y+1)+\theta_1(x,y+1)-\theta_0(x,y)-\theta_1(x,y)]\sin[\theta_0(0,0+1)+\theta_1(0,0+1)-\theta_0(0,0)-\theta_1(0,0)]}\\
         &=\frac{\tilde J_s^2c^2}{2}\left(\frac{a^2}{x^2+v_0^2\tau^2+a^2}\right)^{\frac{1}{2K_0}+\frac{1}{2K_1}}\delta_{y,0}.
    \end{aligned}
 \end{equation}
 Therefore
 \begin{equation}
     \Pi_{\bot}(\bq\to0,\omega)=\int dxdt e^{i\omega t}\sum_y \Pi_{\bot}(x,y,t)\sim \tilde J_s^2c^2\omega^{1/K_0+1/K_1-2}
 \end{equation}
and 
\begin{equation}
    \text{Re}\sigma(\omega)\sim \tilde J_s^2c^2\omega^{1/K_0+1/K_1-3} \to \text{Re}\sigma(T)\sim \tilde J_s^2c^2T^{1/K_0+1/K_1-3}.
\end{equation}

\subsubsection{2. charge-4e superconductivity}
Like the charge-2e superconductivity case, the tunneling current is determined by the coupling Hamiltonian,
\begin{equation}
     \begin{aligned}
         H_J&=\tilde J_s\sum_{\braket{y,y'}} \int dx \cos\left[2\theta_0(x,y)-2\theta_0(x,y')\right].
     \end{aligned}
 \end{equation} 
The resulting current operator is 
 \begin{equation}
     j_y(x)=\tilde J_s c \sin[2\theta_0(x,y+1)-2\theta_0(x,y)].
 \end{equation}
The calculation of conductivity is exactly the same as the 2e-SC case, with the modification of $\theta_0\to2\theta_0$ and hence $1/K_0\to4/K_0$, and the final result is
\begin{equation}
  \text{Re}\sigma(T)\sim \tilde J_s^2c^2T^{4/K_0-3}
\end{equation}
\bibliography{tBG}

\end{widetext}